\pdfoutput=1
\documentclass[PAPER, atlasdraft=false, coverpage=false, texlive=2023, UKenglish, cernpreprint, texmf, orcidlogo]{atlasdoc}
\usepackage[subcaption=true,block=none]{atlaspackage}
\usepackage{atlasbiblatex}

\usepackage[particle=true]{atlasphysics}
\graphicspath{{logos/}{figures/}}

\usepackage{ANA-EGAM-2021-01-PAPER-defs}
\usepackage{chngpage}

\AtlasTitle{Electron and photon efficiencies in LHC Run 2 with the ATLAS experiment}

\AtlasAbstract{%
Precision measurements of electron reconstruction, identification, and isolation efficiencies and photon identification efficiencies are presented.
They use the full Run 2 data sample collected by the ATLAS experiment in $pp$ collisions at a centre-of-mass energy of 13 TeV during the years 2015--2018, corresponding to an integrated luminosity of 139~$\mathrm{fb}^{-1}$.
The measured electron identification efficiencies have uncertainties that are around 30\%--50\% smaller than the previous Run 2 results due to an improved methodology and the inclusion of more data. A better pile-up subtraction method leads to electron isolation efficiencies that are more independent of the amount of pile-up activity.
Updated photon identification efficiencies are also presented, using the full Run 2 data.
When compared to the previous measurement, a 30\%--40\% smaller uncertainty is observed on the photon identification efficiencies, thanks to the increased amount of available data.
}

\author{The ATLAS Collaboration}

\AtlasRefCode{EGAM-2021-01}

\PreprintIdNumber{CERN-EP-2023-182}

\AtlasDate{\today}

\AtlasJournalRef{JHEP 05 (2024) 162}
\AtlasDOI{10.1007/JHEP05(2024)162}

\hypersetup{pdftitle={ATLAS document},pdfauthor={The ATLAS Collaboration}}

\begin{document}

\maketitle

\tableofcontents
\clearpage

\section{Introduction}

The ATLAS detector~\cite{PERF-2007-01,ATLAS-TDR-2010-19,PIX-2018-001} at the Large Hadron Collider (LHC)~\cite{Evans:2008zzb} is used to investigate proton--proton collisions at the highest energies. After the discovery of the Higgs boson in 2012 using four-lepton and diphoton final states~\cite{HIGG-2012-27,CMS-HIG-12-036} and following the deeper understanding of the detector and the hadron collider environment,
precision measurements of Standard Model parameters are increasingly viable. Measurements of the weak mixing angle~\cite{STDM-2011-34}, the $W$ boson mass~\cite{STDM-2014-18} and related cross-sections~\cite{STDM-2012-06,STDM-2016-04} have proven to be competitive to prior determinations at the LEP or Tevatron colliders. In addition, precision measurements of the Higgs boson properties have become an active area of research~\cite{HIGG-2022-04}. These experimentally challenging precision measurements rely on understanding and controlling the systematic uncertainties of the object reconstruction in the detector extremely well. Electrons and photons are pivotal final state objects in these measurements. This paper describes the performance of electron and photon identification achieved for precision measurements in the LHC Run~2 data taking period (2015--2018).

The methodology to determine the efficiency of electron identification and isolation selection has been significantly updated with respect to what was used for previous publications~\cite{PERF-2016-01,PERF-2017-01,EGAM-2018-01}, as have the related factors to correct the detector simulation to accurately reflect the data efficiencies. The electron reconstruction efficiencies have been measured using the full Run 2 data. The measurements of the photon efficiencies have been updated to profit from the increased statistics. The results presented here are intended to be the final and most precise for the $pp$ collision data taken by the ATLAS experiment during LHC Run~2. They cover the full $\sqrt{s}=13$ \TeV{} data sample and also include the 2018 data, which is 70\% larger than 2015--2017 data sample used for the previous results~\cite{PERF-2017-01,EGAM-2018-01}.

Data samples of electrons and photons are recorded using electron and photon triggers whose performance during LHC Run~2 is detailed in Ref.~\cite{TRIG-2018-05}. Reconstruction algorithms and energy calibrations are described in Refs.~\cite{PERF-2017-03,EGAM-2018-01}. The work presented here is based on these triggering, reconstruction and calibration algorithms. Section~\ref{sec:detector} introduces the ATLAS detector. A short reminder of how electron and photon objects are reconstructed from the detector signals is given in Section~\ref{sec:shortreco}, followed by an overview of the data samples used in Section~\ref{sec:data_sets}. The general definition of the efficiency measurement along with a description of the changes in methodology are found in Section~\ref{sec:eff_measurement}, with results on electron reconstruction and identification efficiencies given in Section~\ref{sec:ele_reco} and~\ref{sec:ele_id}, respectively.
Section~\ref{sec:rej} reports the rejections against backgrounds achieved with the electron reconstruction and identification requirements.
Improvements to the pile-up subtraction for the calorimeter isolation and the resulting electron isolation efficiencies are described in Section~\ref{sec:ele_iso}.
The photon reconstruction efficiency has not been remeasured and is reported in Ref.~\cite{EGAM-2018-01}. The photon identification efficiencies can be found in Section~\ref{sec:ph_id}. This paper concludes with Section~\ref{sec:conclusion}.


%

%
\section{ATLAS detector}
\label{sec:detector}

\newcommand{\AtlasCoordFootnote}{%
ATLAS uses a right-handed coordinate system with its origin at the nominal interaction point (IP)
in the centre of the detector and the \(z\)-axis along the beam pipe.
The \(x\)-axis points from the IP to the centre of the LHC ring,
and the \(y\)-axis points upwards.
Cylindrical coordinates \((r,\phi)\) are used in the transverse plane,
\(\phi\) being the azimuthal angle around the \(z\)-axis.
The pseudorapidity is defined in terms of the polar angle \(\theta\) as \(\eta = -\ln \tan(\theta/2)\).
Angular distance is measured in units of \(\Delta R \equiv \sqrt{(\Delta\eta)^{2} + (\Delta\phi)^{2}}\). The transverse energy is $\et = E/\cosh(\eta)$.}

The ATLAS detector~\cite{PERF-2007-01,ATLAS-TDR-2010-19,PIX-2018-001} at the LHC covers nearly the entire solid angle around the collision point.\footnote{\AtlasCoordFootnote}
It consists of an inner tracking detector surrounded by a thin superconducting solenoid, electromagnetic and hadron calorimeters,
and a muon spectrometer incorporating three large superconducting air-core toroidal magnets.

The inner-detector system (ID) is immersed in a \SI{2}{\tesla} axial magnetic field
and provides charged-particle tracking in the range of \(|\eta| < 2.5\).
The high-granularity silicon pixel detector covers the vertex region and typically provides four measurements per track,
the first hit normally being in the insertable B-layer installed before Run~2~\cite{ATLAS-TDR-2010-19,PIX-2018-001}.
It is followed by the silicon microstrip tracker (SCT), which usually provides eight measurements per track.
These silicon detectors are complemented by the transition radiation tracker (TRT),
which enables radially extended track reconstruction up to \(|\eta| = 2.0\).
The TRT provides electron identification capability through the detection of transition radiation photons.
It consists of small-radius drift tubes (`straws') interleaved with a polymer material creating transition radiation for particles with a large Lorentz factor. This radiation is absorbed by the
Xe-based gas mixture filling the straws, discriminating electrons from hadrons over a wide energy range. Due to gas leaks, some TRT modules are filled with an Ar-based gas mixture.
The ID is surrounded by a superconducting solenoid producing a 2~T magnetic field that ensures an accurate reconstruction of
tracks from the primary $pp$ collision region. It also
ensures the identification of tracks from secondary vertices, permitting an efficient
reconstruction of photon conversions in the ID up to a radius of about 800~mm.

The calorimeter system covers the pseudorapidity range \(|\eta| < 4.9\).
The electromagnetic (EM) calorimeter is a lead/liquid-argon (LAr) sampling calorimeter with an accordion geometry.
It is divided into a barrel section (EMB) covering the pseudorapidity region
$|\eta|<1.475$,\footnote{The EMB is split into two half-barrel modules,
which cover the positive and negative $\eta$ regions.} and two endcap
sections (EMEC) covering $1.375<|\eta|<3.2$.
The barrel and endcap calorimeters are immersed in three LAr-filled cryostats, and are radially segmented into three layers for $|\eta|<2.5$. The first layer, covering $|\eta|<1.4$ and $1.5<|\eta|<2.4$, has a thickness
of about 4.4 radiation lengths ($X_{0}$) and is finely segmented in the $\eta$ direction, typically $0.003\times 0.1$ in $\Delta  \eta \times \Delta \phi$ in the EMB, to provide an event-by-event
discrimination between single-photon showers and overlapping showers from the decays of neutral hadrons.
The second layer, which collects most of the energy  deposited in the calorimeter by photon and electron showers, has a thickness of about $17 X_{0}$  and a granularity of $0.025\times 0.025$ in
$\Delta \eta \times \Delta \phi$. A third layer, which has a granularity of $0.05\times 0.025$ in $\Delta \eta \times \Delta \phi$ and a depth of about 2$X_{0}$, is used to correct for leakage beyond
the EM calorimeter for high-energy showers.
In front of the accordion calorimeter, a thin presampler layer (PS), covering the pseudorapidity interval $|\eta|<1.8$, is used to correct for energy loss upstream of the
calorimeter. The PS consists of an active LAr layer with a thickness of 1.1~cm (0.5~cm) in the barrel (endcap) and has a granularity of $\Delta \eta \times \Delta \phi = 0.025 \times 0.1$.
The transition region between the EMB and the EMEC, $1.37<|\eta|<1.52$, has a large amount of material in front of the first active calorimeter layer ranging from $5 X_{0}$ to almost $10 X_{0}$. This section
is instrumented with scintillators located between the barrel and endcap cryostats, and covering up to $|\eta|=1.6$.

The hadronic calorimeter, surrounding the EM calorimeter, consists of a steel/scintillator tile calorimeter in the range of $|\eta|<1.7$ and two copper/LAr calorimeters spanning $1.5<|\eta|<3.2$. The acceptance is extended by two copper/LAr and tungsten/LAr
forward calorimeters extending up to $|\eta| =$ 4.9, and hosted in the same cryostats as the EMEC.

The muon spectrometer comprises separate trigger and
high-precision tracking chambers measuring the deflection of muons in a magnetic field generated by the superconducting air-core toroidal magnets.
Three layers of precision tracking chambers covers the region \(|\eta| < 2.7\),
complemented by cathode-strip chambers in the forward region, where the background is highest.
The muon trigger system covers the range \(|\eta| < 2.4\).

Events are selected by the first-level trigger system implemented in custom hardware,
followed by selections made by algorithms implemented in software in the high-level trigger~\cite{TRIG-2016-01}.
The first-level trigger accepts events from the \SI{40}{\MHz} bunch crossings at a rate below \SI{100}{\kHz},
which the high-level trigger further reduces to record events to disk at about \SI{1}{\kHz}.

An extensive software suite~\cite{ATL-SOFT-PUB-2021-001} is used in data simulation, in the reconstruction and analysis of real
and simulated data, in detector operations, and in the trigger and data acquisition systems of the experiment.


%

%
\section{Electron and photon objects}
\label{sec:shortreco}

\subsection{Reconstruction of electron and photon candidates}

The reconstruction of electron and photon objects starts from variable-sized topological clusters combining cells from the EM and hadronic calorimeters~\cite{ATL-LARG-PUB-2008-002,PERF-2014-07}. The energies of the EM calorimeter cells in the cluster are used to reconstruct the EM energy of the cluster.
Those clusters that have an EM energy above 400 \MeV{} and a ratio of EM energy to the total cluster energy above 0.5~\cite{EGAM-2018-01} are selected as so-called EM topo-clusters. These EM topo-clusters are considered for further use in electron and photon reconstruction~\cite{EGAM-2018-01}.

Tracks are reconstructed in the ID~\cite{PERF-2017-01,EGAM-2018-01} using either the standard track reconstruction~\cite{ATL-SOFT-PUB-2007-007} or an alternative procedure that accounts for energy losses due to bremsstrahlung in the pattern recognition~\cite{PERF-2017-01}. Using a sliding window algorithm on EM calorimeter cells~\cite{ATL-LARG-PUB-2008-002}, tracks can be loosely matched to fixed-sized clusters. They are then refitted with a Gaussian sum filter algorithm to improve the accuracy of the fitted track parameters. These tracks are then further used for electron and photon reconstruction.

Initial electron candidates are reconstructed from the EM topo-clusters described above that are matched to one or several tracks. The primary track of the electron is chosen using a metric that depends on the extrapolation of the track from the perigee to the second layer of the EM calorimeter and uses either the measured track momentum or the magnitude of its momentum that has been rescaled to match the cluster energy. Details of the matching criteria and the ranking in case more than one track can be matched are described in Ref.~\cite{PERF-2017-01}.

Initial photon candidates can either be reconstructed as converted or unconverted photon candidates.\footnote{In ATLAS, around 30\% of photons convert in the detector material before the LAr calorimeter.
Because of the very different shower development, the identification, isolation and calibration of converted or unconverted photons is done separately.}
A converted photon object is based on an EM topo-cluster matched to one or several conversion vertices. The tracks described above are used to reconstruct the conversion vertices following the procedure described in Refs.~\cite{EGAM-2018-01,PERF-2017-02,PERF-2013-04}. Two types of vertices are employed: two-track conversion vertices using opposite-charged tracks and single-track conversion vertices using single tracks without hits in the innermost sensitive layers. The matching of the conversion vertices to the EM topo-cluster is based on the extrapolation of the conversion candidate to the second sampling layer of the calorimeter. The difference of the extrapolated $\eta$ and $\phi$ coordinates of the conversion and the EM topo-cluster centre is required to be smaller than 0.05 or 0.1 for double- or single-track conversions, respectively. An unconverted photon is reconstructed from an EM topo-cluster that is not matched to an electron track or a conversion vertex.

Using these initial electron and photon candidates, the final EM clusters are built from all clusters matched to the candidates. If an electron candidate has undergone bremsstrahlung radiation, which mostly occurs in the bending plane, it might be reconstructed as two clusters. This happens for about 10\% of the clusters, mainly in the forward and the transition regions~\cite{EGAM-2018-01}. Therefore for all electron and photon candidates, nearby EM topo-clusters are merged to form the final EM cluster. Starting from the highest-energy EM cluster, nearby clusters within $\Delta \eta \times \Delta \phi = 0.075 \times 0.125$ of their respective barycentres are merged with the initial cluster unless they were already used.
For electrons and converted photons, additional clusters within $\Delta \eta \times \Delta \phi = 0.125 \times 0.3$ are considered if their best matched track or conversion vertex is the same as for the original cluster.
These final EM clusters are retained as part of the electron and photon objects and can also be referred to as superclusters~\cite{EGAM-2018-01}.
After applying initial position corrections and energy calibrations to the merged EM clusters, the track and conversion vertex matching procedures are repeated to build electron and photon objects. Their energies are again calibrated using the procedure described in Refs.~\cite{PERF-2017-03,EGAM-2018-01}.

\subsection{Selection of electron and photon candidates}\label{sec:shortreco_selection}

As a final step, discriminating variables are constructed to separate prompt electrons or photons from background objects. In this context, prompt electrons or photons are understood to be stemming from the hard-scattering vertex or from the decay of heavy resonances such as Higgs, $W$, and $Z$ bosons, whilst background objects are jets misidentified as electrons or photons, non-prompt electrons from decays of $b$- and $c$-hadrons, or even electrons misreconstructed as photons and vice versa. The discriminating variables can be grouped into three types that are either related to properties of the track, the shower development in the different layers of the calorimeter, or the matching between the track and the calorimetric cluster. For both electrons and photons, these identification selection requirements are applied as a function of $|\eta|$ and transverse energy (\et). A detailed overview of their definition is given in Table~1 of Ref.~\cite{EGAM-2018-01}. For the identification of electrons, these variables are combined into a likelihood classifier, as detailed in Ref.~\cite{EGAM-2018-01}. Three sets of identification selection requirements are defined (Loose, Medium, and Tight), which correspond to increasingly restrictive selections~\cite{PERF-2017-01} and hence increasingly suppressed background, at the cost of also reducing signal acceptance. These identification requirements are valid in the pseudorapidity range $|\eta| < 2.47$.

Photons are identified in the pseudorapidity range $|\eta| < 2.37$ using sets of selection criteria~\cite{PERF-2017-02} on EM shower shape variables.
Three identification selection criteria exist, named Loose, Medium and Tight. The former two are employed in the trigger, whilst the Tight criterion is the primary photon identification selection employed for ATLAS analyses. They are defined in Ref.~\cite{EGAM-2018-01}.

A number of different isolation variables exist. The photon isolation has not been updated with regards to Ref.~\cite{EGAM-2018-01} and is not discussed further in this publication. The electron calorimeter isolation~\cite{PERF-2017-01}, \etcx, is calculated by summing the transverse energy of positive-energy topological clusters\footnote{
The LAr signal is electronically shaped to suppress noise contributions~\cite{Buchanan:2008zza} which in some cases cause fluctuations to lead to cells being reconstructed with negative energies. As the signal-over-noise criterion for topological cluster formation uses the absolute energies~\cite{ATL-LARG-PUB-2008-002}, clusters can be reconstructed with negative energy but are not considered here.} whose barycentre falls within a cone centred around the electron cluster barycentre. Here, $\mathrm{XX}$ refers to the size of the employed cone, $\Delta R = \mathrm{XX}/100$. Frequently, XX is set to 20 for calorimeter isolation. The energy of the electron itself, as well as contributions from pile-up activity, are subtracted using the methodology described in Section~\ref{sec:ele_iso}. The track isolation variable (\ptcx) is defined as the scalar sum of the transverse momenta of selected tracks within a cone of size $\Delta R = \mathrm{XX}/100$ centred around the electron track direction. For track isolation, XX is often set to 20 or 30. The tracks are required to originate from the primary vertex of the event\footnote{The primary vertex is defined to be the one with the largest scalar sum of transverse momenta of the associated tracks.} and to fulfil minimum quality requirements as detailed in Ref.~\cite{EGAM-2018-01}. Tracks matched to the electron cluster are in general excluded from the calculation; however, refining the procedure documented in Ref.~\cite{EGAM-2018-01}, tracks that are further than $|\Delta \eta| > 0.01$\footnote{$\Delta \eta$ is evaluated after the tracks have been extrapolated to the EM calorimeter.} away from the primary electron track are still added to the isolation cone. This improves the background rejection for $\et > 100$~\GeV{} significantly compared with previous publications. In simulated $t \bar{t}$ events,\footnote{Simulated using the \POWHEGBOX[v2]~\cite{Frixione:2007nw,Nason:2004rx,Frixione:2007vw,Alioli:2010xd}
generator at next-to-leading-order QCD with the \NNPDF[3.0nlo]~\cite{Ball:2014uwa} PDF set
and the $h_{\mathrm{damp}}$ parameter set to 1.5 times the top quark mass~\cite{ATL-PHYS-PUB-2016-020}. The events were interfaced
to \PYTHIA[8.230]~\cite{Sjostrand:2014zea} to model the parton shower,
hadronisation, and underlying event, with parameters set according
to the A14 tune~\cite{ATL-PHYS-PUB-2014-021} and using the \NNPDF[2.3lo]
set of PDFs~\cite{Ball:2012cx}. The decays of bottom and charm hadrons
were performed by \EVTGEN[1.6.0]~\cite{Lange:2001uf}.} the rejection is larger by more than a factor of two, depending on the electron identification and isolation selection, without degrading the prompt-electron efficiency. For prompt electrons produced in cascade decays of high-momentum heavy particles where other decay products can be very close to the electron direction, an alternative track isolation is defined with a variable cone size (\ptvcx). In this case, the cone size shrinks for larger transverse momentum of the primary electron candidate:

\begin{equation}\label{eq:varcone}
\Delta R = \min \left( \frac{10}{\et[\gev]}, \Delta R_{\mathrm{max}}\right),
\end{equation}

where $\Delta R_{\mathrm{max}}=\mathrm{XX}_{\mathrm{max}}/100$ is the maximum cone size.


%

%
\section{Datasets and simulated-event samples}
\label{sec:data_sets}

\subsection{Selection of the data}

The studies documented in this paper use the full
$pp$ collision data recorded by the ATLAS detector between 2015 and 2018 (Run 2),
with the LHC operating at a centre-of-mass energy of $\sqrt s = 13$~\TeV{} and a bunch spacing of 25 ns.
In this period, the LHC delivered colliding beams with a peak instantaneous luminosity up to $L=\SI{2.1e34}{cm^{-2}s^{-1}}$, achieved in 2018,
and an average number of $pp$ interactions per bunch crossing \muhat\ of 13, 25, 38, and 34 for 2015, 2016, 2017, and 2018 data, respectively.
After applying beam, detector, and data-quality criteria
the total integrated luminosity of the data is \SI{139}{fb^{-1}}~\cite{DAPR-2018-01}.
The uncertainty in the combined 2015--2018 integrated luminosity is 1.7\%~\cite{ATLAS-CONF-2019-021},
obtained using the LUCID-2 detector~\cite{LUCID2} for the primary luminosity measurements.\footnote{The final luminosity for Run 2 was determined be \SI{140}{fb^{-1}} with an uncertainty of 0.83\%~\cite{DAPR-2021-01}. The impact on the presented efficiency measurement is fully negligible. Therefore the presented results have not been updated using the new integrated luminosity.}

The measurements of the electron reconstruction, identification and isolation efficiencies
use a large sample of \Zee\ events selected with single-electron triggers.
The single-electron high-level trigger has an \et threshold ranging from 24 \gev\ in 2015 and most of 2016 to
26 \gev\ at the end of 2016, and during 2017 and 2018; it requires a Tight identification and loose tracking-based
isolation criteria~\cite{EGAM-2018-01}.

A sample of \Jee{} events
was collected for studies with low-\et\ electrons
using dedicated prescaled dielectron triggers with asymmetric electron \et thresholds ranging from 4 to 14~\gev. Each of
these triggers requires Tight trigger identification and \et above a certain threshold for one trigger object,
while only requiring the electromagnetic cluster \et to be higher than some other (lower) threshold for
the second object.

Samples of $Z \to \ell^+\ell^-\gamma$ events (with $\ell$ being an electron or a muon), used to measure photon identification
efficiency at low \et, are selected with the same non-prescaled single-lepton triggers as the \Zee\ events described above. Additionally, dielectron triggers are used with electron \et\ thresholds ranging from 12 to 24~\GeV{} and Loose identification. Events with $Z$ boson decays into muons are selected using a single muon with a loose tracking-based isolation criterion and a transverse momentum threshold of 26~\GeV{} and a dimuon trigger with \pT\ thresholds at 14 \GeV~\cite{TRIG-2018-05,TRIG-2018-01}. The muons are reconstructed as described in Ref.~\cite{MUON-2018-03}.

Inclusive QCD samples and inclusive-photon production samples,
used to measure photon identification efficiency at high \et,
are selected with prescaled and non-prescaled single-photon triggers~\cite{TRIG-2018-05} with Loose identification.
The lowest transverse energy threshold of these triggers is 10~\gev.

\subsection{Monte Carlo event simulation}

Various Monte Carlo (MC) event generators were employed to simulate the signal and background samples
(\Zee, \Jee, $Z \to \ee \gamma$, $Z \to \mu^+\mu^- \gamma$, inclusive-photon production and tree-level $2\rightarrow 2$ QCD production including also top-quark pair and weak vector-boson production). These were used to perform the studies discussed in this paper. Simulation details additional to those provided below can be found in Ref.~\cite{EGAM-2018-01}.

The generated signal and background events were processed through the full ATLAS detector simulation~\cite{SOFT-2010-01} based on \textsc{Geant4}~\cite{Agostinelli:2002hh}.
The MC events were simulated with additional interactions in the same or neighbouring bunch crossings to match the pile-up conditions
during LHC operations.
The overlaid $pp$ collisions were generated with the soft QCD processes of \textsc{Pythia8}~\cite{Sjostrand:2014zea}
using the A3 set of tuned parameters~\cite{ATL-PHYS-PUB-2016-017} and the \NNPDF[2.3lo] PDF set~\cite{Ball:2012cx}.
Although this set of tuned parameters improves the modelling
of minimum-bias data relative to the set used in the past (A2~\cite{ATL-PHYS-PUB-2012-003}), it overestimates
by roughly 3\% the hadronic activity as measured using charged-particle tracks. Simulated events
were reweighted to reproduce the distribution of the average number of interactions per bunch crossing in data, scaled
down by a factor 1.03~\cite{STDM-2015-05}.

Studies presented throughout this paper using MC simulation select electrons originating from \Zee\ or \Jee\ decays,
or photons from $Z \to \ell^+\ell^-\gamma$, inclusive-photon production samples, or other processes
using generator-level information.
The matching of the reconstructed and generated electrons is based on their ID track~\cite{PERF-2015-08}. The reconstructed electrons are classified as genuine if this track is reconstructed from the primary electron, or from secondary particles produced
in a material interaction of the primary electron or of final state radiation emitted collinearly.
Reconstructed and generator-level photons are matched based on their distance in $\eta$--$\phi$ space between the EM cluster barycentre and the extrapolation of the photon direction to the 2nd layer of the EM calorimeter.\footnote{In an elliptical cone of $\eta\times\phi=0.025\times0.5$ the highest \et{} generator-photon is chosen as a match. If none is found, the leading particle in the cone is matched instead independent of particle type. If no particle is found, the closest particle in a cone of $\Delta R = 0.1$ is used.}


%

%
\section{Electron efficiency measurements}
\label{sec:eff_measurement}

The efficiency of reconstructing, triggering on, and selecting a genuine, prompt electron can be factorized into different terms:
\begin{equation}
\varepsilon_{\operatorname{total}} = \varepsilon_{\operatorname{EMclus}} \times \varepsilon_{\operatorname{reco}} \times \varepsilon_{\operatorname{id}} \times \varepsilon_{\operatorname{iso}} \times \varepsilon_{\operatorname{trigger}}.
\label{eq:ele_total_eff}
\end{equation}

In Equation~\ref{eq:ele_total_eff}, $\varepsilon_{\operatorname{EMclus}}$ is the probability to reconstruct an EM-cluster given a genuine electron.
This efficiency is computed as the ratio of the number of reconstructed EM clusters, $N_{\operatorname{cluster}}$, and the number of produced electrons, $N_{\operatorname{all}}$.
It is evaluated entirely from MC simulation, where the reconstructed cluster is associated to a genuine electron produced at generator level,
and is found to be above 99\% for electrons with $\et > 10$~\GeV~\cite{EGAM-2018-01}.

The reconstruction efficiency, $\varepsilon_{\operatorname{reco}}$, reflects the performance of the track reconstruction and its association with the cluster.
It is obtained from the number of reconstructed electron
candidates with a track, $N_{\operatorname{reco}}$, divided by the number of EM cluster candidates, $N_{\operatorname{cluster}}$.
The identification efficiency, $\varepsilon_{\operatorname{id}}$, is the probability of a reconstructed electron to
satisfy a certain identification criterion.
It is defined as the number of identified and reconstructed electron candidates, $N_{\operatorname{id}}$, divided by $N_{\operatorname{reco}}$.
The isolation efficiency, $\varepsilon_{\operatorname{iso}}$,
is the probability of an identified electron to pass requirements on the track or calorimeter isolation.
It is calculated as the number of identified
electron candidates satisfying the isolation, identification, and reconstruction requirements, $N_{\operatorname{iso}}$, divided by
$N_{\operatorname{id}}$.
Only $\varepsilon_{\operatorname{reco}}$, $\varepsilon_{\operatorname{id}}$, $\varepsilon_{\operatorname{iso}}$
efficiencies are further discussed in this paper.

Finally, the trigger efficiency, $\varepsilon_{\operatorname{trigger}}$, is calculated as the number of
(reconstructed, identified, and isolated) electron candidates passing the trigger requirements, $N_{\operatorname{trig}}$, divided by
$N_{\operatorname{iso}}$~\cite{TRIG-2018-05}.

All efficiencies except $\varepsilon_{\operatorname{EMclus}}$
are estimated directly from data using tag-and-probe methods
that select from unbiased samples of electrons produced in
the decays of well-known resonances, such as \Zee\ or \Jee, unbiased samples of electrons produced in the particle’s decay.
One of the electrons must satisfy strict selection requirements and serves as a tag, while the second electron serves as a probe.
The efficiency of a given requirement is then computed
by applying it to the probe sample after accounting for residual background contamination~\cite{PERF-2016-01,PERF-2017-01,EGAM-2018-01}. If both the electrons satisfy the tag requirements, each of them are used as tag and the event provides two probes.
This is necessary in order not to bias the probe selection.

The biggest challenge in the efficiency measurements is the estimation of probes that
originate from background rather than signal processes. This background is largest for the sample of
cluster probes, but the fraction of such events is reduced with each efficiency step given in Equation~\ref{eq:ele_total_eff}, from left to right.
Various data-based background estimation methods are used as described in the relevant sections below.

The accuracy of the observed electron efficiency plays an important
role when using MC simulation to predict physics processes.
In order to achieve reliable results, the MC simulated events need to be corrected to reproduce
as closely as possible the efficiencies measured in data.
This is achieved by applying a multiplicative correction factor in the MC simulation
in addition to the other event weights that are used e.g to scale to luminosity, or to accurately reflect the detector resolution. The correction factor to improve the modelling of the efficiency is defined as the ratio of the efficiency measured in data to that determined from MC events where the same efficiency measurement method is used in both the data and MC simulation.
A dedicated correction factor is measured for each of the efficiencies in Equation~\ref{eq:ele_total_eff} separately.
The advantage of using these correction factors is that they are universally applicable whilst absolute efficiencies might differ from physics process to physics process.
As shown below, these correction factors are normally close to unity with deviations arising from a mismodelling in the simulation of tracking
properties, shower shapes in the calorimeters, or the isolation from other energy deposits.
As their application brings data and MC simulation closer together, this helps to remove biases from predictions and to avoid constraints and shifts in likelihood fits when extracting results.

Systematic uncertainties in the correction factors are evaluated by varying the requirements on the selection of the candidate sample, as well as varying the details of the background-subtraction~\cite{PERF-2016-01,PERF-2017-01,EGAM-2018-01}.
How these variations are used to construct the central values and their uncertainties depends on the method used and is described in the respective method sections below.

The statistical uncertainty of the efficiency in a single variation of the measurement is calculated assuming a binomial distribution. The statistical uncertainty related to the subtracted background is also included in the overall statistical uncertainty.


%

%
\section{Electron reconstruction}
\label{sec:ele_reco}

The electron reconstruction efficiency, $\varepsilon_{\operatorname{reco}}$, is measured in bins of electron \et\ and $\eta$ with bin edges at $\et = \{$15, 20, 25, 30, 35, 40, 45, 50, 60, 80, 150, $\infty$$\}$~\GeV\ and $\eta = \pm \{0$, 0.1, 0.6, 0.8, 1.15, 1.37, 1.52, 1.81, 2.01, 2.37, 2.47$\}$. The binning in $\eta$ is driven by the detector geometry, e.g. the transition region between EMB and EMEC ($1.37<|\eta|<1.52$), the acceptance of the presampler ($|\eta|<1.81$) and the region where the strip granularity changes to be coarser ($|\eta|=2.37$).

The electron reconstruction efficiency is only measured for $\et > 15$~\GeV.
Below 15~\GeV, the large backgrounds make the measurement unreliable and using low-mass resonances like $J/\psi$ is equally challenging. The electron reconstruction efficiency is defined as the efficiency to reconstruct a prompt electron with a good quality track\footnote{Defined as a track that has at least 7 precision hits in one of the silicon trackers and at least one pixel hit.} and with a good quality cluster\footnote{Defined as a cluster that is not affected by the presence of a dead front-end board in the first or second sampling, or by the presence of a dead HV region affecting the three layers of EM calorimeter, or by the presence of a masked cell in the core.} from all reconstructed EM clusters.
Including track and cluster quality requirements in the measurement of the reconstruction efficiency reduces it by about 0.5\%
compared to the previous publications~\cite{EGAM-2018-01}.

The measurements are performed using a tag-and-probe method based on the \Zee\ invariant mass to estimate both the signal and background as described in Ref.~\cite{ATLAS-CONF-2016-024}. Events are recorded using the single-electron triggers described in Section~\ref{sec:data_sets}.
An event must have at least one
tag electron with $\et > 25$~\GeV\ outside the calorimeter transition region $1.37 < |\eta| < 1.52$. It must pass the Tight identification. To estimate systematic uncertainties, two alternative tag selection requirements are used that vary the amount of backgrounds and thus test the robustness of the background estimate: either the Medium or Tight identification, both in combination with an isolation criterion of $E_{\mathrm{T}}^{\text{cone40}}< 5$~\GeV. The tag must be associated with the object that fired the trigger. Additionally, a probe object must be present, defined as a reconstructed EM cluster that is isolated by $\Delta R > 0.4$ from any jet reconstructed with the anti-$k_{\mathrm{T}}$ algorithm~\cite{Cacciari:2008gp, Fastjet} using a radius parameter of 0.4 with $\pt > 20~\GeV{}$. Candidates that are likely to stem from a photon conversion are discarded~\cite{EGAM-2018-01}. The invariant mass of the tag and the probe is restricted to be in the range $m_{ee} \in [60, 250]~\GeV{}$ and only the highest-mass pair is chosen in case there is more than one. No charge requirements are applied.

The measurement of the reconstruction efficiencies uses separate methods to estimate backgrounds from data for EM clusters\textit{ with and without} an associated track:
\begin{enumerate}
\item For the sample of EM cluster with associated tracks, non-electron backgrounds are estimated using a template created by inverting identification and isolation selection requirements on the probe to select a pure background sample~\cite{ATLAS-CONF-2016-024,PERF-2016-01}. Two different variations of this background selection are used, each of which is normalised simultaneously in the low and high mass region of the invariant mass distribution ([60, 70]~\GeV{} and [120, 250]~\GeV{} respectively).
\item For the sample of EM clusters without associated tracks, it is difficult to construct a data-driven template. Therefore, backgrounds from photons or jets are determined using a third order polynomial fit in sidebands regions around the invariant mass of the $Z$ boson. This fit function was found to provide a numerically stable description of the background with a good fit quality. Four different sets of sideband regions are employed: [70, 80]~\GeV{} and [100, 110]~\GeV{}; [60, 80]~\GeV{} and [100, 120]~\GeV{}; [50, 80]~\GeV{} and [100, 130]~\GeV{}; [55, 70]~\GeV{} and [110, 125]~\GeV{}. Signal contributions in these sidebands are subtracted using estimates from MC simulation~\cite{ATLAS-CONF-2016-024,PERF-2016-01}.
\end{enumerate}

\begin{figure}[h!]
\centering
\subfloat[]{
\includegraphics[width=0.45\textwidth]{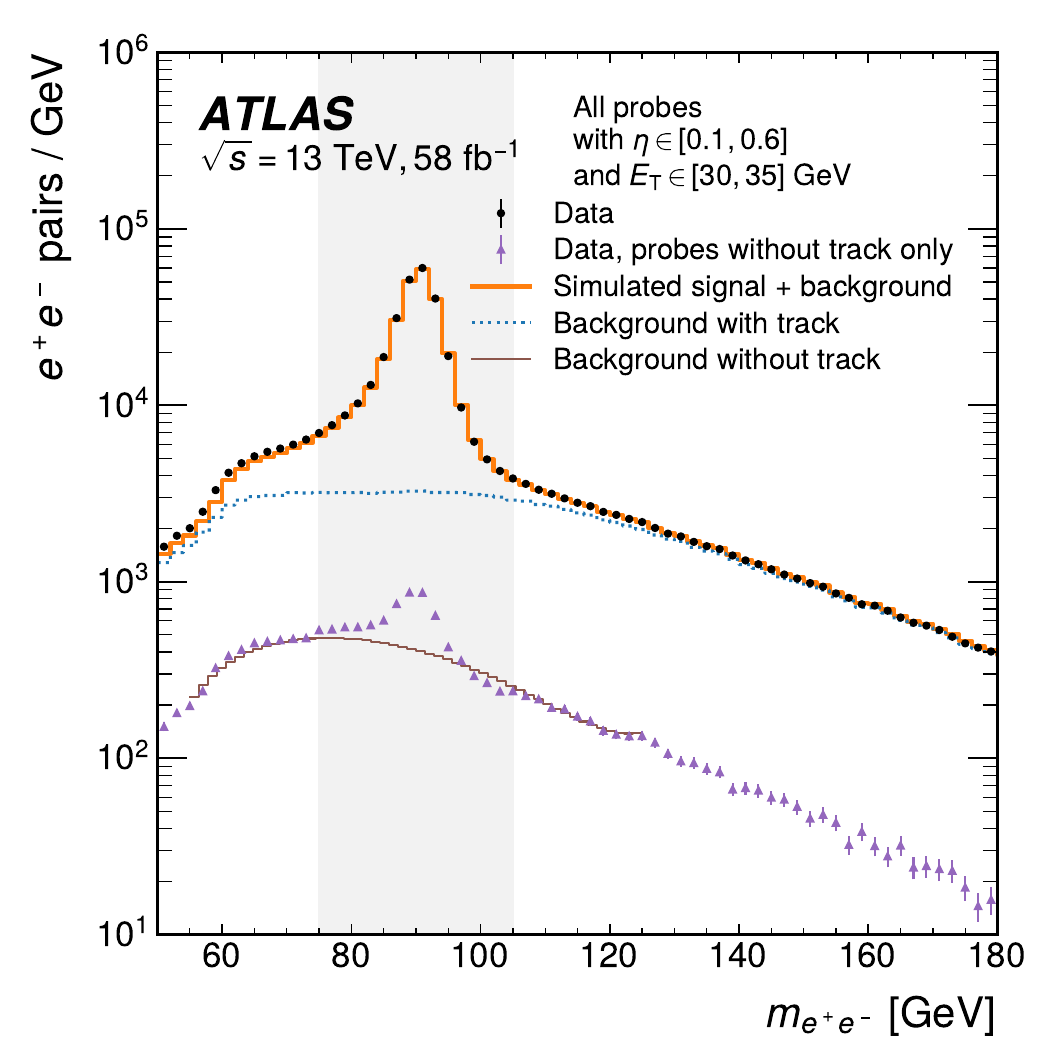}
\label{fig:mll_reco_sub1}
}
\subfloat[]{
\includegraphics[width=0.45\textwidth]{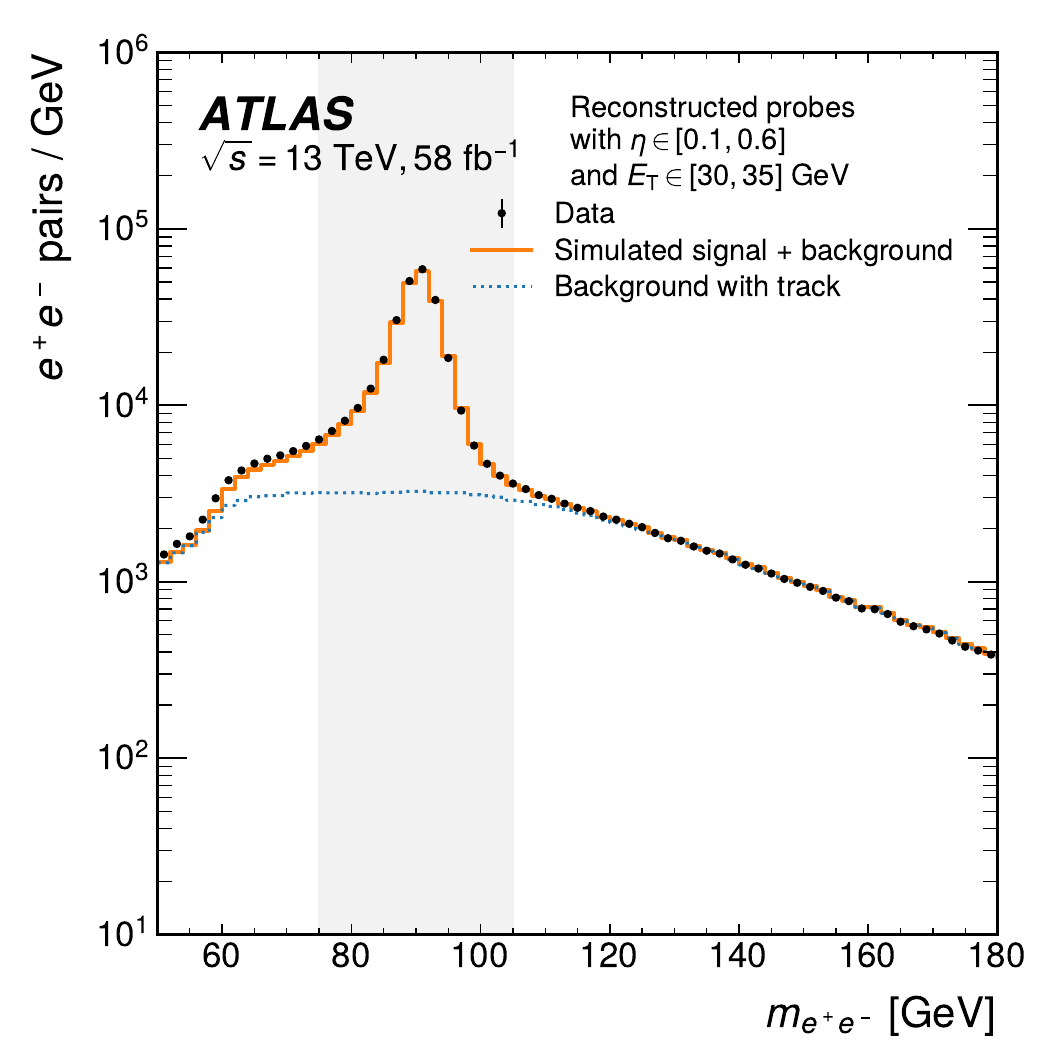}
\label{fig:mll_reco_sub2}
}
\caption{Example of the \mee{} distribution used to determine the electron reconstruction efficiency, using 2018 data and probes with $\et \in [30, 35]~\GeV{}$ and $|\eta| \in [0.1, 0.6]$. The shaded band indicates the nominal mass range used for efficiency extraction. The Monte-Carlo-simulated signal added to the total background is shown for illustration purposes; it is not used in the determination of the data efficiencies.
\protect\subref{fig:mll_reco_sub1} Including all probe clusters, whether reconstructed as an electron or not. The background with a track is shown separately. Also shown is the data component without track, as is an example background without track determined by fitting a polynomial to it in the sidebands [55, 70]~\GeV{} and [110, 125]~\GeV{}.
\protect\subref{fig:mll_reco_sub2} Including all reconstructed probes. Since these all have a track, there is only one background component.
}\label{fig:mll_reco}
\end{figure}

The signal yields are extracted by counting the background-subtracted data events in a window around the $Z$-boson mass. The efficiency is calculated directly from the background-subtracted data yields. As systematic variations, the ranges [70, 110]~\GeV{}, [75, 105]~\GeV{} and [80, 100]~\GeV{} are used. Varying the selection of the tag as well as of the background template selections and the definition of the sideband regions and mass windows, yields in total 72 systematic variations of the measurement. The efficiency is taken to be the mean value of these whilst the root mean square of them is taken as systematic uncertainty. Example \mee{} distributions  are shown in Figure~\ref{fig:mll_reco}.

\begin{figure}[h!]
\centering
\subfloat[]{
\includegraphics[width=.49\textwidth]{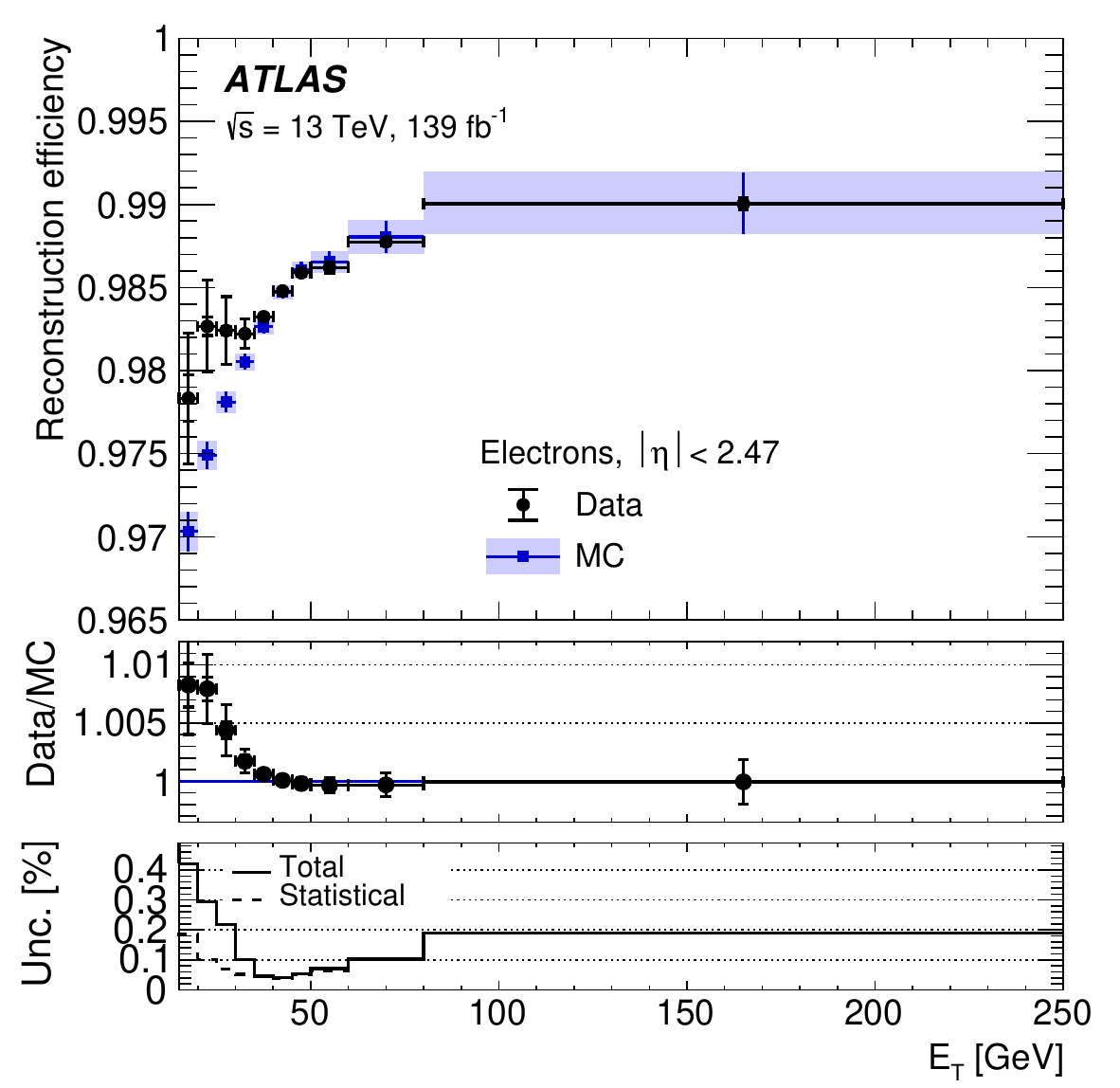}
\label{subfig_reco_pt}
}
\subfloat[]{
\includegraphics[width=.49\textwidth]{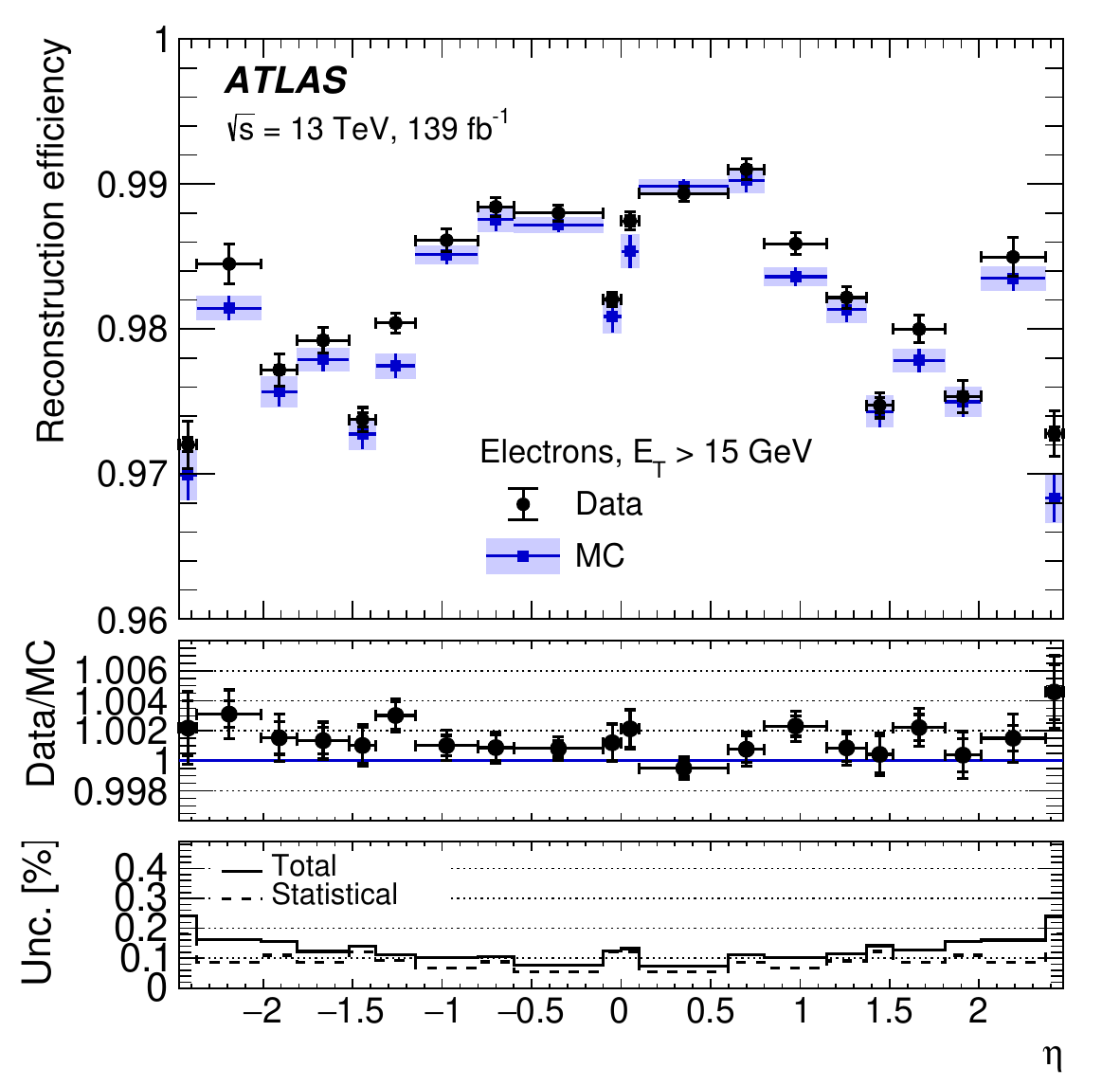}
\label{subfig_reco_eta}
}
\caption{Reconstruction efficiencies of electrons from $Z \to e^+ e^-$ decays as a function of the electron \et{} \protect\subref{subfig_reco_pt} and $\eta$ \protect\subref{subfig_reco_eta} respectively. The top panels show the efficiencies obtained in data and simulation with their statistical and total uncertainties displayed as inner and outer error bars. The middle panels show their ratio which is applied as correction factor in analyses. The bottom panels show the relative statistical and total uncertainties on the data/MC ratio.}
\label{fig:reco_pt}
\end{figure}

The electron reconstruction efficiency is very high with values between 98-99\%; however, it drops at higher $\arrowvert\eta\arrowvert$ and in the calorimeter transition region.
Electron reconstruction efficiencies corresponding to the kinematics of simulated $Z \to e^+ e^-$ events are shown in Figure~\ref{fig:reco_pt}
as a function of the electron transverse energy and pseudorapidity, respectively.
The correction factors are shown in the middle panels of the figures. The systematic uncertainties on the correction factors are less than 0.5\% over most of the kinematic range and less than 0.1\% for $\et>30$~\GeV. They are generally within 1\% from unity.
The correction factors for electrons with energies below $\et<15$~\GeV{} are not accessible in data measurements. They are therefore taken to be 1 with uncertainties of 2\% and 5\% for electron within $|\eta|<1.37$ and $|\eta|>1.37$ respectively to conservatively cover the values and uncertainties of the correction factors measured at low \et{}. The efficiencies measured as function of $\eta$ are symmetric within uncertainties between the negative and positive ranges of electron pseudo-rapidity.

\begin{figure}[h!]
\centering
\includegraphics[width=0.65\textwidth]{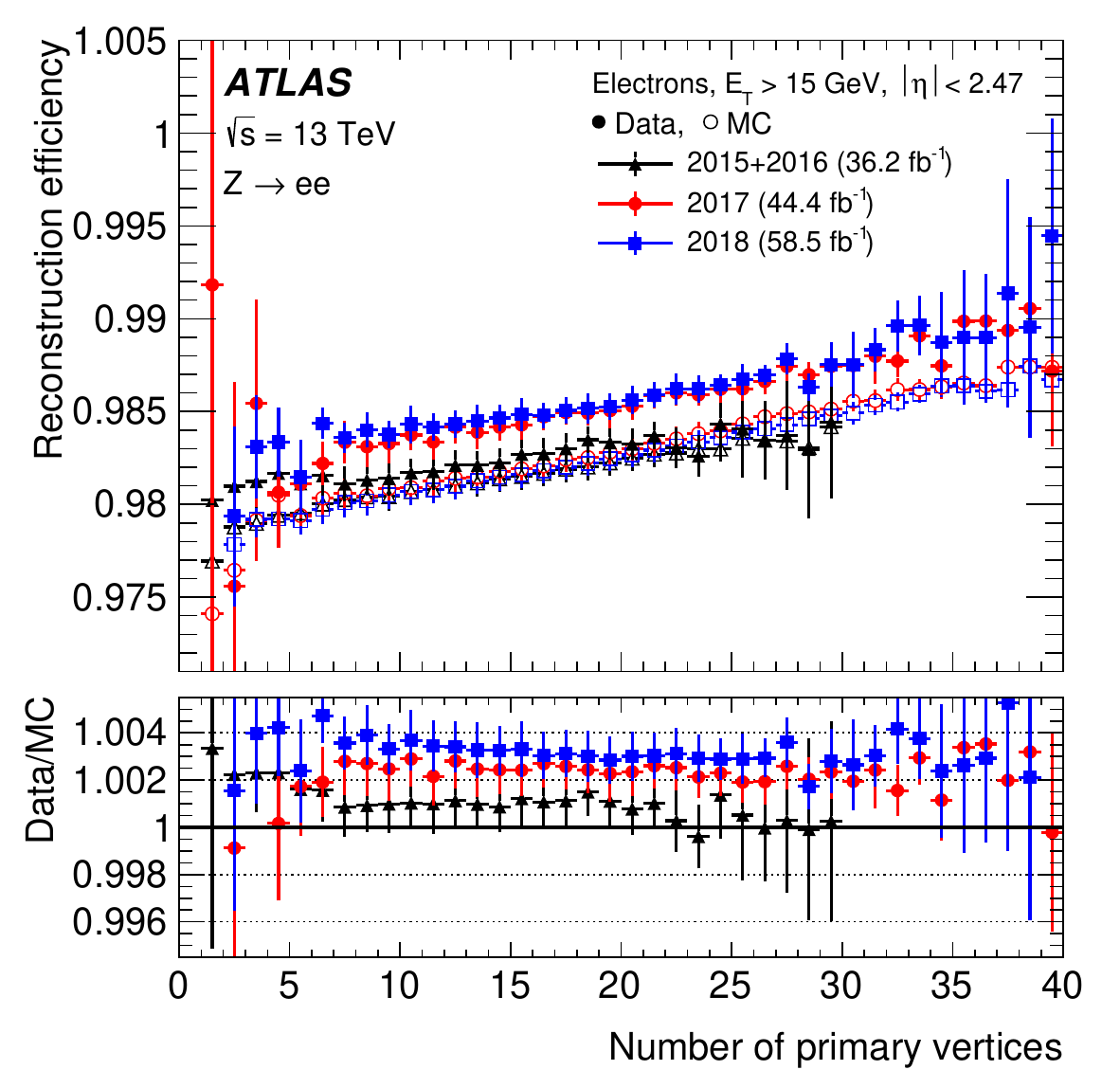}
\caption{Reconstruction efficiencies of electrons from $Z \to e^+ e^-$ decays as a function of number of primary vertices. The top panel shows the efficiencies obtained in data and simulation with their respective combined systematic and statistical uncertainties. The lower panel shows their ratios. }
\label{fig:reco_PU}
\end{figure}

Figure~\ref{fig:reco_PU} shows the reconstruction efficiencies for different years of data taking as a function of the number of primary vertices in the events, which is closely correlated with the number of additional $pp$ interaction in an event. The efficiency generally rises as a function of pile-up.
For example, there is a larger probability to match tracks with the clusters reconstructed from genuine electrons which otherwise might have only been reconstructed as clusters without tracks. The purity of the sample of reconstructed electrons also decreases with pile-up as further discussed in Section~\ref{sec:rej}. Generally, these effects are well-described in the MC simulation as evidenced by the flat data/MC ratio displayed in the middle panel of the figure. This indicates that correction factors have no strong dependence on the amount of pile-up.

%

%
\section{Electron identification}
\label{sec:ele_id}

Electron identification efficiencies are measured in 234 bins of electron $\et$ and $\eta$. The bin edges lie at $\et = \{4.5$, 7, 10, 15, 20, 25, 30, 35, 40, 45, 50, 60, 80, 150, $\infty$$\}~\GeV$ and $\eta = \pm \{0$, 0.1, 0.6, 0.8, 1.15, 1.37, 1.52, 1.81, 2.01, 2.37, 2.47$\}$. To enhance statistics, some $\eta$ bins are merged for $\et < 20~\GeV$ and the absolute value of $\eta$ is used for $\et > 150~\GeV$. For precision measurements, a finer granularity in $\eta$ is available that has 52 bins instead of 20.\footnote{The bin edges for this finer granularity lie at $\eta = \pm \{0$, 0.1, 0.2, 0.3, 0.4, 0.5, 0.6, 0.7, 0.8, 0.9, 1, 1.1, 1.2, 1.3, 1.37, 1.52, 1.6, 1.65, 1.7, 1.8, 1.9, 2, 2.1, 2.2, 2.3, 2.4, 2.47$\}$.}

For electrons with $4.5~\GeV{} < \et < 20~\GeV{}$, identification efficiencies are measured in $J/\psi \to e^+e^-$ events.
In the bin $\pt \in [15, 20]~\GeV$, the correction factors from the $J/\psi$ measurement are statistically combined with the ones from the $Z$ measurements discussed below. Events with
$J/\psi \to e^+e^-$ decays are recorded with the triggers described in Section~\ref{sec:data_sets} and are required to contain at least two oppositely charged electron candidates with $\et > 4.5$~\GeV{} and $|\eta| < 2.47$. If more than two electrons pass the selection criteria, the pair with invariant mass closest to the $J/\psi$ mass is considered. The tag electron must pass Tight identification, and it is required to be outside the calorimeter transition region, $1.37 < |\eta| < 1.52$.
Both the tag and the probe electrons must be matched within $\Delta R < 0.07$ to the trigger objects that prompted the recording of the event.
The trigger efficiency for electromagnetic clusters is high and was found not to bias the properties of the probe electron.
Unlike previously, no isolation requirements are applied to the electrons in the nominal case.
To suppress background from photon conversions and hadronic jets, the tag and the probe candidates must be separated from each other by $\Delta R > 0.15$.

For electrons with $\et > 15$~\GeV, identification efficiencies are measured using $Z \to e^+e^-$ events. Events are recorded using the single-electron triggers and must contain two electron candidates with $|\eta| < 2.47$ and opposite charge.
The tag electron must have $\et > 27$~\GeV, and lie outside the calorimeter transition region $1.37 < |\eta| < 1.52$. It must pass the Tight identification and a track isolation requirement of $p_{\mathrm{T}}^{\mathrm{cone}20} / \pt < 0.1$. It must be associated with the object that fired the trigger.
The efficiency is extracted by counting the background-subtracted data yields in the range $m_{e^+e^-} \in [75, 105]~\GeV{}$.

The background due to objects other than prompt electrons, e.g. light- and heavy-flavour hadrons or photon-conversions and Dalitz decays, cannot be simulated reliably using MC simulation. It is therefore determined from data. In $Z \to e^+e^-$ events, two methods are used to measure the background and their results are statistically combined. The first method, called $Z$ mass, uses the electron--positron invariant mass, \mee, as the discriminating variable between signal and background. The second, called $Z$ isolation, uses the amount of energy in an isolation cone around the probe electron. Both methods are described in Refs.~\cite{PERF-2016-01,PERF-2017-01,EGAM-2018-01} and here the improvements from the previous results are discussed.

\subsection{$J/\psi \to e^+e^-$ measurement}

Similarly to the methodology in Refs.~\cite{PERF-2016-01,PERF-2017-01,EGAM-2018-01}, the invariant-mass distribution of the two $J/\psi$ electron candidates in the range [1.8, 4.6]~\GeV{} is fitted with functions to extract three contributions: $J/\psi$ events, $\psi(2S)$ events, and the background from hadronic jets, heavy flavour decays, and electrons from conversions. The $J/\psi$ and $\psi(2S)$ contributions are each modelled with a Crystal Ball function convoluted with a Gaussian function. The background is determined by fitting a third-order Chebyshev polynomial that was chosen for its numerical stability. In addition, where possible, combinatorial backgrounds from two random electron candidates that fall into the mass window are also considered. This is achieved by defining a same-charge region where the tag and the probe electron have the same charge and otherwise pass the normal criteria. If this region contains a sufficiently large number of events, which is on average more than 25 events per bin, then another component is added to the signal-region fit: a fifth-order Chebyshev polynomial that is fitted simultaneously in the same-charge and signal region. This treatment allows a better modelling of the background by adding an additional constraint on the combinatorial background. It is based on the assumption that the shape of the mass distribution is similar for the same charge and the opposite charge sign regions, but is only possible if there is sufficient statistics. This improved the agreement of the fit model with the data with respect to the previous publication.

$J/\psi \to e^+e^-$ events come from a mixture of prompt and non-prompt $J/\psi$ production, with relative fractions depending both on the triggers used to collect the data and on the \et{} of the probe electrons. Non-prompt $J/\psi$ production can occur e.g.~in $b$-hadron decays.
Only the prompt $J/\psi$  production yields prompt electrons, which are expected to behave similarly as those electrons from physics processes of interest to analyses.
The non-prompt $J/\psi$ production is therefore considered to be a source of background.
To suppress this background, the pseudo-proper lifetime\footnote{The pseudo-proper lifetime is defined as $t_{0}=L_{x y} \cdot m_{\mathrm{PDG}}^{J / \psi} / p_{\mathrm{T}}^{J / \psi}$, where $L_{x y}$ is the displacement of the $J / \psi$ vertex from the primary vertex projected onto the flight direction of the $J / \psi$ in the transverse plane, $m_{\mathrm{PDG}}^{J / \psi}$ is the nominal $J / \psi$ mass~\cite{PDG2020} and $p_{\mathrm{T}}^{J / \psi}$ is the $J / \psi$-reconstructed transverse momentum. The ratio  $m_{\mathrm{PDG}}^{J / \psi} / p_{\mathrm{T}}^{J / \psi}$ is the transverse component of the inverse of the relativistic factors, $\frac{1}{\beta \gamma}$.} $t_{0}$ of the reconstructed $J/\psi$ is required to be $-1~\text{ps} < t_{0} < 0.2~\text{ps}$.
The residual non-prompt fraction in bins of probe electron $\et$ and $\eta$ is calculated by using MC simulation and the ATLAS measurement using $J/\psi \to \mu^+\mu^-$~\cite{BPHY-2012-02} of the non-prompt fraction in bins of  $J/\psi$ $\pt$ and $\eta$. It is subtracted from the fitted $J/\psi$ signal. The use of the more recent measurement~\cite{BPHY-2012-02} compared to the previous methodology~\cite{BPHY-2010-01} gives an improved estimate of the non-prompt fraction that has an uncertainty that is up to a factor of five smaller than previously.

\begin{figure}[h!]
\centering
\subfloat[]{
\includegraphics[width=0.45\textwidth]{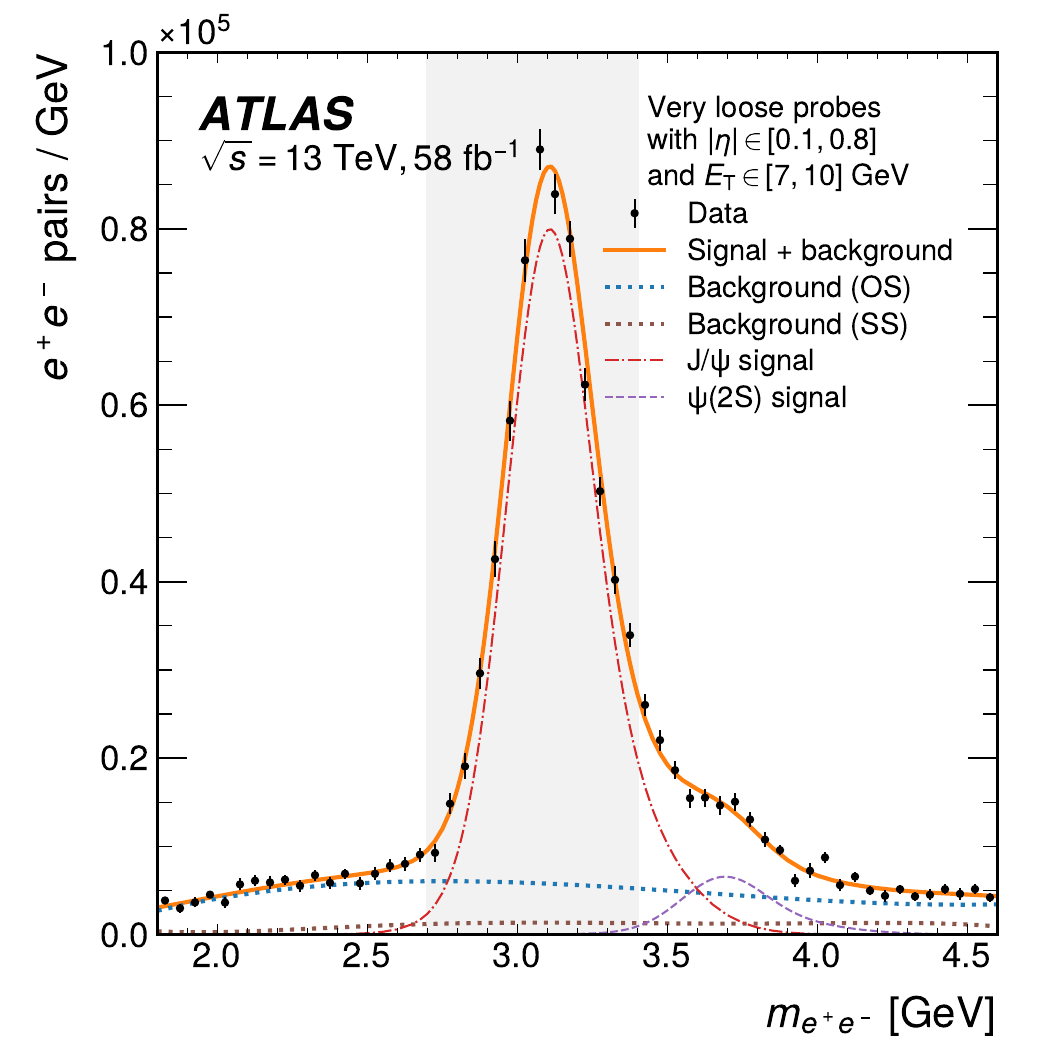}
\label{fig:mll_taucut_den_sub1}
}
\subfloat[]{
\includegraphics[width=0.45\textwidth]{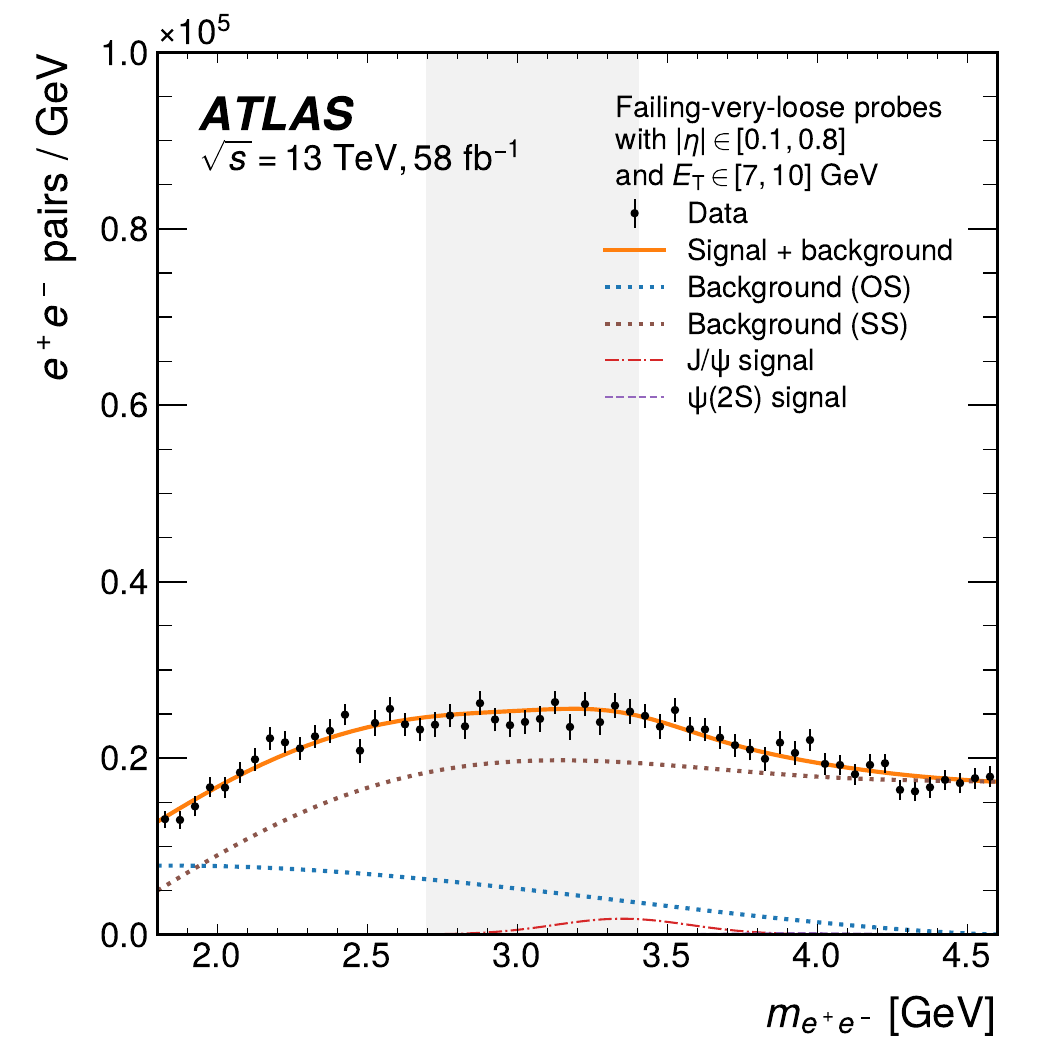}
\label{fig:mll_taucut_den_sub2}
}
\caption{Examples of nominal signal and background fits to the $m_{e^+e^-}$ distribution used to determine the efficiency denominator in the $J/\psi \to e^+e^-$ measurement, using 2018 data and probes with $\et \in [7, 10]~\GeV{}$ and $|\eta| \in [0.1, 0.8]$. The shaded band indicates the nominal mass range used for efficiency extraction. \protect\subref{fig:mll_taucut_den_sub1} Including all reconstructed probes that pass the very loose likelihood identification. The $J/\psi$ meson peak around 3.1~\GeV{} is clearly visible. \protect\subref{fig:mll_taucut_den_sub2} Including all reconstructed probes that fail the very loose likelihood identification. There is only a small signal component in a large background. Due to poor momentum modelling and combinatorial effects, the observed $J/\psi$ meson peak is shifted to about 3.3~\GeV{}. The $\psi(2S)$ component is so small that it is barely visible in the figure.}\label{fig:mll_taucut_den}
\end{figure}

\begin{figure}[h!]
\centering
\subfloat[]{
\includegraphics[width=0.45\textwidth]{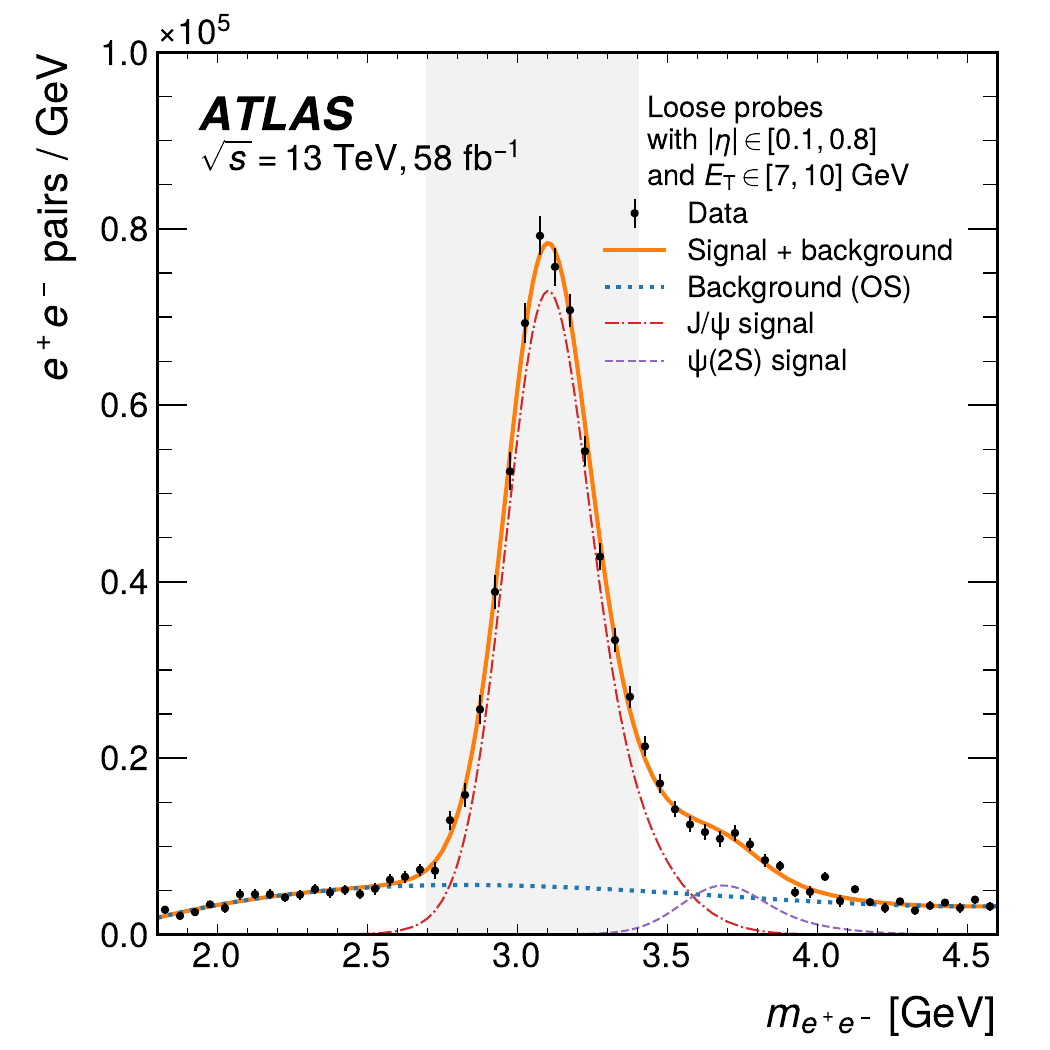}
\label{fig:mll_taucut_num_sub1}
}
\subfloat[]{
\includegraphics[width=0.45\textwidth]{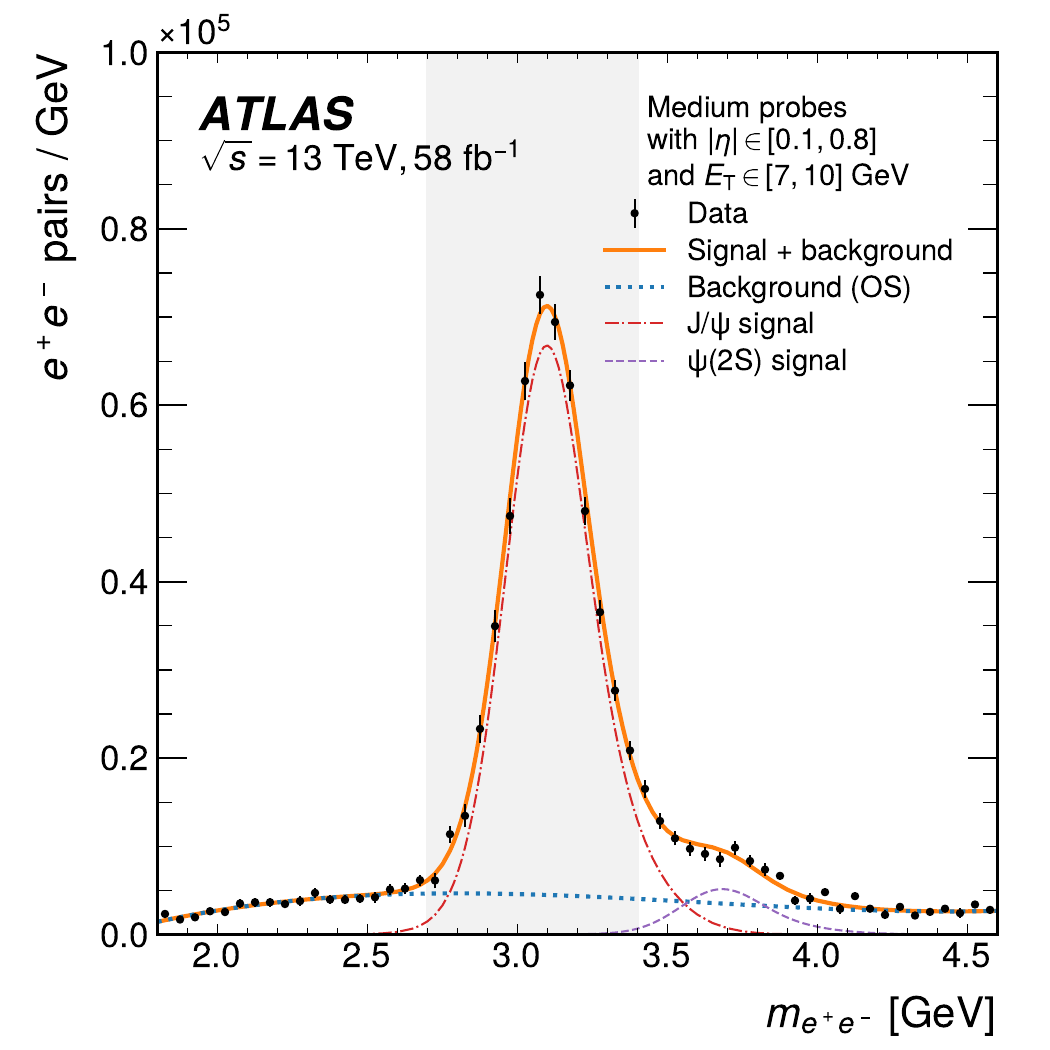}
\label{fig:mll_taucut_num_sub2}
} \\
\subfloat[]{
\includegraphics[width=0.45\textwidth]{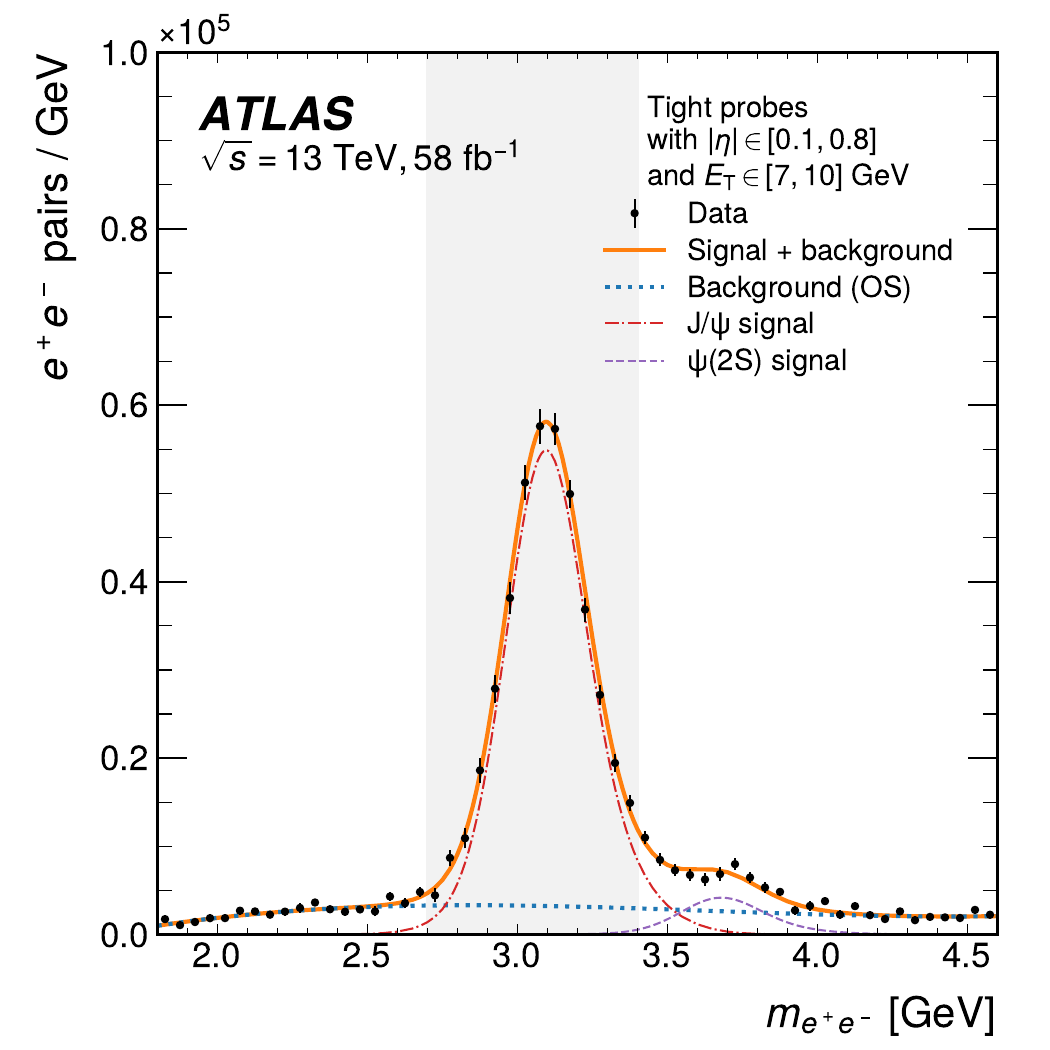}
\label{fig:mll_taucut_num_sub3}
}
\caption{Examples of nominal signal and background fits to the $m_{e^+e^-}$ distribution used to determine the efficiency numerators in the $J/\psi \to e^+e^-$ measurement, using 2018 data and probes with $\et \in [7, 10]~\GeV{}$ and $|\eta| \in [0.1, 0.8]$. The shaded band indicates the nominal mass range used for efficiency extraction. The probes passing Loose, Medium, Tight identification are included in \protect\subref{fig:mll_taucut_num_sub1}, \protect\subref{fig:mll_taucut_num_sub2}, \protect\subref{fig:mll_taucut_num_sub3}, respectively, yielding the amount of signal in the numerator of the corresponding efficiency. For electrons candidates passing identification criteria, there is often not enough events to determine the combinatorial background using the same charge region.}\label{fig:mll_taucut_num}
\end{figure}

After background subtraction, the efficiency is extracted in the range $m_{e^+e^-} \in [2.7, 3.4]~\GeV{}$. To improve the modelling of signal and background in the selection that enters the denominator of the efficiency, the selection of these probes is split into two orthogonal selections: into probes that fail or pass a very loose likelihood identification which is looser than the usual loose criterion. Hence, those probes that fail it only contribute to the denominator of the efficiency. An example of the \mee{} distribution for each of these selections, along with the various fitted signal and background contributions, is shown in Figure~\ref{fig:mll_taucut_den}. Example \mee{} distributions with probes passing the various likelihood identification working points are shown in Figure~\ref{fig:mll_taucut_num}.

The following variations are performed to establish systematic uncertainties of the low-\pt{} efficiency measurement:
\begin{itemize}
\item Changing the mass window for efficiency extraction to $m_{e^+e^-} \in [2.8, 3.3]~\GeV{}$.
\item Raising the upper limit of the pseudo-proper lifetime to $t_0 < 0.4$~ps.
\item Adding an isolation criterion on the probe: $p_{\mathrm{T}}^{\text{cone20}} / \pt{} < 0.02$, $p_{\mathrm{T}}^{\text{cone20}} / \pt{} < 0.15$, or $E_{\mathrm{T}}^{\text{cone20}} < 0.2$. This helps to vary the amount of non-prompt $J/\psi$ background without impacting the efficiency of the prompt $J/\psi$ sample significantly as tested on simulation. Applying the isolation to the probe gives a larger variation than applying it on the tag.
\item Adding an isolation criterion on both the tag and the probe:  $p_{\mathrm{T}}^{\text{cone20}} / \pt{} < 0.15$.
\item Changing the Chebyshev polynomial for opposite-charge background to second order.
\end{itemize}

Unlike for the methods using $Z \to e^+e^-$ events, the systematic variations are not combined by type. Instead, all variations for which the signal-region fit had a goodness of fit of $\chi^2$ per number of degrees of freedom of less than 4 are included and their root mean square distance from the nominal is taken to be the uncertainty. The reason for this different treatment is that the $J/\psi$ region suffers from much larger statistical fluctuations as well as modelling issues and individual systematic variations can yield a quite different efficiency value. Therefore this treatment can be understood as a regularisation procedure that reduces the effect of outliers. Less than 5\% of the systematic variations across all measurements are discarded due to this requirement.

\subsection{$Z$-mass method}
\label{subsec:ID_Zmass}

The $Z$-mass method uses reference $\mee$ distributions for signal and background,
referred to as templates,
obtained respectively in $Z \to e^+e^-$ MC simulation and using control regions in data.
Previously, the background under the $Z$-boson peak was estimated by normalising a background template in sidebands of the \mee\ distribution. A long-standing limitation of this method was a tension between normalising the background in the low-mass compared to the high-mass sideband, which was reflected in a large systematic uncertainty. The reason for this tension was found to be the considerably different levels of signal contamination in the low- and high-mass sidebands and a non-optimal method to describe and normalize this signal contamination when creating the background template, biasing its shape. A new iterative procedure was devised to reliably subtract the background
and to reduce dependence on the MC predictions. This yields a significantly more reliable efficiency measurement.
The following steps are performed as part of the $Z$-mass method:
\begin{itemize}
\item The background templates are defined using a background control region where the probe electron is required to fail a relaxed version of the Loose likelihood identification and to have a minimum amount of energy deposited close-by, $E_{\mathrm{T}}^{\text{cone30}} / \pt{} > 0.1$.
\item These background templates still contain signal contamination, which is subtracted in a step called \emph{template cleaning}. The template cleaning consists of fitting the background template using a suitable polynomial for the background and a template for $Z \to e^+e^-$ contamination obtained from MC simulation to extract the relative contributions of the background and the contamination of the signal to the background template. The scaled signal contamination is then subtracted to create a pure background template.
\item The pure background template and the MC signal template are fitted to the data in the signal region to extract the background normalization. Post-fit, the scaled background template is then subtracted from the data. As a result of residual miscalibrations and inaccuracies of the MC description of the $Z$ line shape, a discrepancy between the data and the MC simulation of a few percent can be observed.
\item Therefore weighting factors are extracted to describe the signal invariant mass distribution better in data. These weighting factors are obtained only for probes passing the Tight identification and are calculated as the ratio of the background-subtracted data to the signal MC prediction in bins of \mee. The Tight probes only have small background and therefore give a pure, unbiased signal shape with the greatest confidence.
\item The weighting factors are applied to all signal \mee\ distributions taken from the MC simulation, independent of the requirements applied to the probe electrons, and for both the background control region and the signal region.
\item This procedure is iterated once more, starting from the \emph{template cleaning}, the first step in the procedure.
\end{itemize}

The systematic uncertainties of the $Z$-mass method are estimated by varying choices in the method and taking the resulting efficiencies as systematic variations of the nominal one. The variations are:
\begin{itemize}
\item The background control region is varied by changing the very loose likelihood identification the candidate probes need to fail or alternatively the isolation requirement is varied to $E_{\mathrm{T}}^{\text{cone40}} / \pt{}> 0.07$.
\item The signal region fit is repeated using only $\mee$ ranges below or above the $Z$-boson peak, respectively.
\item In all fits, the fit uncertainty is applied to the normalisations of the templates as a systematic variations.
\item The tag electron isolation is tightened by adding the requirement $E_{\mathrm{T}}^{\text{cone}40} < 5~\GeV$.
\item The efficiency is extracted by counting the background-subtracted data yields in the \mee\ ranges $[70, 110]~\GeV{}$ and $[80, 100]$~\GeV{}.
\end{itemize}

The systematic uncertainties evaluated as described above are summed in quadrature.

\begin{figure}[h!]
\centering
\subfloat[]{
\includegraphics[width=0.45\textwidth]{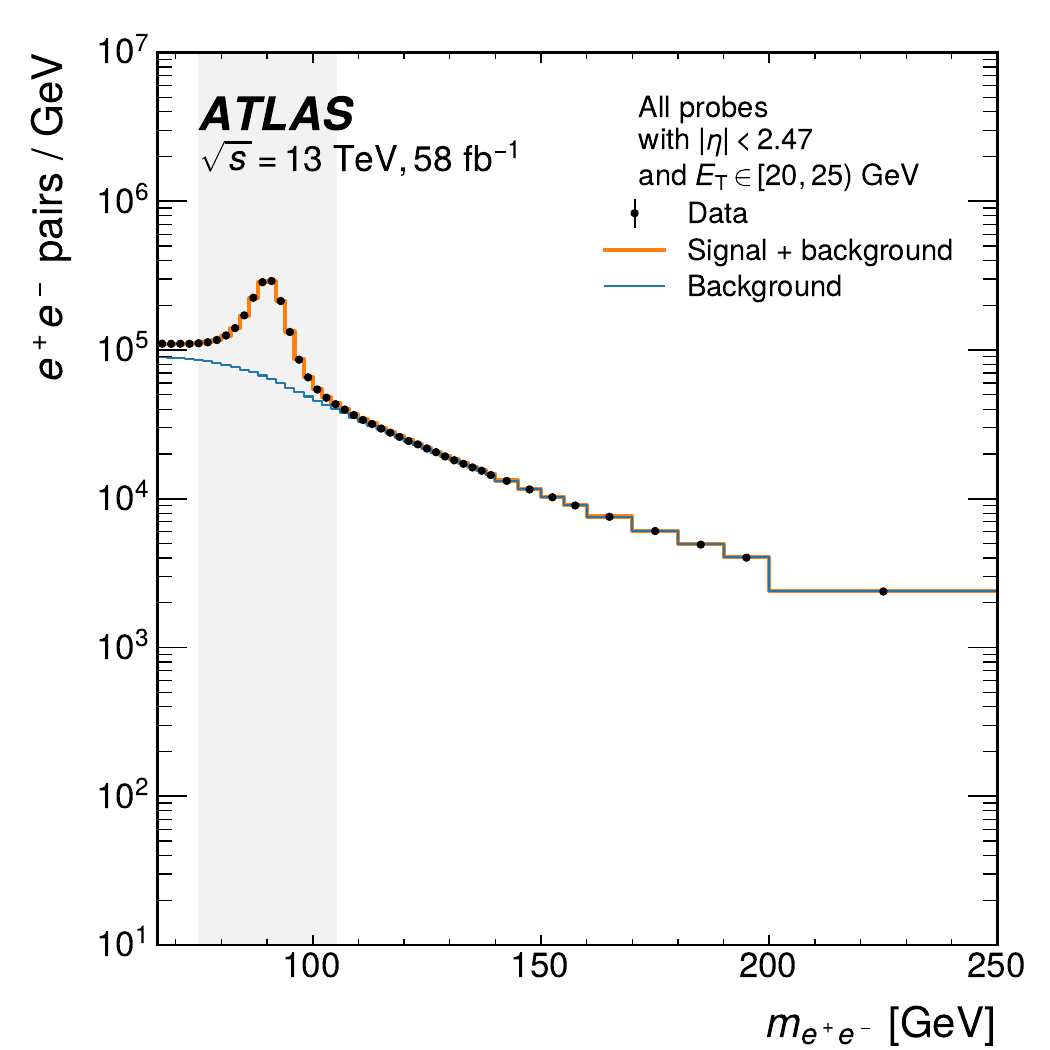}
\label{fig:zmass_fits_sub1}
}
\subfloat[]{
\includegraphics[width=0.45\textwidth]{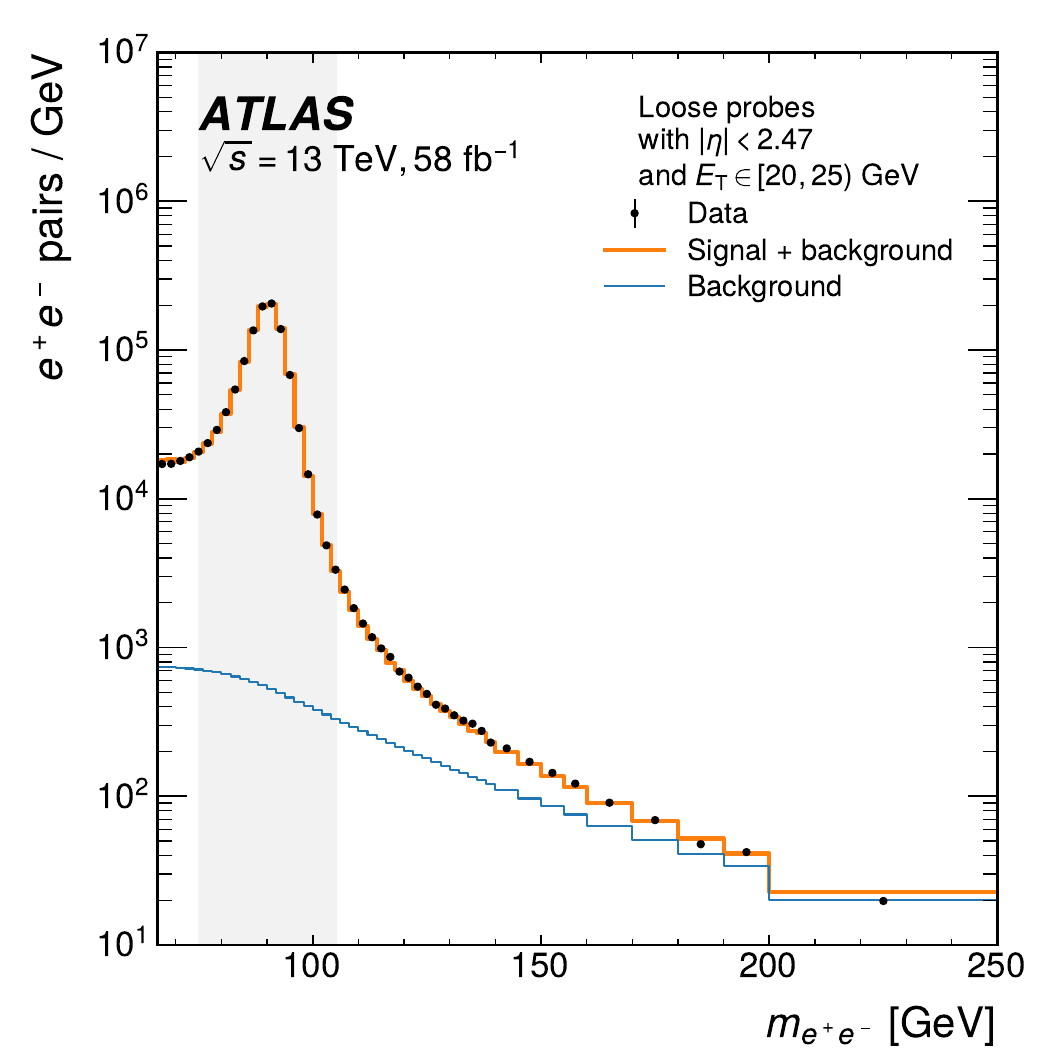}
\label{fig:zmass_fits_sub2}
} \\
\subfloat[]{
\includegraphics[width=0.45\textwidth]{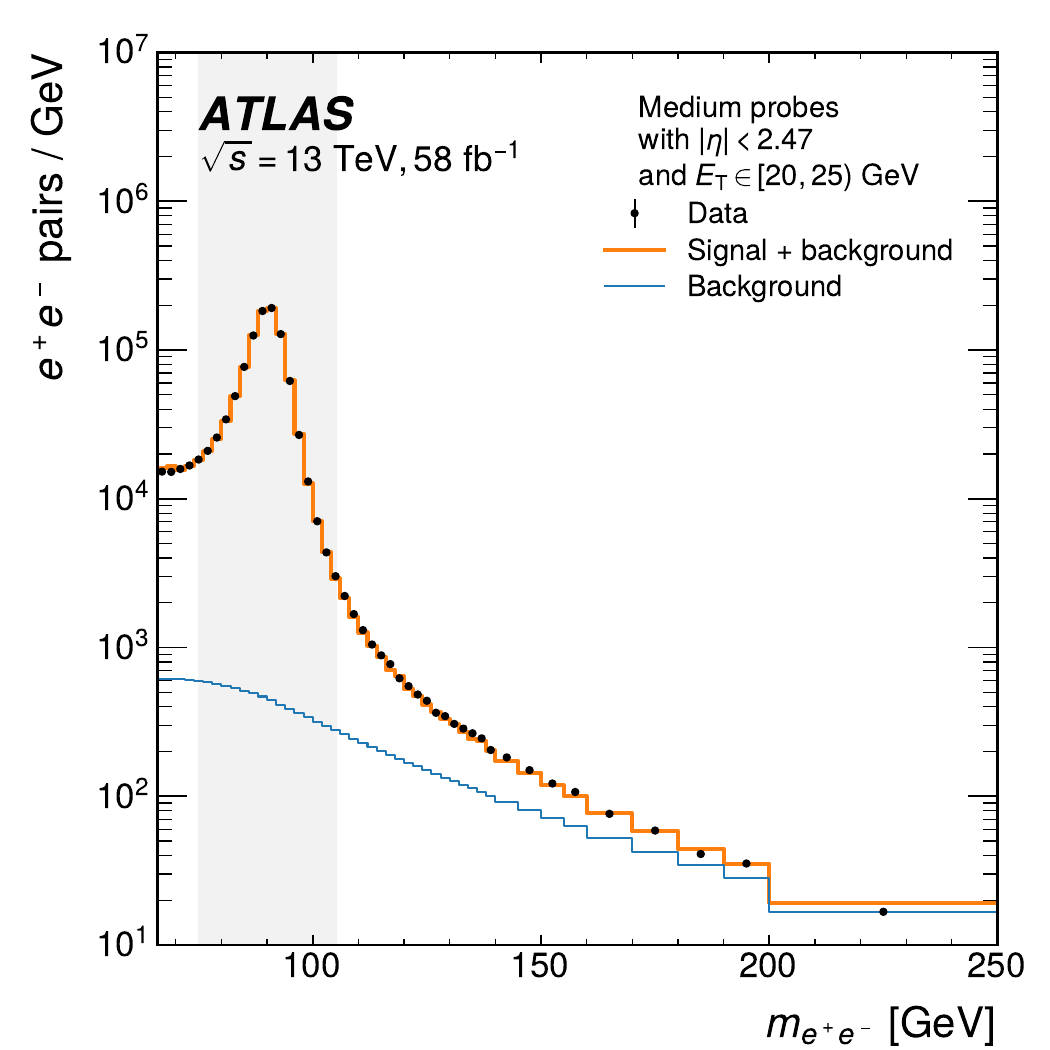}
\label{fig:zmass_fits_sub3}
}
\subfloat[]{
\includegraphics[width=0.45\textwidth]{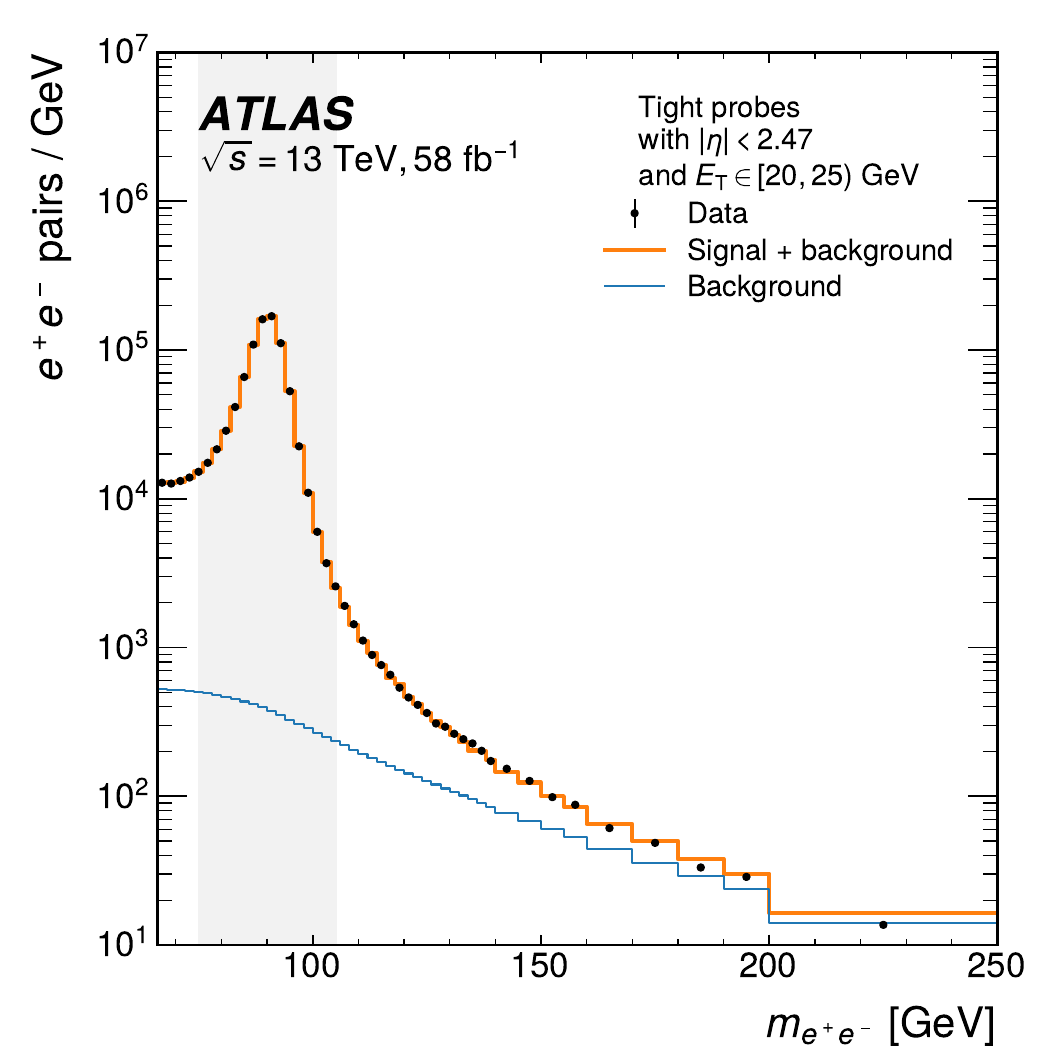}
\label{fig:zmass_fits_sub4}
}
\caption{Examples of nominal signal and background template fits to the $m_{e^+e^-}$ distribution in the $Z$-mass method, using probes with $\et \in [20, 25]~\GeV{}$. The $Z$-boson peak around 91~\GeV{} is clearly visible. The shaded band indicates the nominal mass range used for efficiency extraction. \protect\subref{fig:zmass_fits_sub1} Including all the reconstructed probes, yielding the amount of signal in the denominator of the efficiency. Including probes passing the \protect\subref{fig:zmass_fits_sub2} Loose, \protect\subref{fig:zmass_fits_sub3} Medium, \protect\subref{fig:zmass_fits_sub4} Tight identification, respectively, yielding the amount of signal in the numerator of the corresponding efficiency. }\label{fig:zmass_fits}
\end{figure}

Examples of $\mee$ distributions used as the sensitive variable to establish the amounts of signal and background in the $Z$-mass method are shown in Figure~\ref{fig:zmass_fits} for the case including all reconstructed probes (as in the identification efficiency denominator) and including probes passing the different identification working points (as in the numerator of the corresponding efficiency). For simplicity, the figures include probes over the full used pseudorapidity range $|\eta| < 2.47$, whereas the actual measurement is done in bins of $\eta$.

\subsection{$Z$-isolation method}

The $Z$-isolation method relies on the amount of transverse energy in an isolation cone of radius $\Delta R = \mathrm{XX}/100$ around the probe electron, $E_{\mathrm{T}}^{\text{coneXX}}$, to distinguish between signal and background. Background templates are defined in data in a background-enriched region where the charges of the tag and probe candidate are required to have the same sign and the probes must fail cuts on various identification variables. The requirements are designed to be as uncorrelated with $E_{\mathrm{T}}^{\text{coneXX}}$ as possible so as to avoid bias. The signal contamination is subtracted from the background templates using MC simulation.

The exact selection of the background templates, one nominal and one systematic variation, is done using the MC-simulated QCD sample by choosing template definitions in each $\et$-$\eta$ bin that minimise the bias defined as
\begin{equation}\label{eq:ziso_bias}
\text{bias} = \left| \, \frac{N^{\mathrm{template}}(E_{\mathrm{T}}^{\text{cone30}} < 7.5~\GeV)}{N^{\mathrm{template}}(E_{\mathrm{T}}^{\text{cone30}} > 7.5~\GeV)}
\,
\frac{N^{\mathrm{all}}(E_{\mathrm{T}}^{\text{cone30}} > 7.5~\GeV)}{N^{\mathrm{all}}(E_{\mathrm{T}}^{\text{cone30}} < 7.5~\GeV)}
\; -\; 1\, \right|,
\end{equation}
where $N(*)$ signifies a yield of MC-simulated background probes in the tag-and-probe region with requirement $*$ applied to the probe, and the superscripts ``all'' and ``template'' indicate that either all MC-simulated background probes are used, or only those passing the template selection, respectively.
The bias variable quantifies how well the background template reproduces the true (MC-simulated) ratio of the background in the $E_{\mathrm{T}}^{\text{cone30}}$ high tails and at low $E_{\mathrm{T}}^{\text{cone30}}$.

In most $\et$-$\eta$ bins, the chosen templates contain less than 5\% signal contamination. However, for $|\eta| > 1.81$, $\et > 25~\GeV$ this increases to about 25\%. There are two different template definitions that are used depending mostly on the probe electron $\et$.

The efficiency is extracted by considering the yields in the $Z$-boson peak region with probe isolation $E_{\mathrm{T}}^{\text{cone30}} < 12.5~\GeV$. The background is estimated by normalising the background template in the region $E_{\mathrm{T}}^{\text{cone30}} > 12.5~\GeV$, where the signal contamination has been subtracted using MC simulation.

\begin{figure}[h!]
\centering
\subfloat[]{	\includegraphics[width=0.45\textwidth]{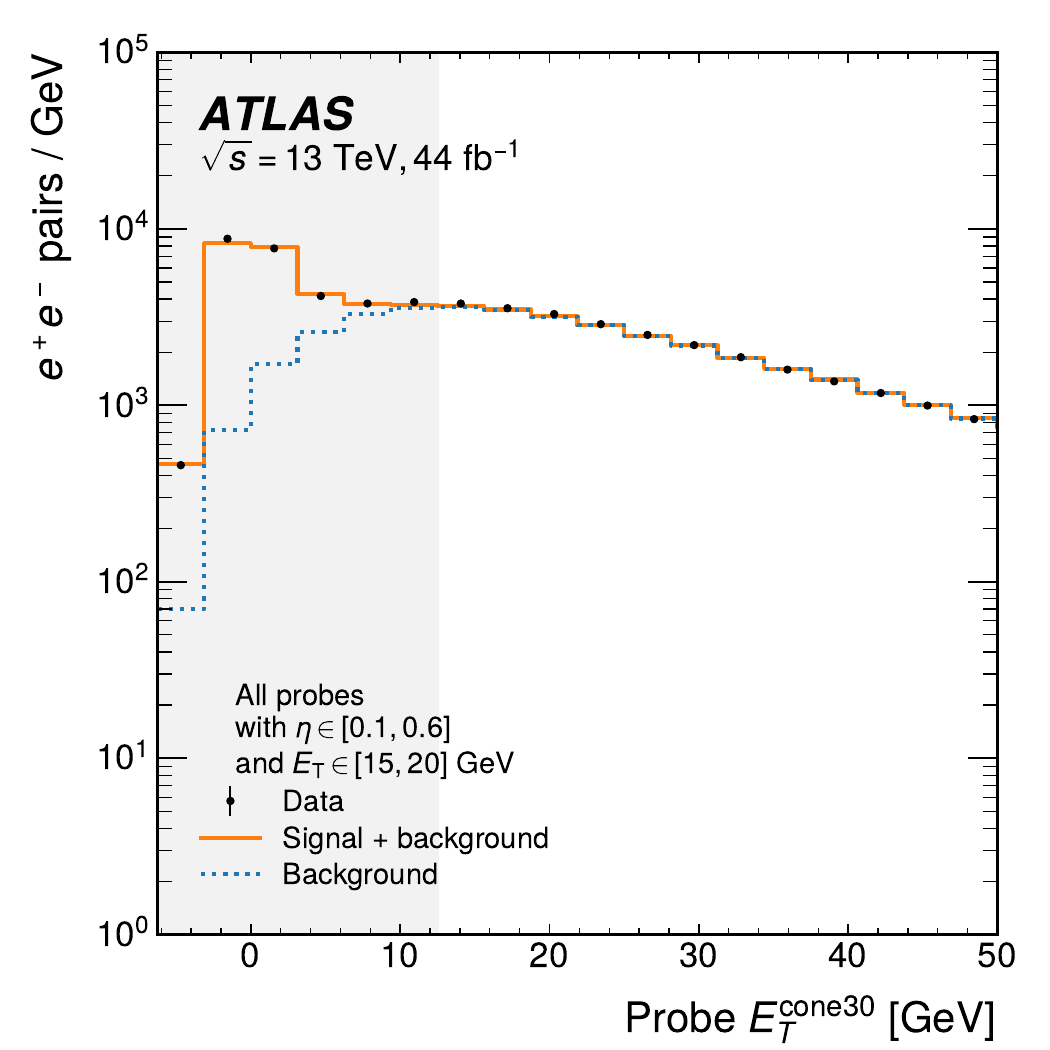} \label{fig:ziso_etcone30_sub1} }
\subfloat[]{	\includegraphics[width=0.45\textwidth]{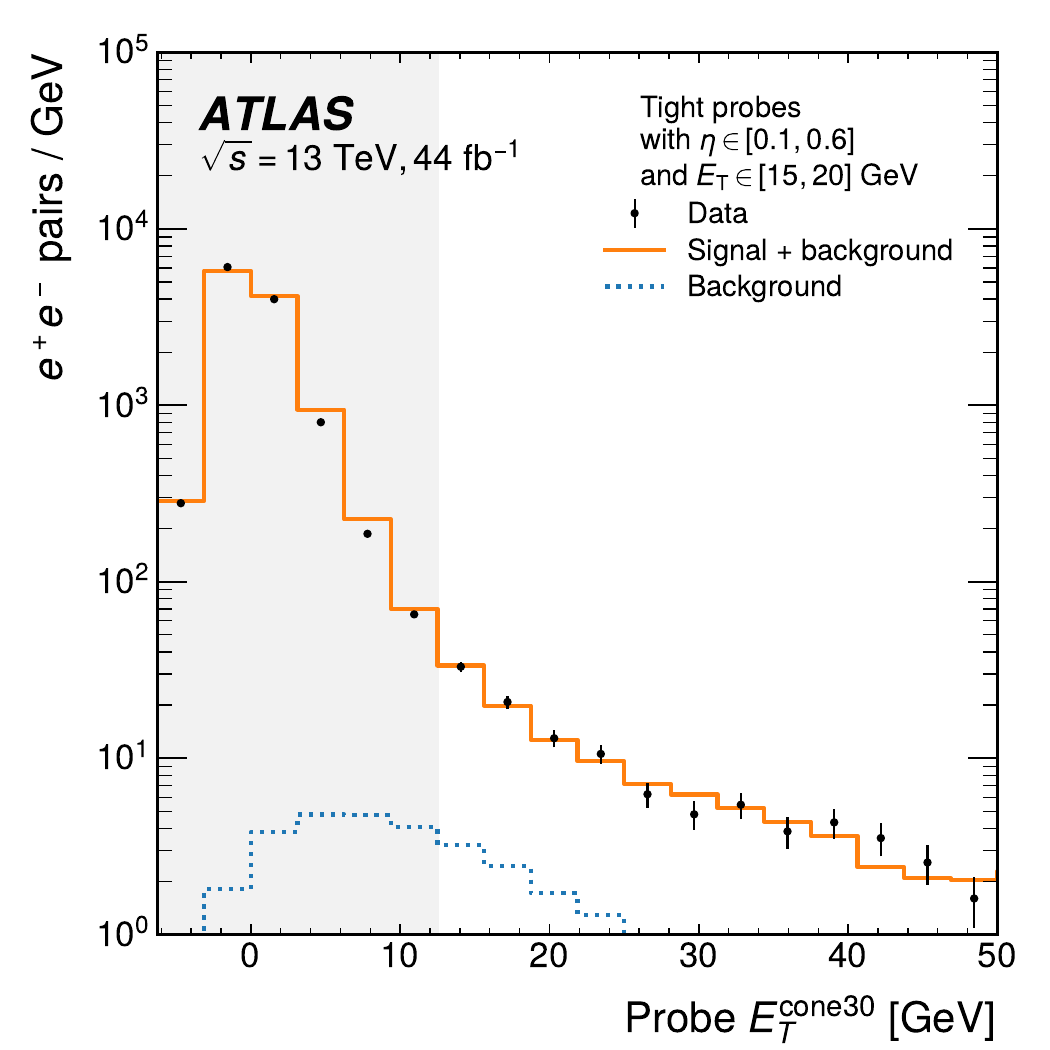}	\label{fig:ziso_etcone30_sub2} }
\caption{Examples of the probe $E_{\mathrm{T}}^{\text{cone30}}$ isolation variable distribution with the corresponding signal and background, using probes with $\eta \in [0.1, 0.6]$ and $\et \in [15, 20]~\GeV{}$. The nominal mass requirement $m_{e^+e^-} \in [75, 105]~\GeV{}$ is already applied. The background template is normalised in the tail $E_{\mathrm{T}}^{\text{cone30}} > 12.5~\GeV$. The shaded area, $E_{\mathrm{T}}^{\text{cone30}} < 12.5~\GeV$, shows the nominal region used for efficiency extraction. \protect\subref{fig:ziso_etcone30_sub1} Including all the reconstructed probes, yielding the amount of signal in the denominator of the efficiency. \protect\subref{fig:ziso_etcone30_sub2} Including probes passing Tight identification, yielding the amount of signal in the numerator of the corresponding efficiency.}\label{fig:ziso_etcone30}
\end{figure}

Figure~\ref{fig:ziso_etcone30} shows examples of the distribution of the probe isolation variable $E_{\mathrm{T}}^{\text{cone30}}$ in data along with the signal and background predictions. It can be seen that signal electrons are more isolated than the background and that the two contributions together describe the observed isolation distributions reasonably well, when considering that there are some modelling problems in the MC simulation.

The following variations are performed to estimate systematic uncertainties of the $Z$-isolation method:
\begin{itemize}
\item Using an alternative background template chosen based on the bias defined in Eq.~\eqref{eq:ziso_bias}. It is the same for the whole kinematic phase space and effectively is the second-least biased template found.
\item Scaling the normalisation of signal contamination subtracted from the background (both the template and the tail region where it is normalised) by an estimated $\pm$30\% to conservatively account for modelling uncertainties.
\item Using $E_{\mathrm{T}}^{\text{cone40}}$ to define the background normalisation and signal regions (instead of $E_{\mathrm{T}}^{\text{cone30}}$).
\item Varying the $E_{\mathrm{T}}^{\text{cone30}}$ boundary defining the background normalisation and signal regions to 10~\GeV{} and 15~\GeV{}.
\item An additional tag electron isolation $E_{\mathrm{T}}^{\text{cone}30}  < 7.5$~\GeV{} is applied.
\item The efficiency is extracted by counting the background-subtracted data yields in the \mee\ ranges $[70, 110]~\GeV{}$ and $[80, 100]$~\GeV{}.
\end{itemize}

The systematic variations are combined in the same way as in the $Z$-mass method, i.e., each of them is interpreted as one symmetric systematic uncertainty. In the case where the systematic uncertainty is evaluated by varying a parameter up and down, the uncertainty is symmetrised and the largest variation is used as a systematic source.

\subsection{Combination}
To calculate the final results for the identification efficiency, the correction factors obtained with the $Z$ methods are combined by averaging them weighted by their uncertainties in each $\et$-$\eta$ bin independently but taking into account their correlations following the procedure in Ref.~\cite{Schmelling:273386}. During the combination, those uncertainties that stem from the modelling of the backgrounds are treated as uncorrelated since they are specific to each of the two $Z$ methods. However, the uncertainty related to the signal definition, estimated by varying the \mee\ window and the uncertainty from varying the amount of background by tightening the isolation requirements on the tag electron are treated as fully correlated. In addition, statistical uncertainties are treated as fully correlated for the signal events between the two methods whilst background events are considered to be uncorrelated. The improvements to the $Z$-mass and $Z$-isolation methods have both shrunk the systematic uncertainties in each method and significantly improved the agreement between the two methods, which was previously driving their combined uncertainty. Whilst there were differences of up to 2\% between the methods, this has now been reduced to less than 0.5\% on average.

In a final step, the $Z$ correction factors are combined with those obtained using the $J/\psi$ resonance in the overlapping $\et$ range, $[15, 20]~\GeV{}$. This combination uses a $\chi^2$ minimization with a profiling of the systematic uncertainties following the method reported in Ref.~\cite{H1:2009jxj}. Thanks to the reported improvements, the uncertainties on the combined corrections factors have been halved compared with previous measurements~\cite{EGAM-2018-01}.

\subsection{Results}
\begin{figure}[h!]
\centering
\subfloat[]{
\includegraphics[width=0.48\textwidth]{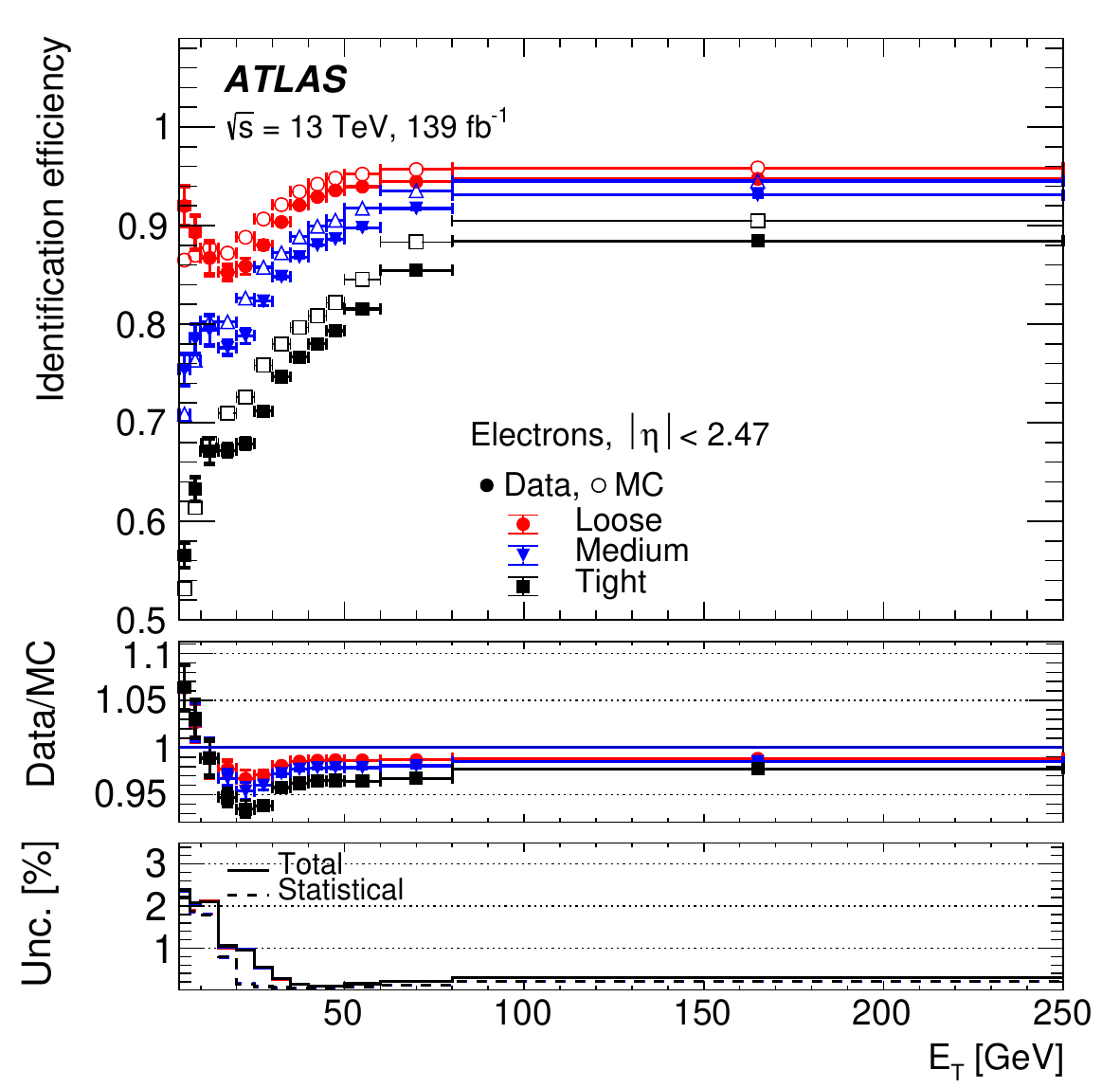} %
\label{fig:id_pt}
}
\subfloat[]{
\includegraphics[width=0.48\textwidth]{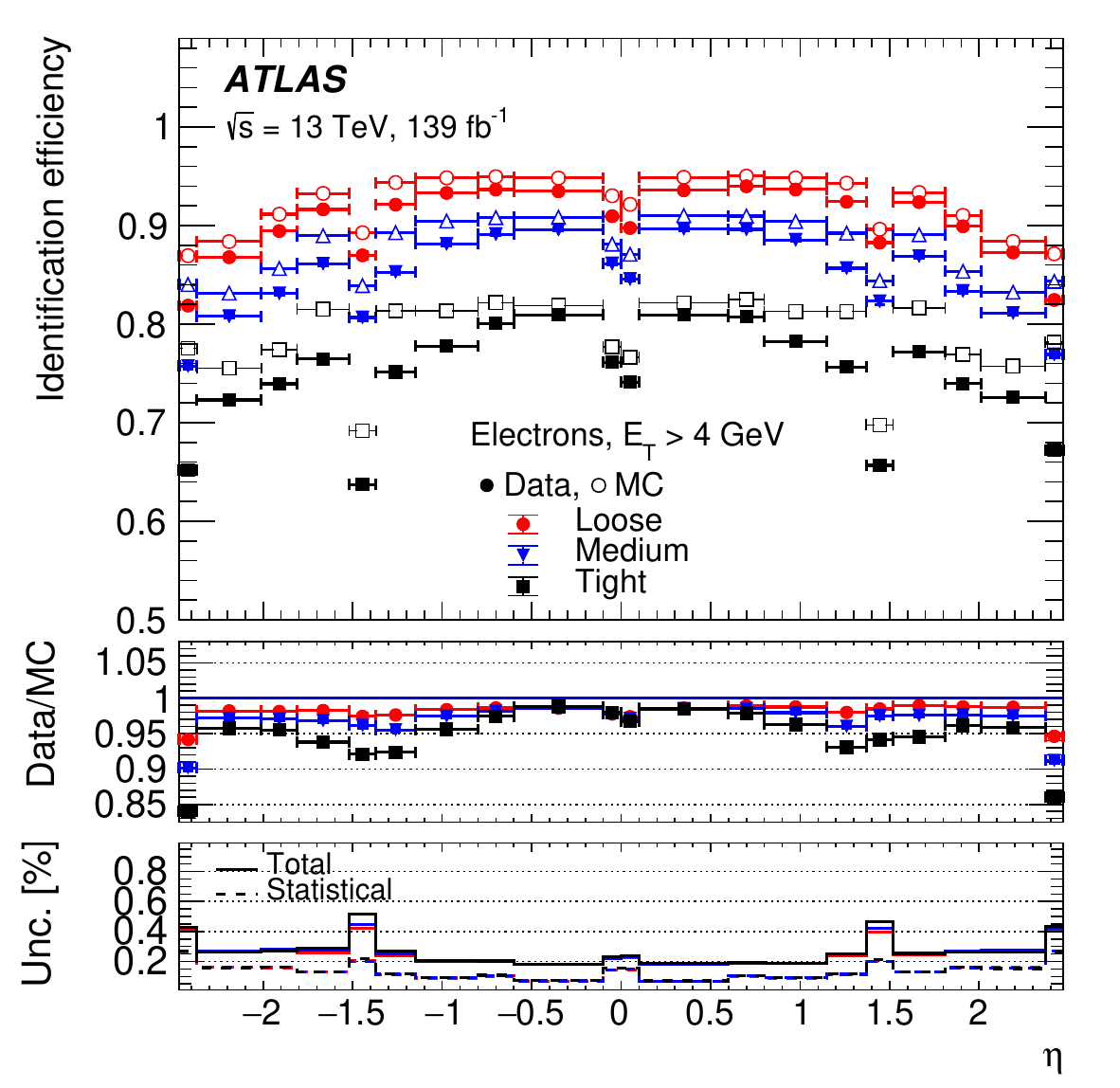}
\label{fig:id_eta}
}
\caption{Identification efficiencies of electrons from $Z \to e^+e^-$ decays as a function of the electron's \protect\subref{fig:id_pt} transverse momentum and \protect\subref{fig:id_eta} pseudorapidity for the different identification working points. The top panels show the efficiencies obtained in data and simulation with their statistical and total uncertainties displayed as inner and outer error bars. The middle panels show their ratio that is applied as correction factor in analyses. The bottom panels show the relative statistical and total uncertainties in the data/MC ratio.}
\end{figure}

The measured identification efficiencies for the various working points are shown in Figure~\ref{fig:id_pt} as a function of $\et$ and in Figure~\ref{fig:id_eta} as a function of $\eta$. In each case, the respective other variable is integrated over. The ratio of efficiencies in data and MC simulation is close to unity, typically within 5\%. Except for $\pt{} < 15$~\GeV{}, the efficiencies are slightly lower in data than in simulation. Looser identification criteria show a better agreement between data and MC simulation. The efficiencies generally decrease for lower transverse momenta.\footnote{For the Loose identification criterion, an updated version of the likelihood was created using more appropriate shower-shapes in the tuning process that has a better efficiency specifically in the $\pt{}$ range of 15--20~\GeV, and it is used for $H\rightarrow 4\ell$ analyses~\cite{HIGG-2018-29}, but is not shown here.} The efficiencies measured as function of pseudo-rapidity are found to be mostly symmetric between the negative and positive ranges of $\eta$, especially when considering uncertainties.

\begin{figure}[h!]
\centering
\includegraphics[width=.65\textwidth]{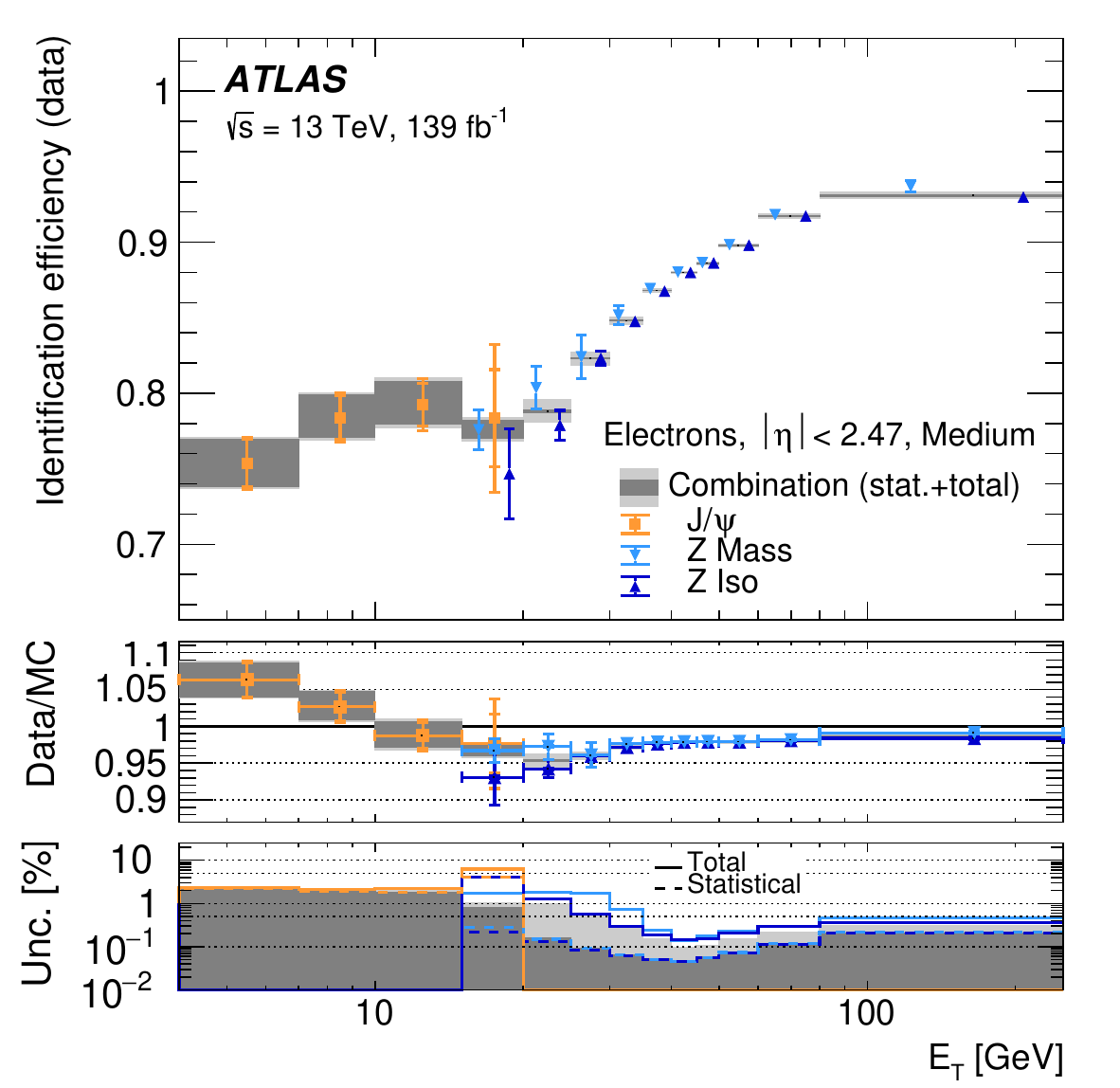}
\caption{Data electron identification efficiencies using the $J/\psi$ and $Z$ resonance are shown as a function of the electron's transverse energy. They are obtained by applying correction factors to the electrons from $Z \to e^+e^-$ events before the integration over $\eta$ and therefore reflect the underlying kinematics of $Z \to e^+e^-$ events. The Medium identification working point is shown for the breakdown of individual measurements that enter the combination as well as the results of the combination. The middle panel contains the ratio of the measured to the MC-simulated efficiencies. The bottom panel gives the relative uncertainties with the horizontal dotted lines indicating uncertainties of 0.1, 0.5, 1, 5 and 10\%. }\label{fig:id_combi}
\end{figure}

In Figure~\ref{fig:id_combi}, the individual measurement methods in their respective range of applicability and the combination are shown. The Medium identification efficiencies are shown for electrons in simulated $Z \to e^+e^-$ events with the corrections factors applied such that they reflect the data efficiencies. They give a general idea of the performance in various kinematic domains that are of interest in ATLAS analysis even if they are based on the $Z \to e^+e^-$ topology and do not necessarily reflect the efficiencies of other processes. The comparison of the relative uncertainties of the individual methods to the results of the combination illustrate the impressive improvements in the energy range above 15~\GeV{} where the systematic uncertainties dominate over the statistical uncertainties.

\begin{figure}[h!]
\centering
\subfloat[]{
\includegraphics[width=0.48\textwidth]{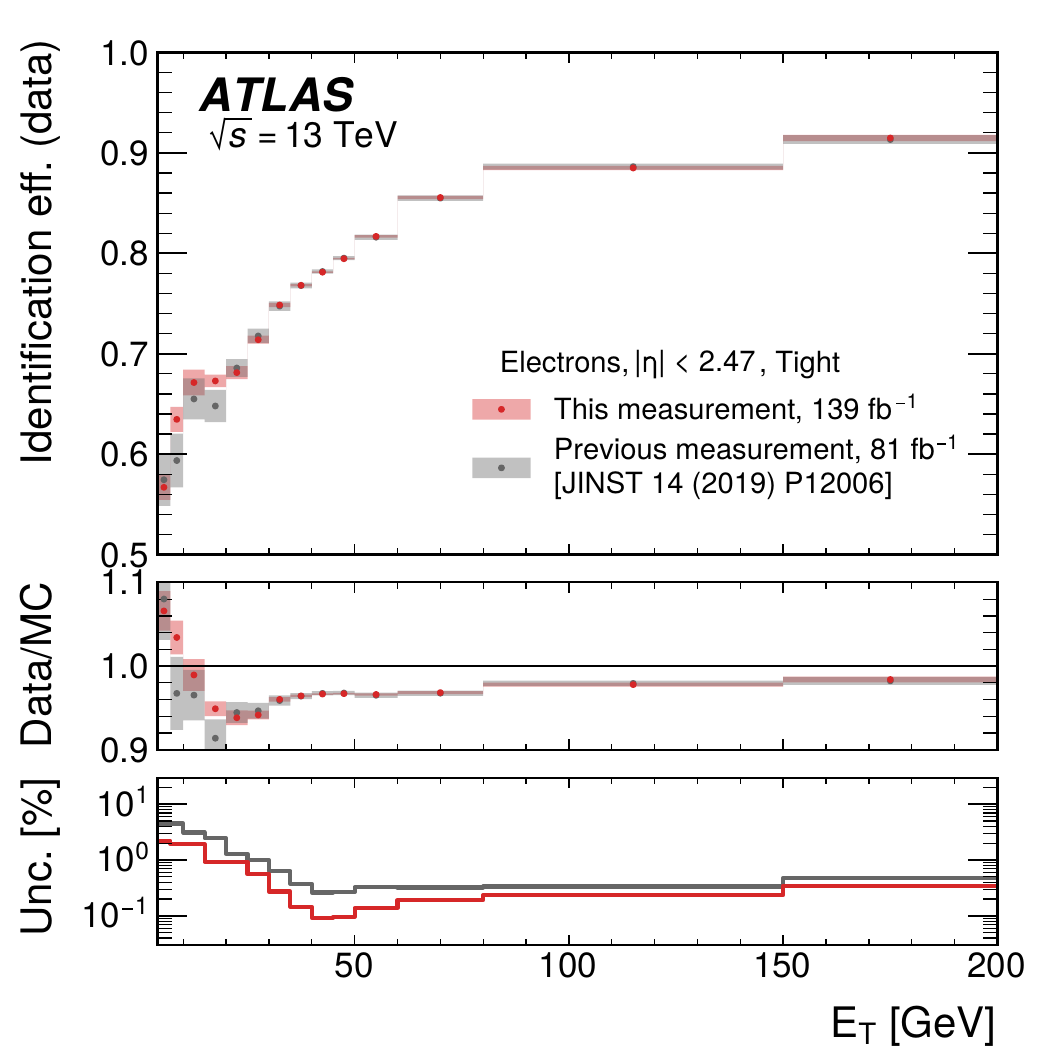}
\label{fig:id_comparison_to_previous_sub1}
}
\subfloat[]{
\includegraphics[width=0.48\textwidth]{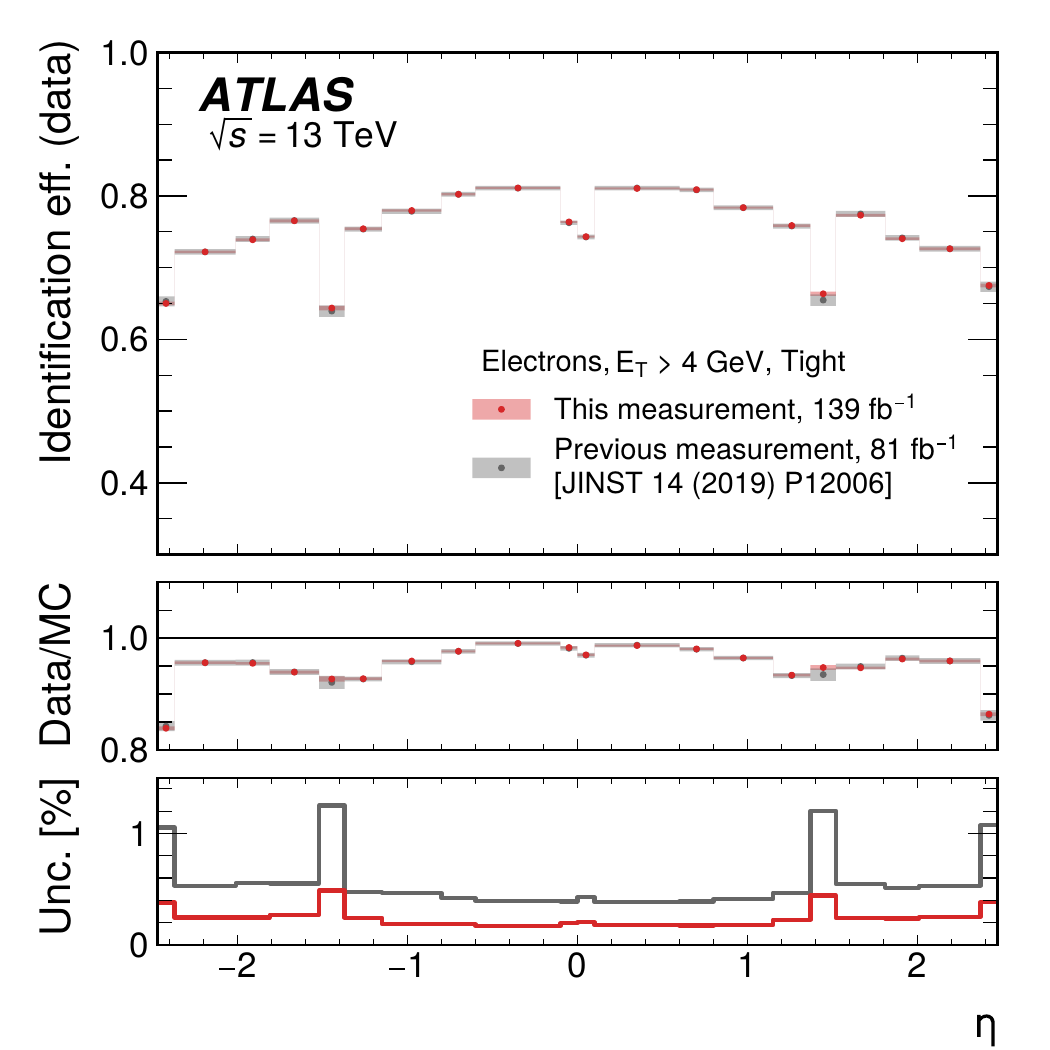}
\label{fig:id_comparison_to_previous_sub2}
}
\caption{Comparison of the new combined Tight identification efficiencies to the previous ATLAS measurement using data from the years 2015–2017~\cite{EGAM-2018-01} as a function of the electron's \protect\subref{fig:id_comparison_to_previous_sub1} transverse energy and \protect\subref{fig:id_comparison_to_previous_sub2} pseudorapidity. The middle panels contain the ratio of the measured to the MC-simulated efficiencies. The bottom panels give the relative uncertainty of the combination of measurements.}\label{fig:id_comparison_to_previous}
\end{figure}

Figure~\ref{fig:id_comparison_to_previous} shows comparisons of the new combined Tight identification efficiencies to the previous ATLAS measurement using data from the years 2015–2017~\cite{EGAM-2018-01}. The new measurement is in good agreement with the previous one. When evaluated by integrating over the electron kinematics in simulated $Z \to e^+e^-$ events, the efficiency uncertainty has decreased by almost one third for the Loose working point and by more than half for the Medium and Tight working points. The uncertainty decrease is quite independent of $\eta$, small for low $\pt$ ($< 15$~\GeV{}) and high \pt{} ($> 80$~\GeV{}), and large for the \pt{} range between these extremes.

\begin{figure}[h!]
\centering
\includegraphics[width=.65\textwidth]{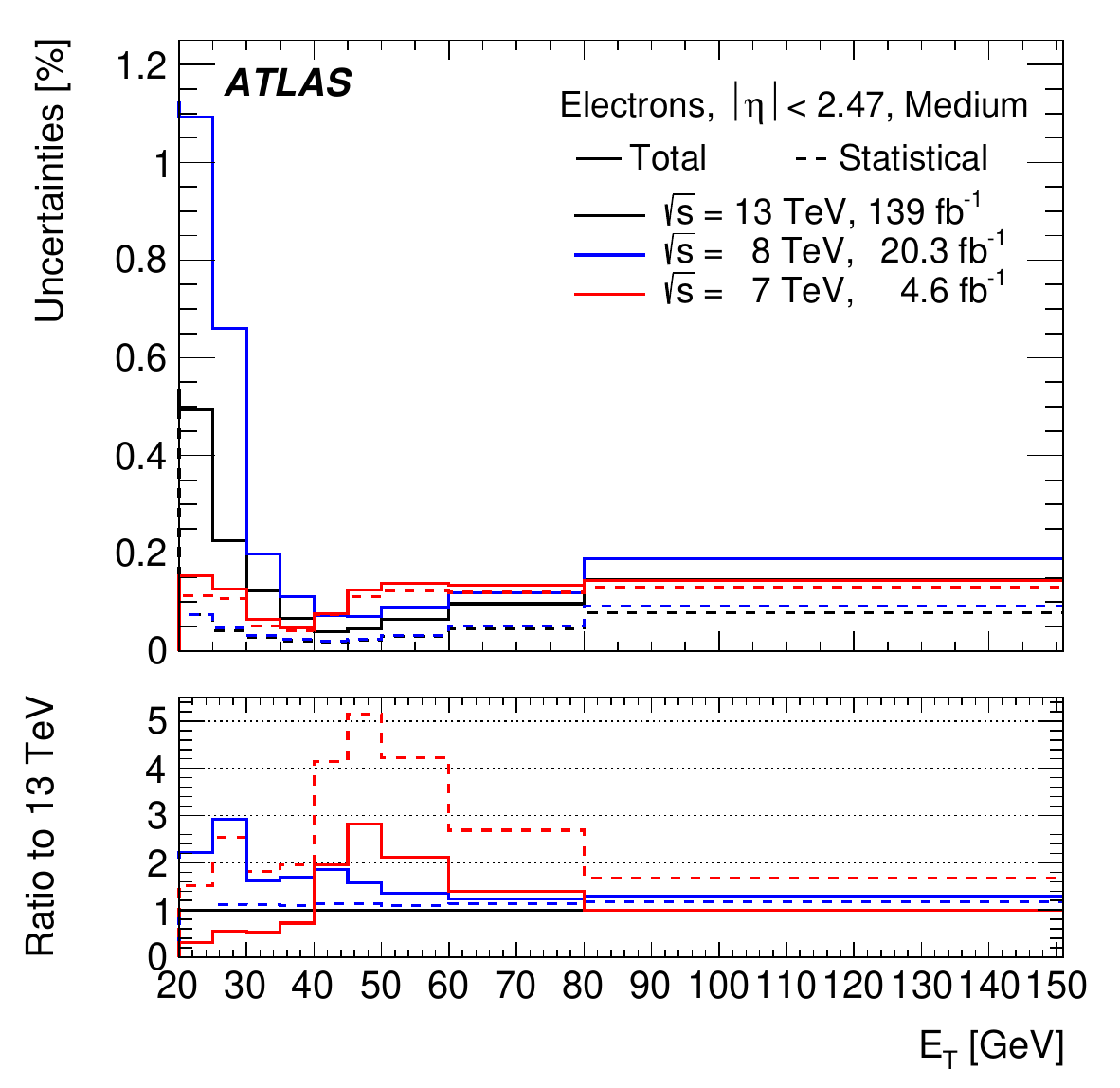}
\caption{Uncertainties on the data-to-MC correction factors shown for the underlying kinematics of $Z \to e^+e^-$ events at 13~\TeV, as a function of the electron's transverse energy. The Medium identification working point is shown and compared for the main $pp$ data-taking periods at the LHC.  }\label{fig:id_allruns}
\end{figure}

Figure~\ref{fig:id_allruns} depicts the uncertainties on the data-to-MC correction factor for the Medium identification efficiencies as a function of the transverse energy of the electron for the main $pp$ run periods at the LHC, namely Run 1 data taking at collision energies of $\sqrt{s}=7$~\TeV{} (2011) and $\sqrt{s}=8$~\TeV{} (2012) as well as Run 2 data taking at $\sqrt{s}=13$~\TeV{} (2015--2018). The experiment has been able to tune the definition of the Medium identification to reach target efficiencies of about 80~\% for each of the data-taking periods despite the significant increase in pile-up for the 13~\TeV{} data. Despite the increased backgrounds, the uncertainties on the data/MC correction factors have been reduced for most of the energy range except for $\et < 40$~\GeV, where the low pile-up environment of the 2011 data-taking period had allowed to employ the $W$-boson resonance for efficiency determination in addition to the methods used in 2015--2018~\cite{PERF-2013-03}. The comparison is made between the respective correction factors used in high precision analyses, which are applicable only down to transverse energies of 20~\GeV.

%

%
\section{Background rejection}
\label{sec:rej}

Section~\ref{sec:ele_reco} and~\ref{sec:ele_id} introduced the measurement of the prompt electron efficiencies and the MC to data correction factors. Of equal interest is the knowledge of the background rejection and of how well the rejection is modelled in the MC simulation. Whilst this is not immediately applicable in physics analyses, it is important for their design. The signature of a cluster with a matched track can be caused by a variety of particles other than electrons, and the fact that electrons and jets are reconstructed independently of each other causes prompt electron candidates to be reconstructed as jets and vice versa. The knowledge of the general rates of jet production combined with rejection rates can therefore give indications of expected backgrounds. In the following, rejection rates of the reconstruction are studied in data by determining how many jets are also reconstructed as electrons. Rejection rates of the identification with respect to the reconstruction can only be studied using MC simulation because of the large rejection rates leading to large amounts of prompt electrons in any data samples.

\subsection{Measurement of the background rejection of the electron reconstruction in data}

The background rejection of the electron reconstruction is measured using the production of $Z$ bosons in association with jets. Jets are reconstructed using the anti-$k_t$ algorithm with a radius parameter of $R = 0.4$ and topological clusters as input~\cite{Cacciari:2008gp,PERF-2014-07}. Jets originating from pile-up are rejected using an algorithm based on the fraction of tracks originating from the primary vertex~\cite{PERF-2014-03} or outside the ID coverage based on missing transverse momentum~\cite{PERF-2016-06}. Jets are calibrated as reported in Ref.~\cite{JETM-2018-05}. Events are selected requiring at least two oppositely charged electrons passing the medium identification criteria and isolated from hadronic activity measured using tracks, $\ptcs /\pt < 0.15$. The invariant mass of the dielectron system is required to lie between $81.18~\GeV{} < \mee < 101.18$~\GeV{} and the pair closest to the $Z$-boson mass
is taken to consist of prompt electrons, whilst all other reconstructed electron candidates in the event are assumed to be backgrounds misreconstructed as electron candidates. These electron candidates are required to be reconstructed and satisfy the general quality requirements described in Section~\ref{sec:ele_reco}. Jets closer than $\Delta R=0.4$ to the electrons forming the pair closest to the $Z$ mass are removed. Only those electrons reconstructed in the central region having $|\eta|< 2.47$ and $\et > 30$~\GeV{} are considered for the rejection measurement. MC studies suggest that electroweak processes with a $Z$ boson and additional prompt isolated electrons contribute less than 0.5\% to the selected events and therefore no attempt was made to subtract prompt isolated electrons. For the study, only the 2017 $pp$ data using 44 \ifb\ with a mean number of interactions per bunch crossing of 37.8 is used. The pile-up in this data sample is slightly larger than in the other years of data taking, but given its relatively small contribution to background electrons of 5-10~\%, can still be taken as representative of the other years.

The selected jets and the electron candidates have very different energy scales, as depicted in Figure~\ref{fig:jetele_diff}. Here the relative difference between jet and associated fake electron, normalised to the jet \pt, $\frac{\pt^{\textrm{jet}} - \pt^{\textrm{e}}}{\pt^{\textrm{jet}}}$, is shown for all fake electrons with a transverse energy above 4~\GeV{} and for those with $\et>18$~\GeV. In most cases, an electron candidate reconstructed from a jet has only about 10\% of the \pt\ of its jets.  The large difference in the \pt\ distributions between jets and their associated reconstructed background electrons explains the breadth of the energy difference distribution.
Figure~\ref{fig:diff_all} is dominated by very low \et\ electrons as without further restrictions any jet with an electromagnetic energy component of more than 4~\GeV{} is likely to be reconstructed as an electron. Once a suitable energy requirement of \et$>18$~\GeV{} is applied to suppress backgrounds (Fig~\ref{fig:diff_18gev}), the peak of the distribution shifts to lower values of $\frac{\pt^{\textrm{jet}} - \pt^{\textrm{e}}}{\pt^{\textrm{jet}}}$, indicating that the reconstructed electrons on average have an \et\ more similar to the jet \pt.

The efficiency to reconstruct an electron from a genuine jet is measured as $\frac{n_\textrm{electrons}}{n_\textrm{all jets}}$ where $n_\textrm{electron}$ is the number of reconstructed electron candidates and ${n_\textrm{all jets}}$ is the number of all jets. The rejection is measured as function of the \pt\ of its associated jet and is shown in Figure~\ref{fig:jetele_eff} where again all fake electrons (\et$>$ 4~\GeV) and those with $\et>18$~\GeV{} are studied separately. In addition to the jet \pt\ an approximation of the electron \et\ is given below the $x$-axis, calculated as $(1-0.88)$ and $(1-0.68)$ times the jet \pt using the most probable values in the distributions in Figure~\ref{fig:jetele_diff}. The reconstruction efficiency of background electron candidates with \et$>4$~\GeV{} rises very quickly and reaches about 60\% to reconstruct a 12~\GeV{} electron from a 100~\GeV{} jet. For background electron candidates with \et$>18$~\GeV{} the efficiency ranges from 0.05$\times10^{-3}$ at a jet \pt\ of 30~\GeV{} %
to 0.03 at a $\pt^{\mathrm{jet}}$ of 50~\GeV{} and steeply rises to almost 0.5 in data above 100~\GeV{} for both jets and electrons.

There is some tension between the measured and predicted efficiencies for low-\pt\ electrons (Fig~\ref{fig:eff_all}). However, when considering electrons with \pt\ above 18~\GeV{} (Fig~\ref{fig:eff_18gev}), the measured efficiencies agree within the systematic uncertainties of the jet calibration with the predictions from the MC simulation except at the highest energy scales where the prediction overshoots the data and the data efficiency drops slightly due to problems in the track-to-cluster matching in the dense environments of these highly boosted jets.

\begin{figure}[t!]
\centering
\subfloat[]{
\includegraphics[width=0.49\textwidth]{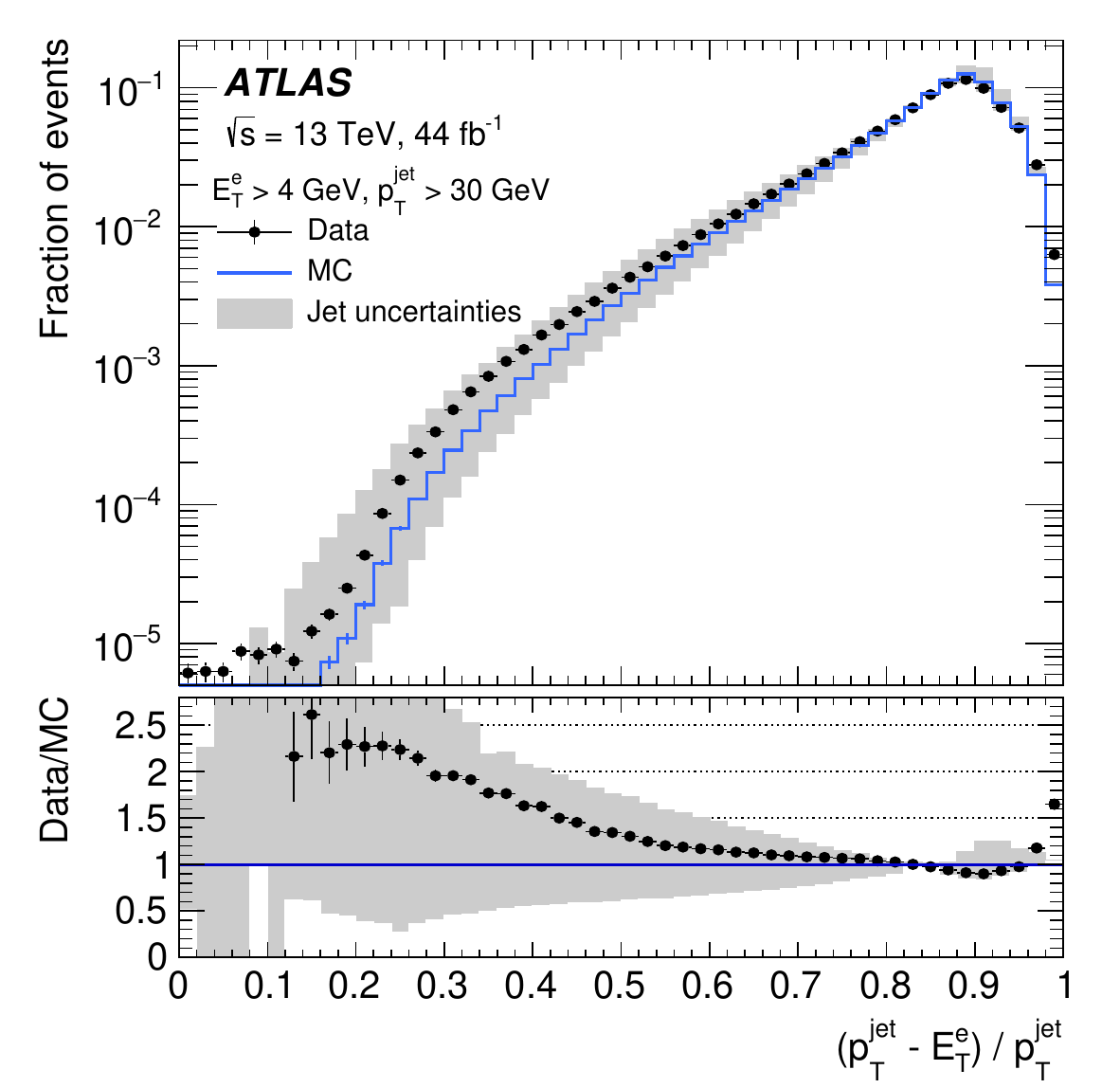}
\label{fig:diff_all}
}
\subfloat[]{
\includegraphics[width=0.49\textwidth]{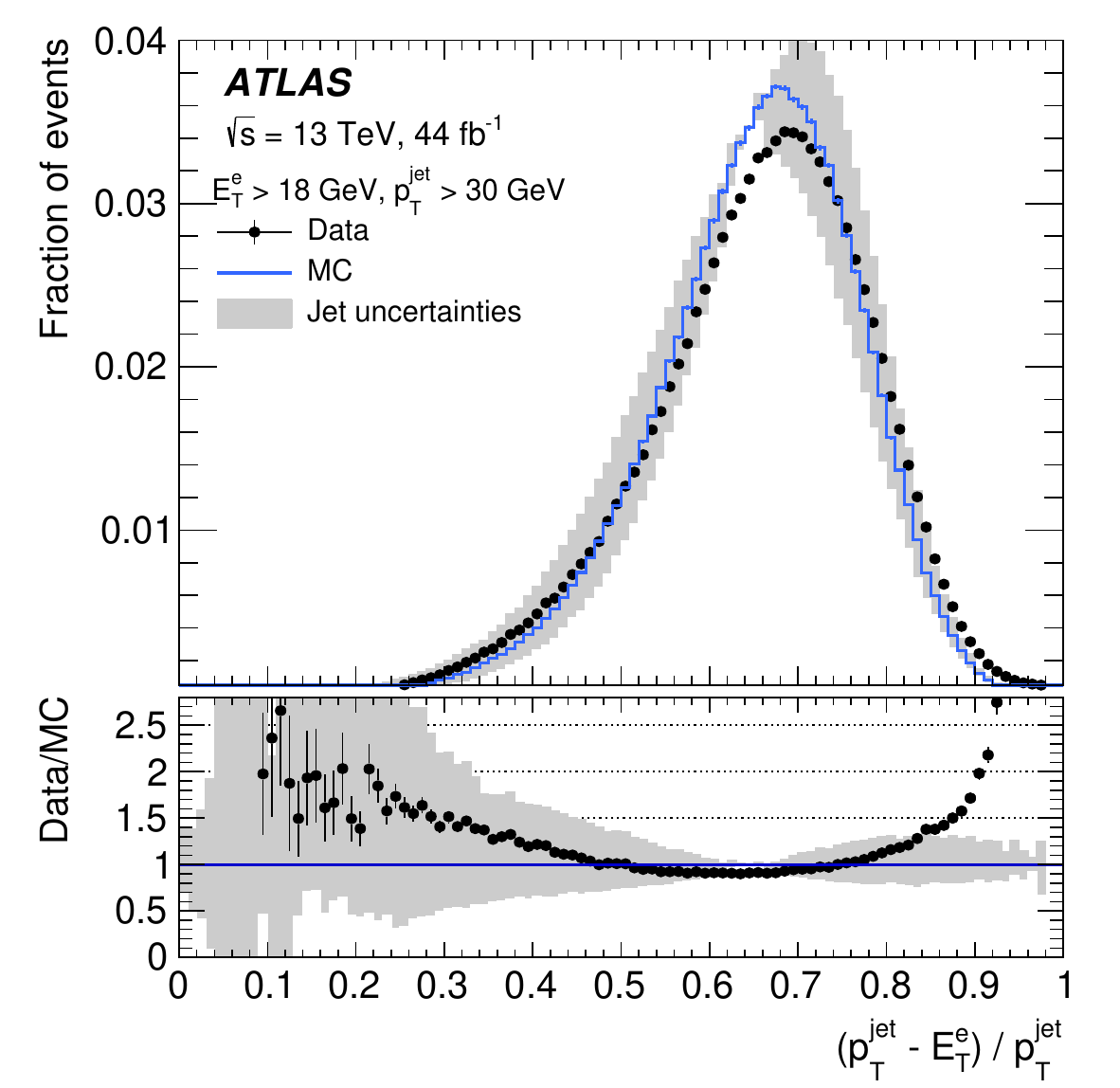}
\label{fig:diff_18gev}
}
\caption{The energies of the jet and its associated electron are compared using their \pt\ difference divided by the jet \pt, $\frac{\pt^{\textrm{jet}} - \et^{\textrm{e}}}{\pt^{\textrm{jet}}}$. The data is shown as markers whereas the MC prediction is shown as a histogram. The grey area indicates the systematic uncertainties of the jet energy calibration. Only additional reconstructed electron candidates in $Z$-boson events are included. The electron pair forming the $Z$ boson as well as their associated jets were removed if closer than $\Delta R<$ 0.4. Jets are required to have $\pt^{\textrm{jet}}>30$~\GeV. \protect\subref{fig:diff_all} shows the comparison for all additional reconstructed electrons with $|\eta|<2.47$ and $\et^{\textrm{e}}>4$~\GeV, whilst \protect\subref{fig:diff_18gev} shows the same comparison for reconstructed electrons with $|\eta|<2.47$ and $\et^{\textrm{e}}>18$~\GeV.}
\label{fig:jetele_diff}
\end{figure}

\begin{figure}[t!]
\centering
\subfloat[]{
\includegraphics[width=0.49\textwidth]{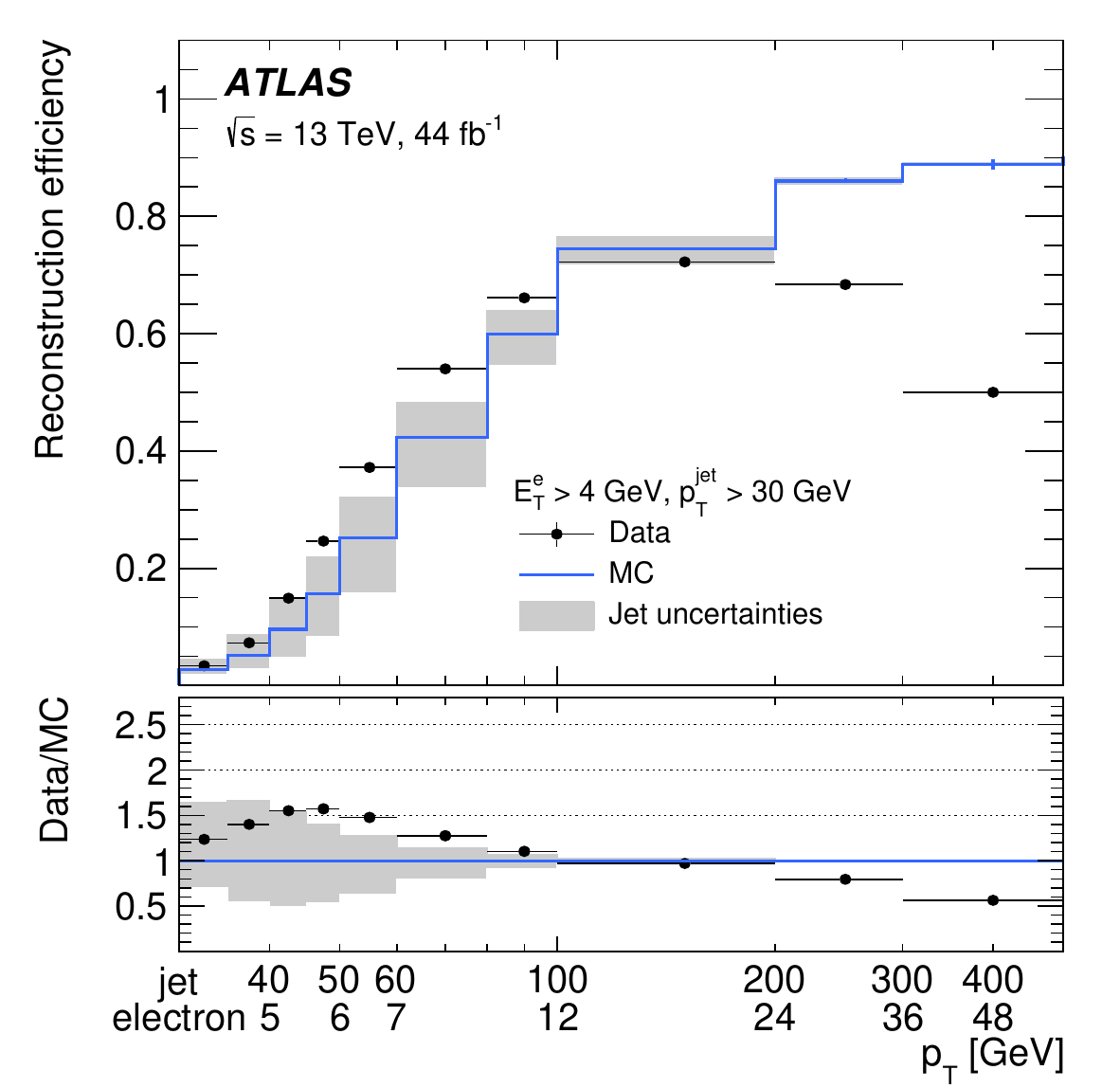}
\label{fig:eff_all}
}
\subfloat[]{
\includegraphics[width=0.49\textwidth]{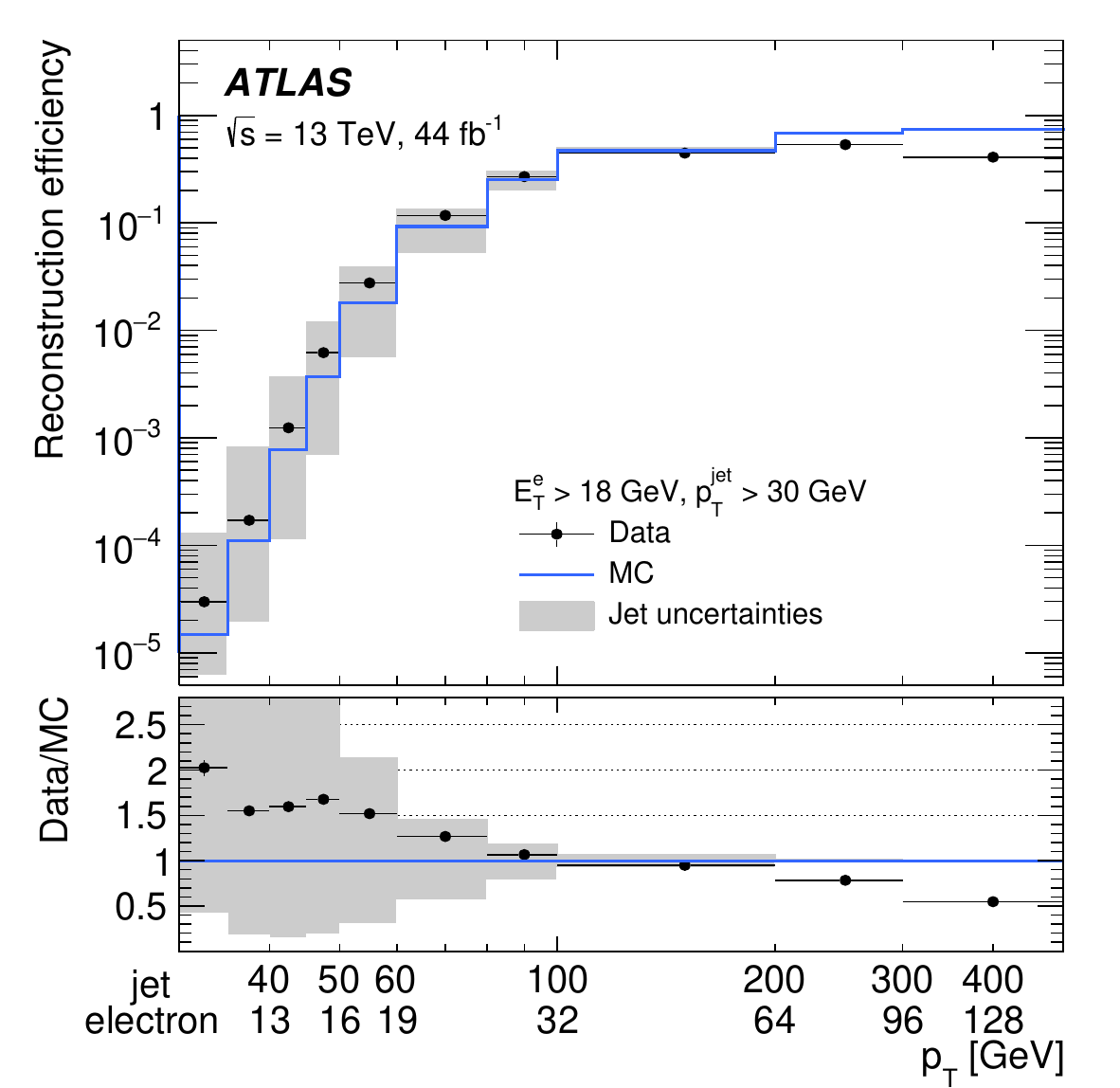}
\label{fig:eff_18gev}
}
\caption{
The efficiencies to reconstruct an electron from a jet are shown as function of the jet transverse momentum. The data is shown as markers whereas the MC prediction is shown as a histogram. The grey area indicates the systematic uncertainties of the jet energy calibration. Only additional reconstructed electron candidates in $Z$-boson events are included. The electron pair forming the $Z$ boson as well as their associated jets were removed if closer than $\Delta R<$ 0.4. Jets are required to have $\pt^{\textrm{jet}}>30$~\GeV. \protect\subref{fig:eff_all} shows the efficiency for all reconstructed electrons with $|\eta|<2.47$ and $\et^{\textrm{e}}>4$~\GeV, whilst \protect\subref{fig:eff_18gev} shows the same comparison for reconstructed electrons with $|\eta|<2.47$ and $\et^{\textrm{e}}>18$~\GeV. Additionally, an approximate electron energy is shown on the horizontal axis that is obtained by scaling the jet transverse momentum according to the most probable value of the relative energy difference as shown in figure~\ref{fig:jetele_diff}. Due to this approximation, in \protect\subref{fig:eff_18gev} efficiencies are shown also for bins with an approximate $\et^{\textrm{e}}<18$~\GeV. These are populated by electrons carrying a larger than average fraction of the jet energy.}
\label{fig:jetele_eff}
\end{figure}

\subsection{Expected background rejection of the electron identification requirements}

The Loose, Medium and Tight identification requirements introduced in Section~\ref{sec:shortreco_selection} are designed to deliver fixed signal selection efficiencies for prompt electrons on top of the reconstruction. For prompt electrons, the background can be classified in three main categories,
namely non-isolated electrons from heavy-flavour (HF) decays, background electrons from photon conversions and Dalitz decays, and light-flavour hadrons (LF). Additionally, electrons coming from pile-up can be identified, comprised of both prompt and non-prompt electrons as well as hadrons. Using MC simulation collision samples of inclusive 2 $\rightarrow$ 2 QCD production, which includes  top-quark pair and weak vector-boson production at leading order (in the following referred to as inclusive QCD production), filtered at particle level to mimic a first-level EM trigger requirement, these three background categories can be investigated. The distinction between the different sources of signal  background is obtained using truth-level information from the MC simulation employing a matching of the reconstructed track hits to energy deposits left by truth-level particles on their way through the detector.

Figure~\ref{fig:bkg_distribution} shows the fraction of the different categories to the total background as a function of \et, $|\eta|$ and number of primary vertices. Only reconstructed electrons in the central region of the ATLAS detector ($|\eta| \leq 2.47$) with $\et > 17$~\GeV{} are considered. The total background is dominated by the LF hadrons when considering reconstructed electrons. Electrons from
photon conversions represent the second most abundant background, while HF decays constitute less than 1\% of the total background and pile-up background contributes about 5~\% on average but with a stronger dependence on the number of reconstructed primary vertices. The composition of the different sources across the \et\ spectrum differs only slightly with less HF electron candidates at high $\et$. Photon
conversions increase with higher values of $|\eta|$ due to the increased material photons traverse and accordingly the higher probability for conversions. As the material distribution of the detector as implemented in the simulation is symmetric around $\eta=0$, the fraction of photon conversions is perfectly symmetric for positive and negative pseudo-rapidity.

\begin{figure}[t!]
\centering
\includegraphics[width=.49\textwidth]{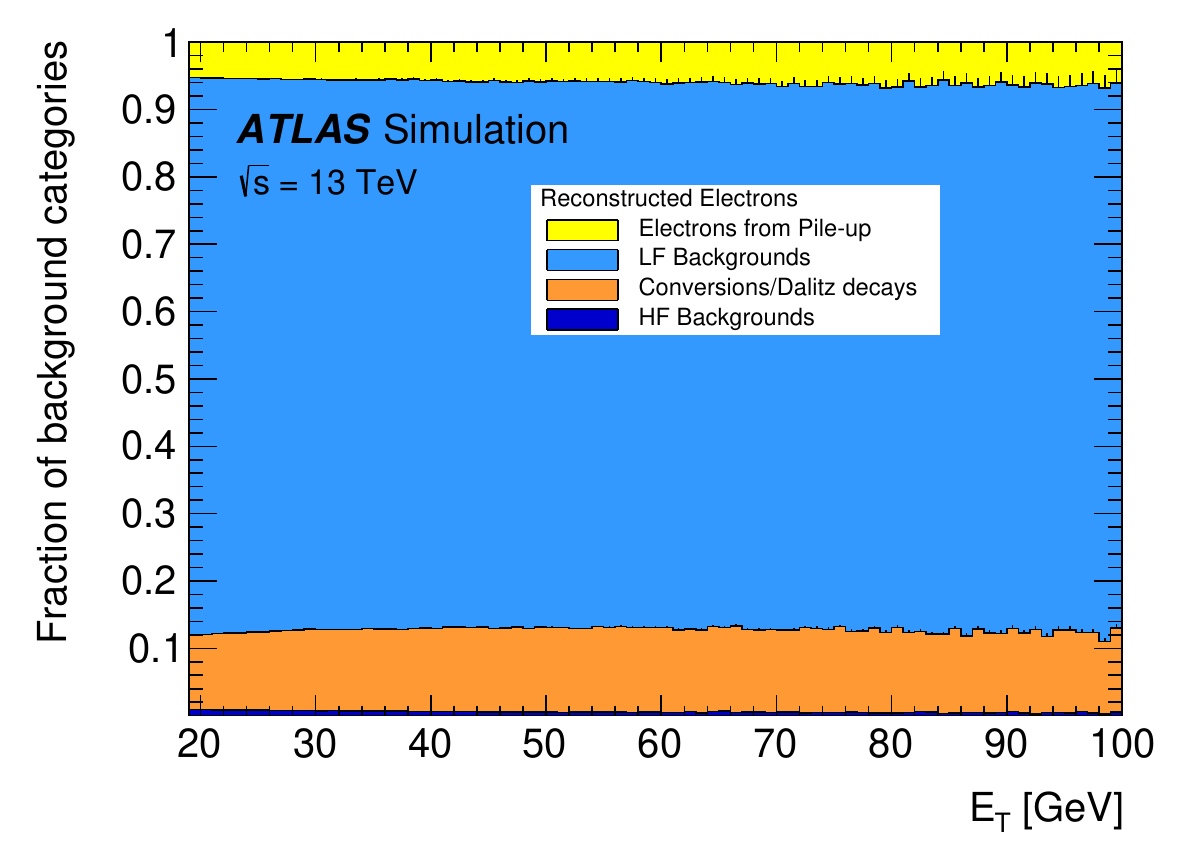}
\includegraphics[width=.49\textwidth]{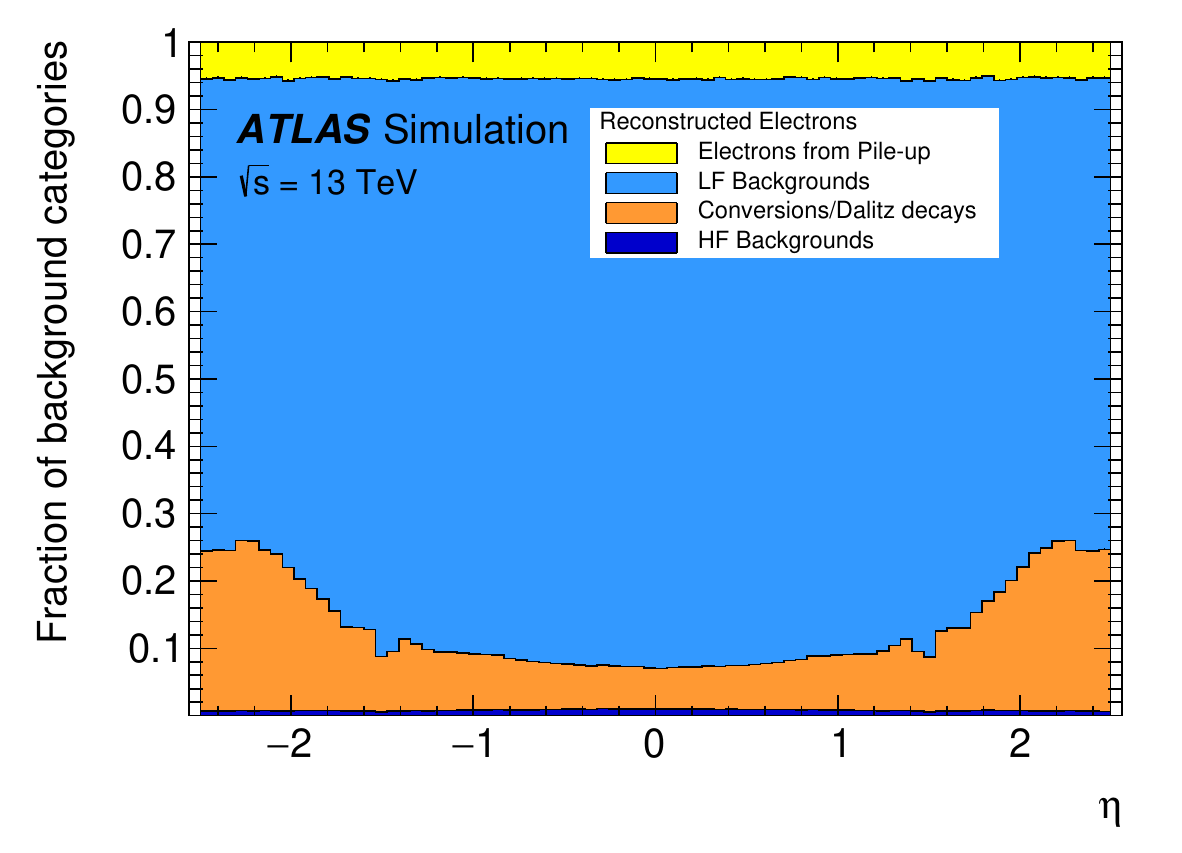}\\
\includegraphics[width=.49\textwidth]{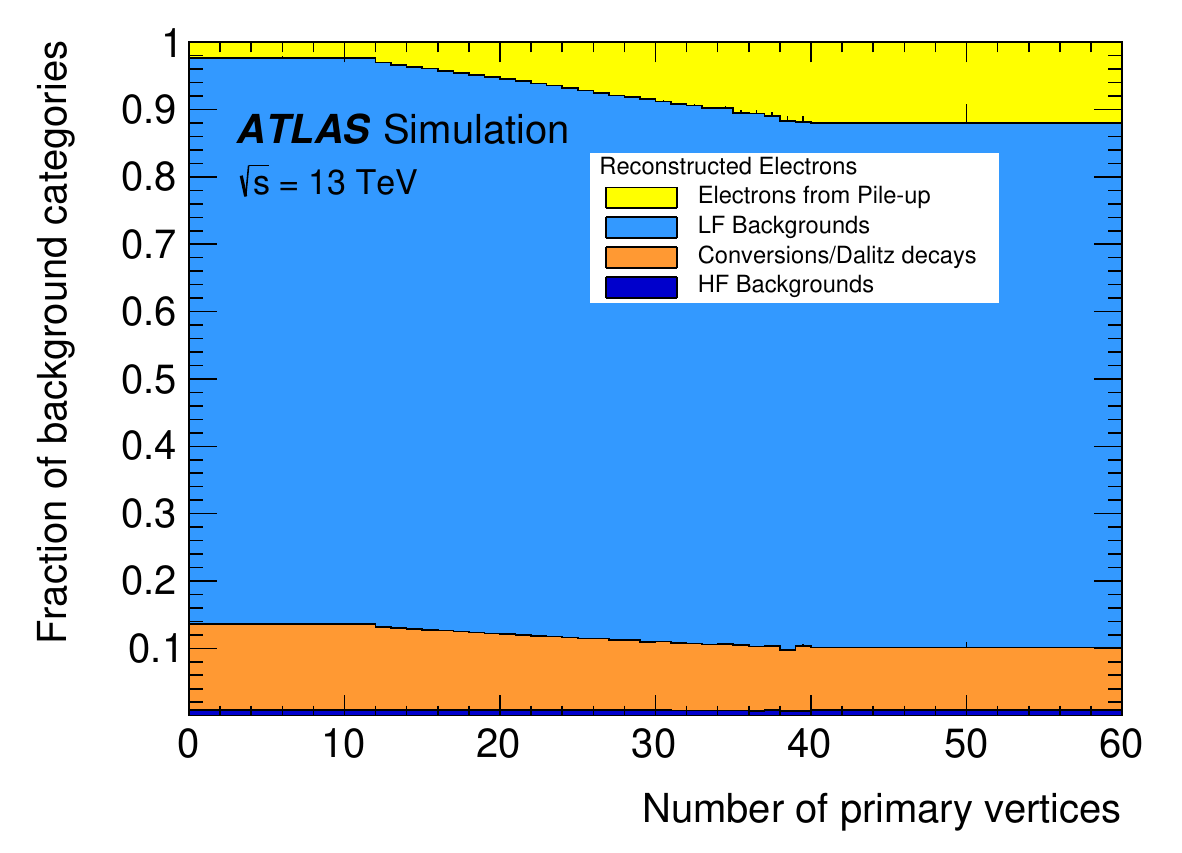}\\
\caption{For the selected sample of reconstructed background electrons, the fraction of four background categories relative to the total background is shown as a function of \et, $|\eta|$ and the number of primary vertices in a MC simulation sample of inclusive QCD production. Only statistical uncertainties are shown.}
\label{fig:bkg_distribution}
\end{figure}

Table~\ref{tab:bkg_effs} details the background rejections in MC simulations for the different categories of backgrounds and their fractions as determined for inclusive QCD production. These numbers are not directly comparable with those published in Ref.~\cite{PERF-2016-01} as the reconstruction algorithms and the identification requirements have changed since the 2012 data taking at $\sqrt{s}= 8$~\TeV. The rejections obtained in Run 2 are worse by a factor of about 1.5 due to the increased pile-up, which makes the calorimetric shower shapes more similar between signal and background, decreasing the achievable rejection at a fixed signal efficiency. The composition of the background is relatively even in Run 2 for the different identification requirements and background categories with about one third consisting of HF background, one half consisting of light flavour backgrounds and the rest made up from background electrons. However, the amount of heavy flavour, light flavour or electron backgrounds depends heavily on the production processes considered and is therefore reflective of the inclusive 2$\rightarrow$2 QCD sample at the higher centre-of-mass energy. It will differ significantly for different event selections (e.g.\ requiring b-tagged jets in the event) and the predominant background processes involved.

The efficiencies to select a light flavour jet or a background electron as a prompt lepton using the Tight identification are less than 1\% whereas about 20\% of heavy flavour decays are identified as prompt leptons. The stark difference between heavy flavour and light flavour efficiencies motivates studies to target the different background categories using specific identification selections in future. Figure~\ref{fig:rej_roc} shows the identification efficiency of signal electrons as determined in MC simulation versus the background rejection.

\begin{figure}[h!]
\centering
\includegraphics[width=.65\textwidth]{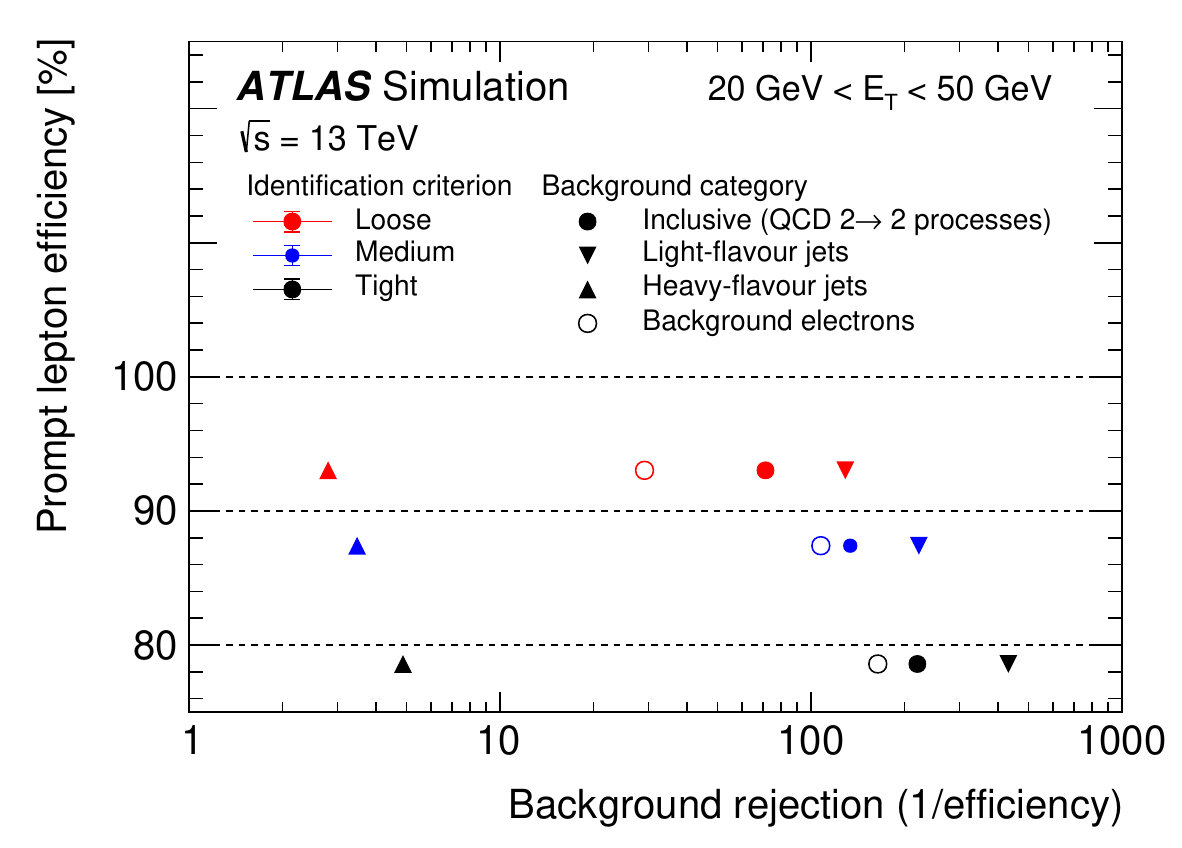}
\caption{The identification efficiency of signal electrons as determined in MC simulation are shown as a function of the background rejection, defined as the ratio of the number of all reconstructed fake electrons to those passing identification criteria. All three identification criteria, Loose, Medium and Tight, are shown for each of the background electron categories as well as for inclusive QCD production.
}\label{fig:rej_roc}
\end{figure}

\begin{table}
\caption{Background efficiency of different identification selections for inclusive QCD production. The electron candidates are required to have transverse energies between \SI{20}{\GeV} and \SI{50}{\GeV} and electrons from $W$- and $Z$-boson decays are removed at particle level. The composition of the sample is categorized according to MC simulation information: non-isolated electrons from heavy-flavour decays, background electrons from photon conversions and Dalitz decays, and light-flavour hadrons. The sample of reconstructed background electrons consists of 0.9\% HF, 12.1\% photon conversions and Dalitz decays (denoted by bkg $e$) and 87.1\% LF\@. The background identification efficiency for each category is quoted. For completeness, the identification efficiency for prompt leptons from $Z \to e^+ e^-$ decays, measured from MC simulation, is also given. The uncertainties are statistical only. }\label{tab:bkg_effs}

\resizebox{\textwidth}{!}{
\begin{tabular}{|r|c|c|c|c|c|c|c|c|}
\hline
\multicolumn{9}{|c|}{$\SI{20}{\GeV}< \et < \SI{50}{\GeV}$ }                                                                                                                                                                    \\ \hline
Selection      & MC efficiency [\%]        & MC efficiency [\%]  & \multicolumn{3}{c|}{Background}       & \multicolumn{3}{c|}{MC efficiency [\%]}                                                                             \\
& $Z \to e^+ e^-$ signal & Background          & \multicolumn{3}{c|}{composition [\%]} & \multicolumn{3}{c|}{for background categories}                                                                      \\
&                           &                     & HF                                    & bkg $e$                                        & LF   & HF               & bkg $e$           & LF                   \\ \hline
Loose          & \(93.03 \pm 0.01\)        & \(1.401 \pm 0.008\) & 22.1                                  & 29.6                                           & 48.3 & \(35.7 \pm 0.5\) & \(3.43 \pm 0.04\) & \(0.776 \pm 0.007 \) \\ \hline
Medium         & \(87.41 \pm 0.01\)        & \(0.748 \pm 0.006\) & 32.3                                  & 15.0                                           & 52.6 & \(28.8 \pm 0.4\) & \(0.93 \pm 0.02\) & \(0.450 \pm 0.005 \) \\ \hline
Tight          & \(78.59 \pm 0.01\)        & \(0.455 \pm 0.005\) & 39.1                                  & 16.2                                           & 44.7 & \(20.5 \pm 0.4\) & \(0.61 \pm 0.02\) & \(0.232 \pm 0.004 \) \\ \hline
\end{tabular}
}
\end{table}


%

%
\section{Electron isolation}
\label{sec:ele_iso}

\subsection{Improved pile-up subtraction for the calorimeter isolation}

As described in Section~\ref{sec:shortreco_selection}, the calorimeter isolation is calculated from energy deposits in the calorimeter cells, $E_{\mathrm{T, raw}}^{\mathrm{isol}}$, and corrected by removing the energy of the electron or photon candidate and contributions from pile-up and underlying event. The explicit calculation for the calorimeter isolation reads as:
\begin{equation}
\etcx = E_{\mathrm{T,raw}}^{\mathrm{isol\mathrm{XX}}} - E_{\mathrm{T,core}} - E_{\mathrm{T,leakage}}(\et,\eta,XX) - E_{\mathrm{T,\textrm{pile-up}}}(\eta,XX),
\label{eq:calo_iso}
\end{equation}

where $\mathrm{XX}$ is the size of the employed isolation cone. $E_{\mathrm{T,core}}$ is the energy of the EM calorimeter cells contained in a $\Delta\eta \times \Delta\phi = 5\times7$ (in EM-middle-layer cell units) rectangular cluster around the barycentre of the EM cluster and is a measure of the energy of the electron or photon candidate. The advantage of this definition is its robustness in the presence of pile-up.
The disadvantage is that it does not subtract all the electron energy and an additional leakage correction, $E_{\mathrm{T,leakage}}$, is needed.
This leakage is parameterised as a function of \et\ and $|\eta|$ of the electron or photon candidate using MC simulation samples of single electrons and single photons without pile-up.
Additionally, a correction for the pile-up and underlying-event contribution to the isolation cone, $E_{\mathrm{T,\textrm{pile-up}}}$, is estimated in situ event per event, based on the ambient energy density~\cite{pusub}, and optimized using a \Zee\ data sample as described below.

The $E_{\mathrm{T,\textrm{pile-up}}}$ term in Equation~\ref{eq:calo_iso} is measured using an improved methodology compared to Refs.~\cite{PERF-2016-01,PERF-2017-01,EGAM-2018-01}. Here,
$E_{\mathrm{T,\textrm{pile-up}}}$ is constructed as the median pile-up energy density, $\rho^{\operatorname{median}}$, multiplied by the area delimited by the isolation cone of $\Delta R$. The core area of the $5\times7$ cluster (see Section~\ref{sec:shortreco}) is subtracted such that $E_{\mathrm{T,\textrm{pile-up}}}$ represents the median energy deposited by pile-up in the isolation cone. The value of $\rho^{\operatorname{median}}$ is measured event-by-event following Ref.~\cite{pusub}. For each event, the $k_t$ jet-clustering algorithm~\cite{Ellis:1993tq,Fastjet} with radius parameter $R=0.5$
is used to construct jets from the topological clusters.
Furthermore, the jet transverse momentum ($p_T^{\operatorname{jet}}$) and the jet area ($A^{\operatorname{jet}}$) are used to
compute the energy density of each jet ($\rho^{\operatorname{jet}} = p_T^{\operatorname{jet}} / A^{\operatorname{jet}}$) that is then used to calculate the median energy density $\rho^{\operatorname{median}}$ from the jets of that event.

The $\rho^{\operatorname{median}}$ value should change monotonically as a function of rapidity; however, detector effects introduce non-monotonic dependences. With increasing $|\eta|$ the granularity of the detector becomes coarser and calorimeter cells become larger, thus noise thresholds are set to higher values and therefore soft energy depositions from pile-up are less likely to form topological clusters~\cite{PERF-2014-07}. As a consequence, $\rho^{\operatorname{median}}$ must be determined as a function of $\eta$. In the past, $\rho^{\operatorname{median}}$ was measured only in two pseudorapidity regions, $|\eta| < 1.5$ and $ 1.5 < |\eta| <3.0$~\cite{EGAM-2018-01} (in the following denoted by ``2-region'' approach), which fails to reproduce the smooth dependency of the average pile-up density as a function of $\eta$.  For the results presented here, this procedure has been updated. The modulation $f_{\eta}$ of the median energy density relative to a reference value $\rho^{\operatorname{median}}_\mathrm{central}$ in a small $|\eta|$-range
is used in combination with the event-by-event value of $\rho^\mathrm{median}_\mathrm{central}$. The value of $\rho^\mathrm{median}_\textrm{central}$ is measured in the region $|\eta| < 1.5$ where pile-up density can be reliably determined per event, whereas measurements of densities in smaller areas require averaging over a large number of events. Measuring the modulation $f_{\eta}$ of the median energy density relative to a central reference $\rho^\mathrm{median}_\textrm{central}$ therefore allows to introduce a smooth variation of the pile-up density as a function of $\eta$ without suffering from the large fluctuations that occur when $\rho^\mathrm{median}$ is determined for small $|\eta|$ ranges event-by-event.

The transverse pile-up energy is then calculated as:

\begin{equation}
E_{\mathrm{T,\textrm{pile-up}}} = \rho^\textrm{median}_\textrm{central}\times  \sum_{\eta}A_{\eta}f_{\eta} = \rho^\textrm{median}_\textrm{central}\times  \zeta(\eta),
\end{equation}

where the $f_{\eta}$ denote the energy densities relative to $\rho^\textrm{median}_\textrm{central}$ in a specific $\eta$-range and the $A_{\eta}$ denotes the area contributing in that bin with the core area of the $5\times7$ EM cluster removed. The relative modulation $\zeta_\eta$ is the total sum of the energy density relative to $\rho^\textrm{median}_\textrm{central}$ calculated for a specific area, e.g.\ an isolation cone of $\etci$, around an axis defined by a given $\eta$.

The $f_{\eta}$ are determined in \Zee\ data events by simultaneously fitting the median isolation energy in two isolation rings, $\etci-\etcs$ and $\etcb-\etci$, as a function of $\eta$ and $\rho^\mathrm{median}_\mathrm{central}$. The areas of the isolation rings defined by $0.2<\Delta R<0.3$ and $0.3<\Delta R<0.4$ have the advantage of being less sensitive to the energy of the electron leaking into the isolation area; therefore on average the isolation energy is dominated by pile-up. Only electrons with 40~$\GeV <\et<45$~\GeV{} are used and backgrounds are negligible. This is referred as the ``smooth correction'' in the following. Figure~\ref{fig:iso_energy_density} shows this median isolation energy in the two isolation rings as measured in data as a function of $\rho^\mathrm{median}_\mathrm{central}$ in one central $\eta$-region ($0.81 < |\eta| <0.86$). The data measurement is compared with the estimate obtained with the previously used 2-region correction and the new method using the smooth correction. Whilst the former overestimates the data measurement, the latter agrees very well. This is also visible in Figure~\ref{fig:run2_zetas} where the modulation relative to $\rho^\textrm{median}_\textrm{central}$ are shown for $f_{\eta}$ as data points.
Additionally, the calculated relative median isolation energy for two specific cone sizes, $\zeta_\eta^{\mathrm{cone20}}$ and $\zeta_\eta^{\mathrm{cone40}}$, are shown as a function of $\eta$. It is also notable that the extracted relative median isolation energies are smaller than one. This indicates that the energy density of $k_t$-jets is not an optimal proxy for the energy density in the fixed sized cones used for the isolation variables, also because $k_t$ -jets are usually clustered around a collimated energy deposit whereas the isolation cones around electromagnetic objects are meant to collect diffuse energy deposits. The $|\eta|$-dependence of the energy from pile-up that is measured as relative modulation is clearly visible.

The performance of the various improvements discussed in this section is illustrated in Figure~\ref{fig:selection_eff},
which shows a comparison between the selection efficiencies achievable using the previous method with that of the improved pile-up subtraction method. To help the comparison, the selections on the isolation are chosen such that similar efficiencies in the $|\eta| < 0.6$ region are achieved. The results are obtained from a $Z\to e^+e^-$ data sample with the method described in Section~\ref{sec:eff_measurement}, using probe electrons with $E_T > 25$~\GeV{} passing the Tight identification selection. As these efficiencies are only for comparison purposes, no background subtraction is performed. It can be seen that the efficiency corresponding to the improved smooth pile-up subtraction method is more stable versus $\eta$ in the $|\eta|>1.2$ region.
When compared with \etcb, a smaller impact is observed for \etcs\ because of the smaller cone size used to define this calorimeter isolation variable.
The improvement leads to better calorimetric isolation at high $|\eta|$ of a few percentage points.

\begin{figure}[t!]
\centering
\subfloat[]{
\includegraphics[width=0.45\textwidth]{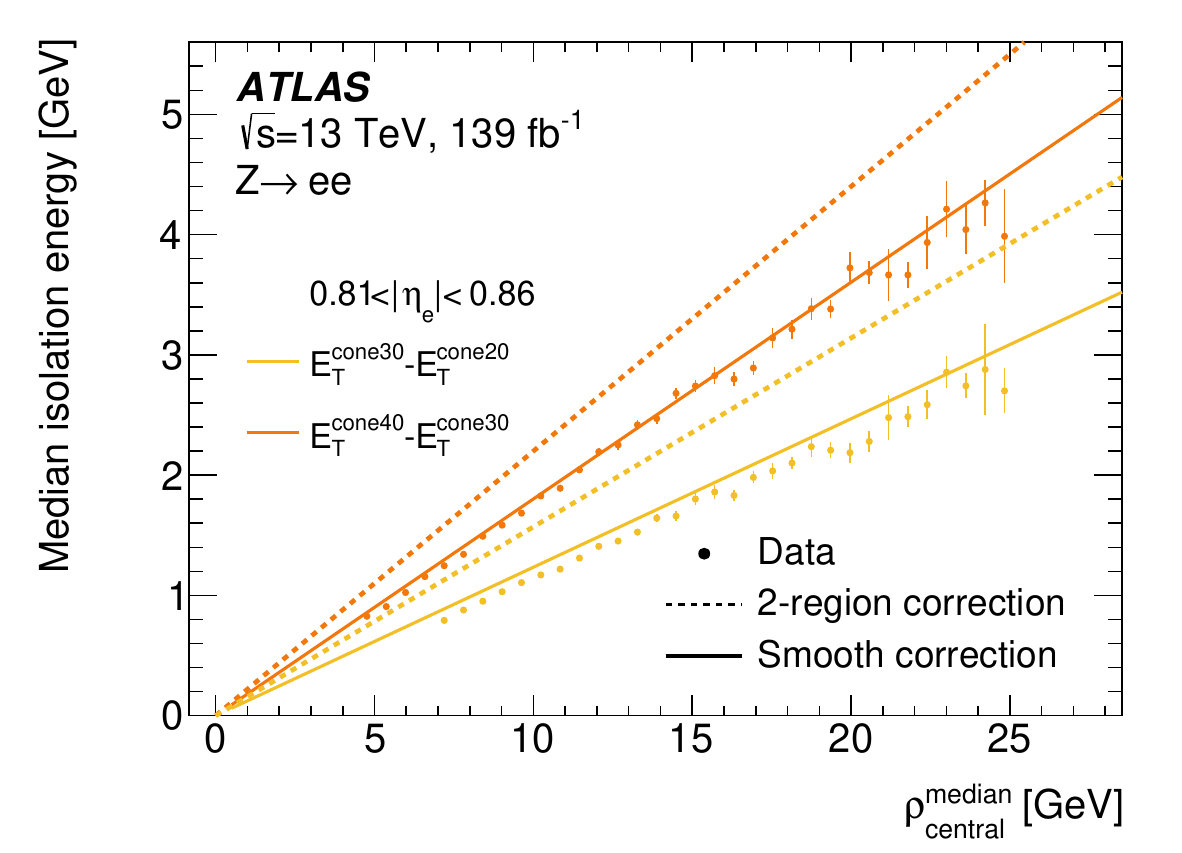}
\label{fig:iso_energy_density}
}
\subfloat[]{
\includegraphics[width=0.45\textwidth]{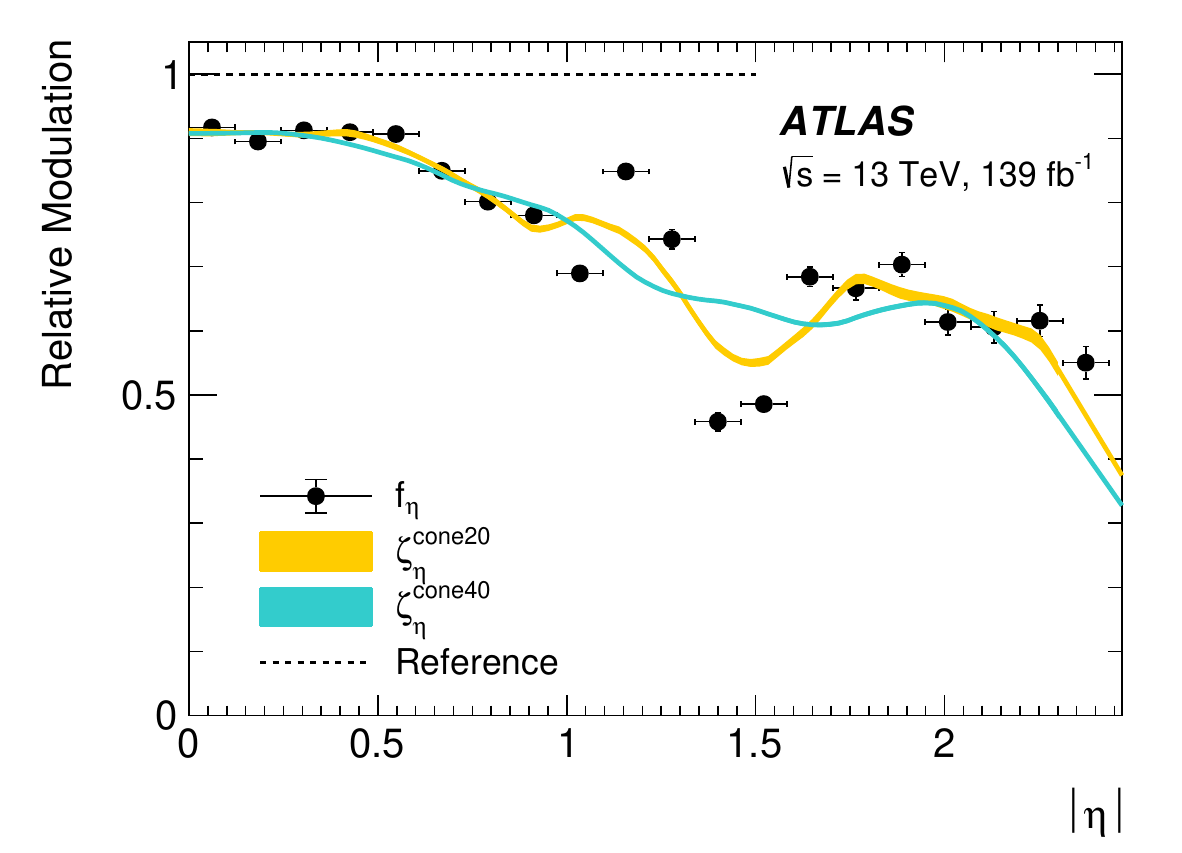}
\label{fig:run2_zetas}
}
\caption{
\protect\subref{fig:iso_energy_density} Median of the isolation distribution without pile-up subtraction as a function of the central ($|\eta| < 1.5$)
energy density, for two isolation rings around the electron, in the $0.81 < |\eta| <0.86$ region.
The data points are compared with the improved smooth pile-up subtraction values (solid line), and to the 2-region correction (dashed line), respectively.
\protect\subref{fig:run2_zetas} Fit parameters $f_{\eta}$ and resulting $\zeta_\eta$ for \etcs\ and \etcb\ versus $|\eta|$.
The statistical uncertainty of the fit parameters $f_{\eta}$ and the propagated uncertainties on $\zeta(\eta_e)$ are shown.
The results are obtained from a $Z\to e^+e^-$ data sample. The range where the reference value of $\rho^\mathrm{median}_\textrm{central}$ is measured on an event-by-event basis is also shown as dashed line.
}
\end{figure}

\begin{figure}[t!]
\centering
\subfloat[]{
\includegraphics[width=0.45\textwidth]{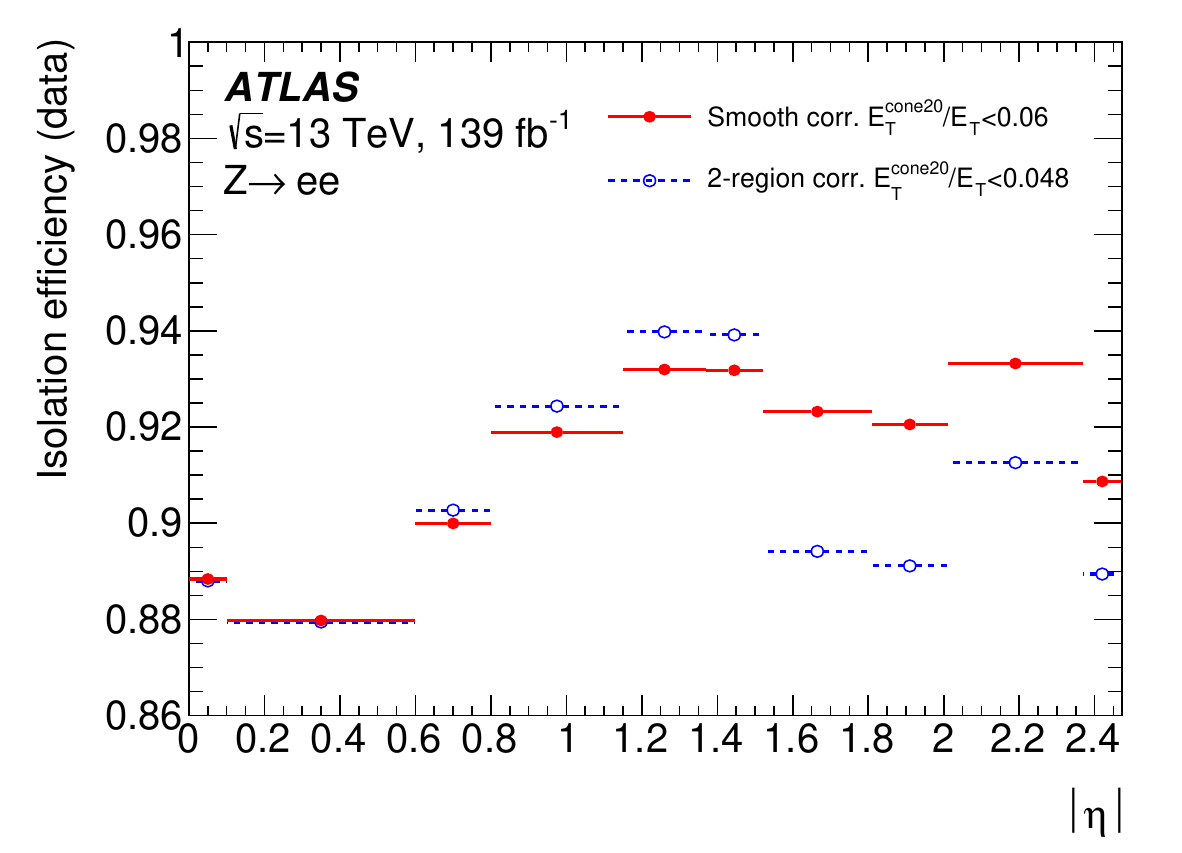}
\label{fig:efficiency_topoetcone20_comp}
}
\subfloat[]{
\includegraphics[width=0.45\textwidth]{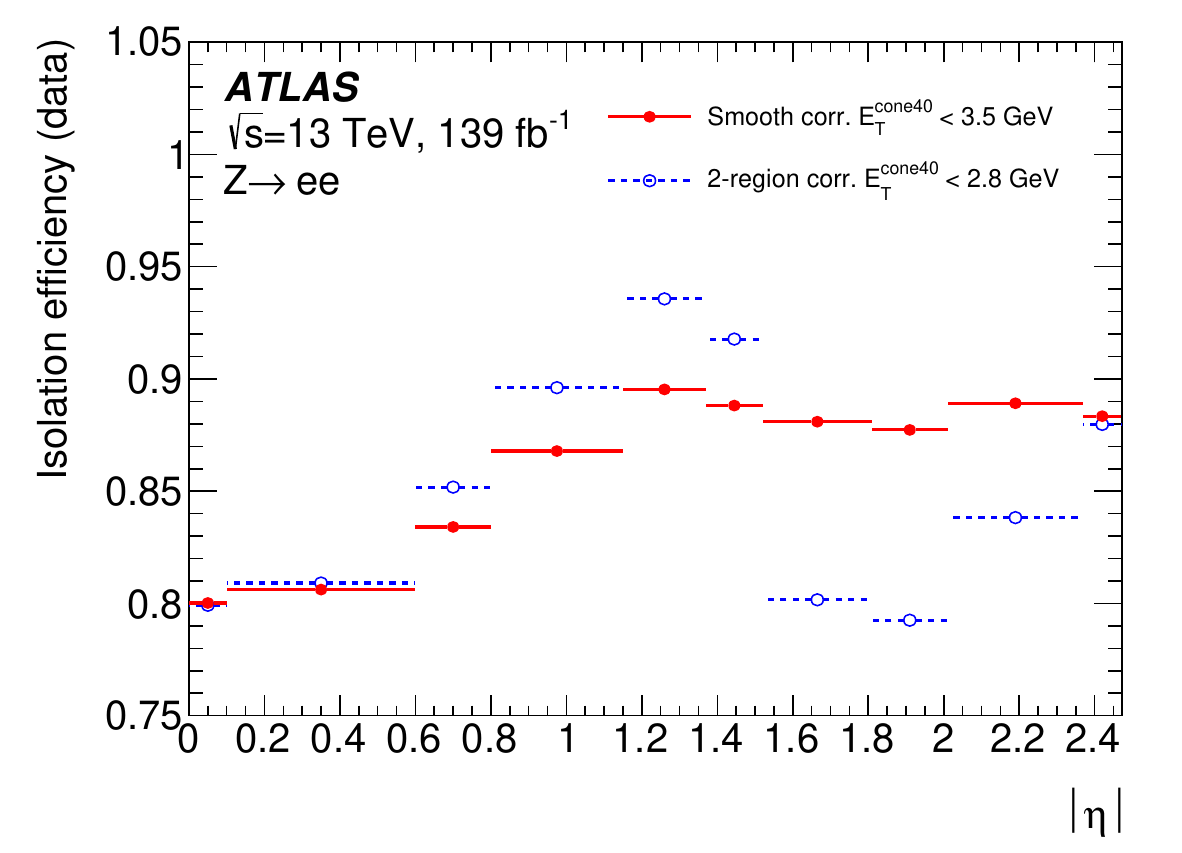}
\label{fig:efficiency_topoetcone40_comp}
}
\caption{
\protect\subref{fig:efficiency_topoetcone20_comp} Efficiency comparison for two calorimeter isolation selection criteria defined with \etcs,
illustrating the impact of using the 2-region and the improved smooth pile-up corrections discussed in the text.
To better see the improvement, the isolation selection criterion defined with the 2-region pile-up correction
is chosen such that it has a similar efficiency in the $|\eta| < 0.6$ region.
\protect\subref{fig:efficiency_topoetcone40_comp} A similar comparison performed for two calorimeter isolation selection criteria defined with \etcb.
The results are obtained from a $Z\to e^+e^-$ data sample.
The efficiencies are computed with the method described in Section~\ref{sec:eff_measurement},
using probe electrons with $E_T > 25$~\GeV{} passing the Tight identification selection.
As these efficiencies are only for comparison purposes, no background subtraction is performed.
Only the (very small) statistical uncertainties are shown.
}\label{fig:selection_eff}
\end{figure}

\subsection{Electron isolation criteria and efficiency measurements}

The implementation of isolation criteria is specific to the physics analysis needs,
and is a compromise between a highly efficient identification of prompt electrons, isolated or produced in a busy
environment, and a good rejection of backgrounds.
The different electron isolation selection criteria used in ATLAS are presented in Table~\ref{tab:ele_wps},
and are defined with fixed requirements on the calorimeter or the track isolation variables.
To increase the rejection of hadronic activity,
the TightTrackOnly\_FixedRad selection criterion use \ptvcs\ for $\et < 50$~\GeV{} and \ptcs\ for $\et > 50$~\GeV.

\begin{table}[!ht]
\caption{
Definition of the electron isolation selection criteria.
All working points use a cone size
of $\Delta R = 0.2$ for calorimeter isolation, and $\Delta R_{\mathrm{max}} = 0.3$ or 0.2 for track isolation. \ptvcs uses a variable cone size and is defined in Equation~\ref{eq:varcone}.
}
\centering
\resizebox{\textwidth}{!}{
\def\arraystretch{1.4}
\begin{tabular}{lll}
\toprule
Selection criteria & Calorimeter isolation  & Track isolation    \\
\midrule

HighPtCaloOnly          & $\etcs < \text{max}(0.015\times \pt, 3.5~ \text\gev)$  & --                                      \\
TightTrackOnly\_VarRad  & --                                                     & $\ptvcs /\pt < 0.06$                    \\
TightTrackOnly\_FixedRad& --                                                     & $\ptvcs /\pt < 0.06$ for $\et < 50$~\GeV \\
&                                                        & $\ptcs /\pt < 0.06$  for $\et > 50$~\GeV \\
Tight\_VarRad           & $\etcs/\pt < 0.06$                                     & $\ptvcs /\pt < 0.06$                    \\
Loose\_VarRad           & $\etcs/\pt < 0.20$                                     & $\ptvcs /\pt < 0.15$                    \\
\bottomrule
\end{tabular}
}
\label{tab:ele_wps}
\end{table}

\begin{figure}[t!]
\centering
\subfloat[]{
\includegraphics[width=0.45\textwidth]{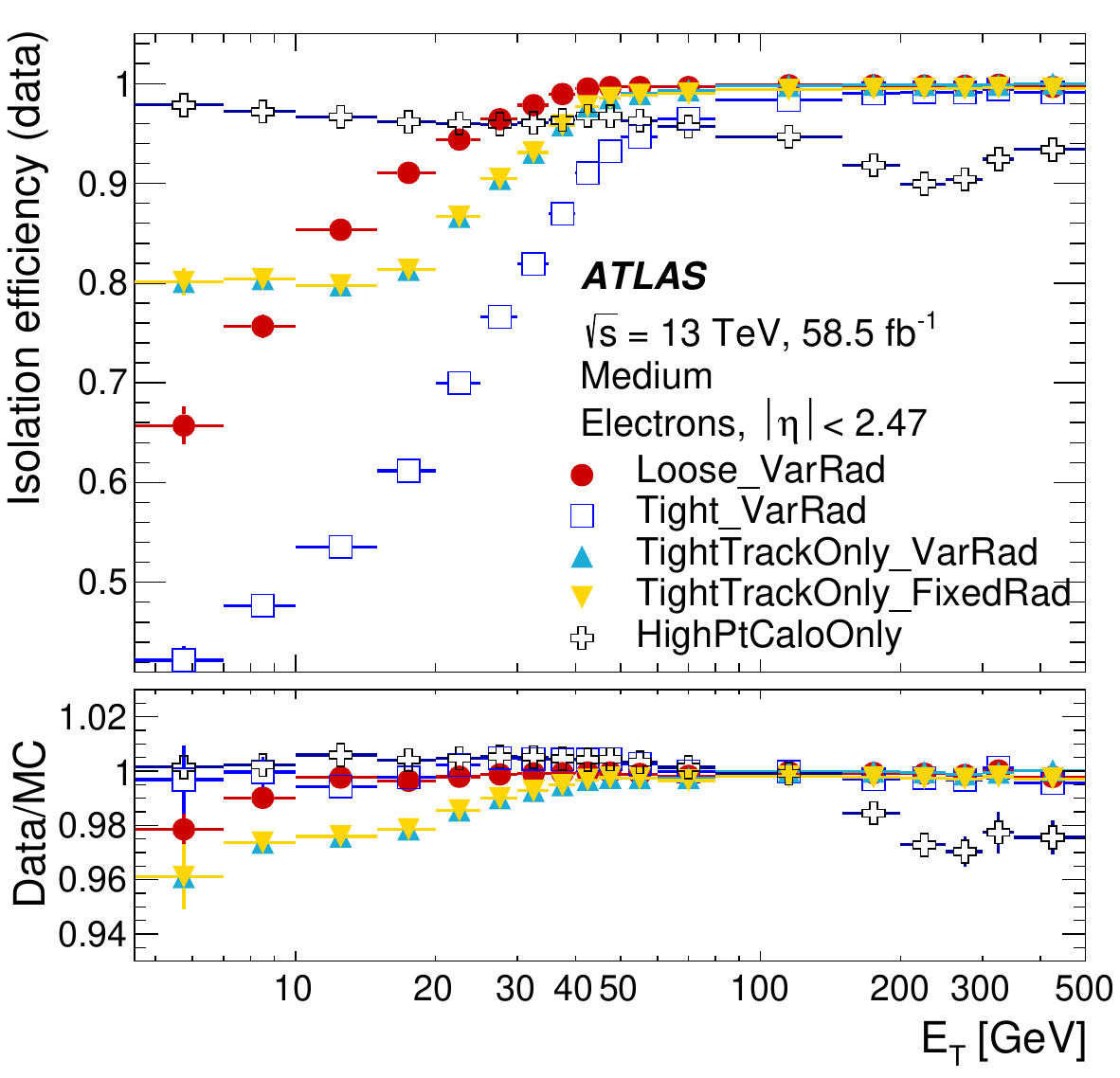}
\label{fig:MediumLLH_d0z0_all_isowps_pt}
}
\subfloat[]{
\includegraphics[width=0.45\textwidth]{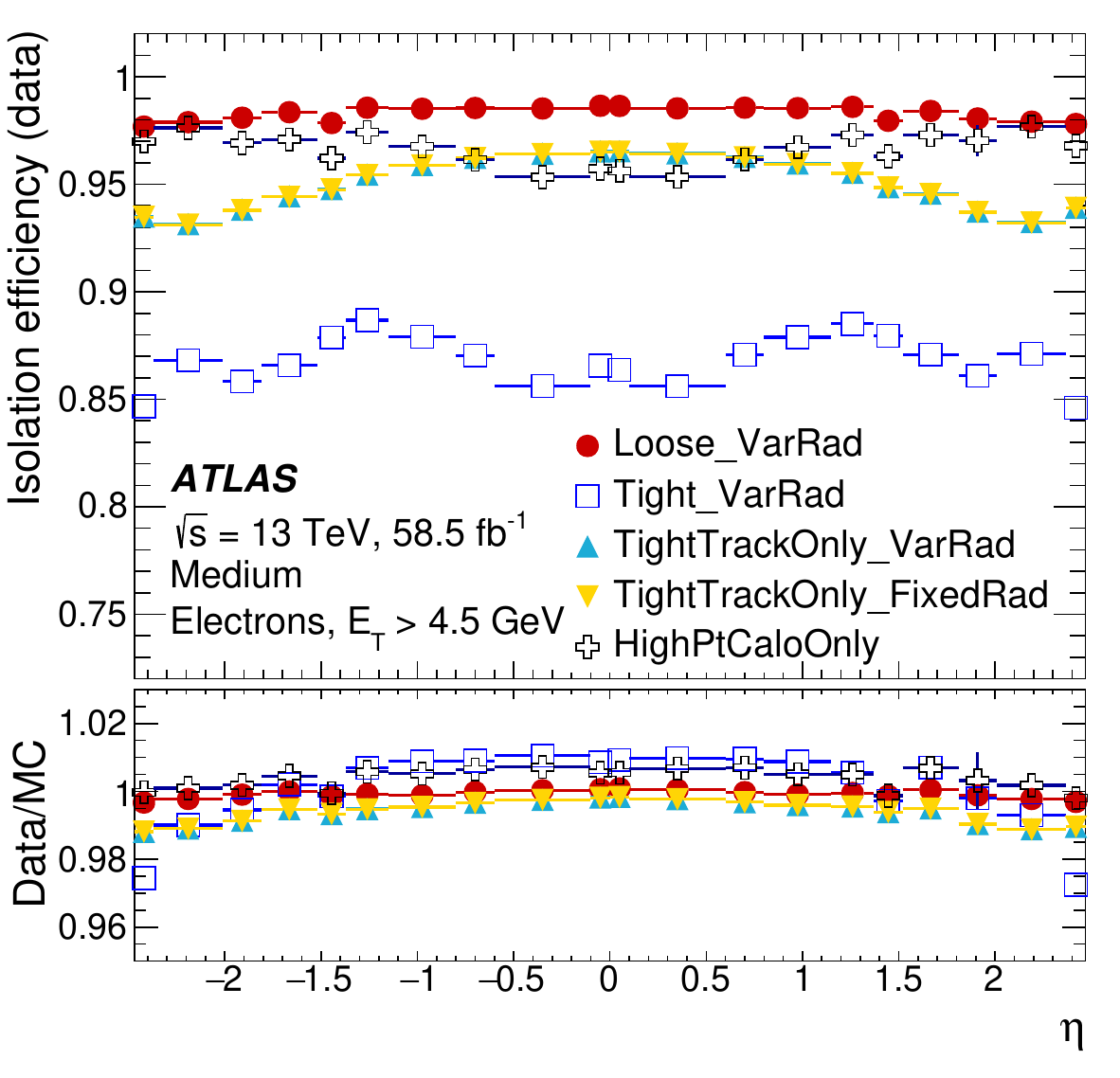}
\label{fig:MediumLLH_d0z0_all_isowps_eta}
}
\\
\subfloat[]{
\includegraphics[width=0.45\textwidth]{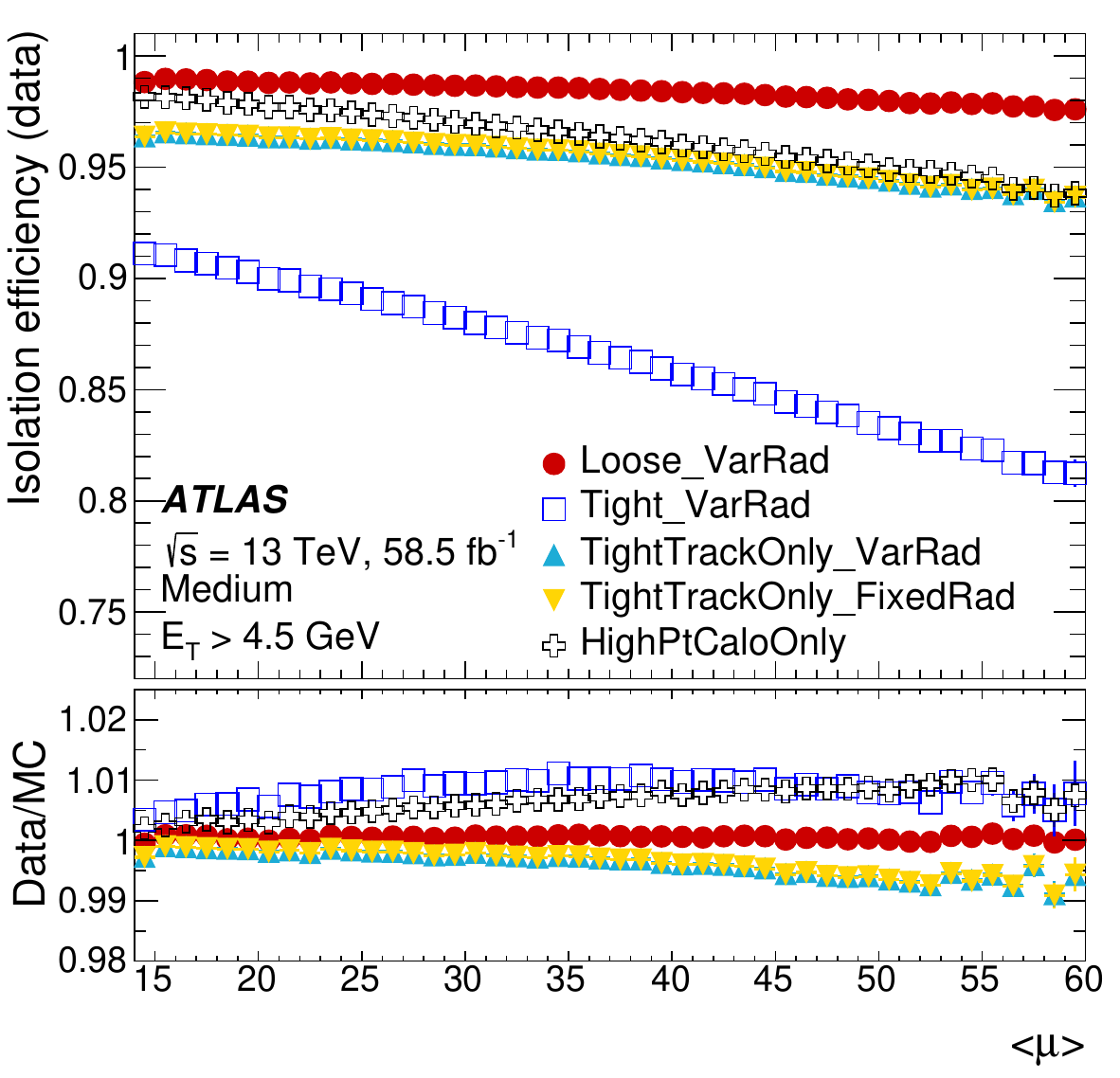}
\label{fig:MediumEff_VS_mu}
}
\caption{
Efficiency of the different isolation working points for electrons from inclusive 2018 data $Z \to e^+e^-$ events as a
function of the \protect\subref{fig:MediumLLH_d0z0_all_isowps_pt} electron \et, \protect\subref{fig:MediumLLH_d0z0_all_isowps_eta} electron $\eta$ and \protect\subref{fig:MediumEff_VS_mu} the average number of interactions per bunch crossing $\muhat$.
The electrons are required to fulfil the Medium selection from the likelihood-based electron identification.
The lower panel shows the ratio of the efficiencies measured in data and in MC simulations. The total uncertainties
are shown, including the statistical and systematic components.
}\label{fig:isolation_eff}
\end{figure}

Figure~\ref{fig:isolation_eff} shows the electron isolation efficiency measured in data recorded in 2018, and the corresponding
data-to-MC simulation ratios as a function of the electron $\et$ and $\eta$, and of the
average number of pile-up interactions \muhat\ for the isolation selection criteria summarized in Table~\ref{tab:ele_wps}.
These results are obtained using a sample enriched in $Z \to e^+e^-$ events recorded in 2018, where the probe electrons
satisfy the Medium identification selection.
Separate measurements were performed also using the Loose and Tight identification selections.
The method used to compute the electron isolation efficiency and the
associated uncertainties is described in Section~\ref{subsec:ID_Zmass}.
The Tight\_VarRad selection criterion gives the highest background rejection below 60~\GeV
and the most significant difference in shape in $\eta$.
The HighPtCaloOnly isolation working point gives the highest rejection in the high-\et\ region ($\et > 100$~\GeV).
The behaviour at high $\et$ for HighPtCaloOnly is expected, as the definition of this working point changes around 233~\GeV{} in electron \et
(the cut-value for $\etcs$ is changed from $0.015\times \pt$ to 3.5~\GeV). The efficiencies are symmetric for positive and negative values of the pseudo-rapidity for all of the considered isolation working points.

The track isolation variables are largely independent of pile-up, due to the
rejection of tracks originating from pile-up vertices or with large transverse impact parameters relative to
the primary vertex.
This is not the case for the calorimeter isolation~\cite{EGAM-2018-01}, as shown in Figure~\ref{fig:MediumEff_VS_mu}.
The Tight\_VarRad selection criterion give the highest
pile-up dependency, the isolation efficiency decreasing from 91\% at low \muhat\ to 82\% when \muhat\ is around 60.

The overall differences between data and MC simulation are less than
5\%--7\% depending on the working point, with the largest difference observed for HighPtCaloOnly isolation criterion. For electrons
with \ET\ higher than $500$ \gev\ no measurement can be performed because of the limited number of data events,
and the results from the [350, 500] \gev\ bin are linearly extrapolated up to 2~\TeV
with uncertainties up to 40\%, depending on the isolation selection criteria.
Overall, the extrapolation uncertainty is dominated by the propagation of the statistical uncertainty in the highest \et measurements.

The efficiency correction factors obtained for the MC simulation are smoothed versus \et, using the SMART smoothing algorithm through ROOT~\cite{SMARTSmoothing,Brun:1997}, taking into account the statistical uncertainties on the corrections. The statistical uncertainties on the smoothed correction values are estimated with pseudo-data, applying the same smoothing procedure to multiple pseudo-data instances, and computing the root-mean-square of the smoothed pseudo-data values.
The total uncertainties on the correction factors range from 5\% to less than 0.1\% in the \et range from 10 to 200 \gev.


%

%
\section{Photon identification}
\label{sec:ph_id}

Photon identification efficiencies are measured separately for converted and unconverted photons in 68 bins of photon $\et$ and $\eta$.
For both the converted and unconverted photons the same binning is used with edges at $\et = \{10$, 15, 20, 25, 30, 35, 40, 45, 50, 60, 80, 100, 125, 150, 175, 250, 350, $\infty\}~\GeV$ and  $|\eta| = \{0$, 0.6, 1.37, 1.52, 1.81, 2.37$\}$. No measurements are carried out for the bin defined by $|\eta| = \{1.37, 1.52\}$ as it coincides with the transition region of the ATLAS calorimeter and is not used in analyses using photons.

Photon identification efficiency measurements are carried out using three different methods that are detailed in Ref.~\cite{PERF-2017-02} and that are combined to yield correction factors for analyses. In all cases, photons are required to satisfy the Loose isolation criterion defined in Ref.~\cite{EGAM-2018-01} and therefore, in contrast to the presented electron efficiencies, the photon efficiencies are measured relative to this isolation criterion.

For the first method, probe photons from a sample of radiative $Z\rightarrow \ell\ell\gamma$ decays are used, collected as described in Section~\ref{sec:data_sets} and selected by requiring an invariant mass of the three-body system, 80~\GeV $< m_{\ell\ell\gamma} <$ 100~\GeV, and of the lepton-pair invariant mass, 40~\GeV $< m_{\ell\ell} <$ 83~\GeV. Backgrounds and signal events are determined using a MC-based template fit to the observed three-body invariant-mass distribution.
The main uncertainty in this method comes from the low statistics especially at high photon \et bins, which is the main weakness of the method. The dominant systematic uncertainty comes from the MC generator modelling uncertainty.

A matrix method~\cite{PERF-2017-02} is used on the second data sample consisting of inclusive photons passing the Loose identification criterion employed in the trigger as described in Section~\ref{sec:data_sets}. This sample is dominated by a background from hadronic jets that must be accounted for. This is achieved using a matrix method that constructs four regions made up of photon candidates that either pass or fail the Tight identification, and pass or fail track-based isolation cuts. For each region, the numbers of signal and
background events are unknown. However, since the isolation efficiencies for signal and background from each region are known, the efficiency for photons passing the Loose identification criteria to also satisfy the Tight identification can be extracted. The isolation efficiencies for signal and background are extracted from MC simulation and a jet-enriched control sample selected by inverting identification criteria respectively. This method is only capable of determining the efficiency for Loose photon candidates to pass the Tight selection. Therefore it is further multiplied with the efficiency of reconstructed photon candidates to satisfy the
Loose identification as determined from simulation.
This correction is however typically less than 5\%, and even smaller at high \et where the matrix method has the most impact in the combination of the three measurements.
The dominant systematic uncertainty comes from the method's non-closure related to the imperfect assumption that isolation and identification efficiencies employed in the method are perfectly independent of each other.

The third method is called ``electron extrapolation'' and uses the electrons from \Zee\ decays employing the same data sample as in the electron efficiency measurements. In the selection, loose photon isolation requirements are applied additionally to the electron candidates. Then, the electron shower shape variables are modified by applying Smirnov transformations to resemble those of photons based on information from simulated \Zee\ events
and the inclusive-photon production sample. After the shower shapes have been modified, the efficiencies are measured using a similar method as for electrons (described in Section~\ref{sec:eff_measurement}), using the invariant mass distribution of the electron pair to distinguish signal and backgrounds.
The method closure uncertainty is the dominant uncertainty at low \et in most of $\eta$ regions with uncertainties up to 20\%. It comes from the method's non-closure connected to ignorance of possible correlations between the different shower shapes during the transformation procedure. However, it decreases as a function of \et and for $\et > 100$~\GeV{} other sources of uncertainty become important such as limited MC statistics, and variations of the applied Smirnov transformations.

The measurement done using radiative $Z$-boson decays
and the one where the matrix method is applied to an inclusive sample of photons were performed using the full Run 2 data, whilst the measurement using electron extrapolation uses the 2015--2017 data with an integrated luminosity of 81 \ifb. Further details of the measurements and of the determination of systematic uncertainties can be found in Ref.~\cite{EGAM-2018-01}.

Figures~\ref{fig:photon_unconv} and~\ref{fig:photon_conv} show the results of the three efficiency measurements for unconverted and converted photons, respectively, as a function of $\pt$ in the four different regions of $|\eta|$. The upper panels of the figures only show the data efficiencies, whilst the lower panels display the ratio of the efficiencies determined from data and those observed in MC simulation. Before determining the efficiencies in the MC simulation, the photon shower shapes were corrected to better
match the data as described in Ref.~\cite{PERF-2017-02}. Similarly, in analyses, first the photon shower shapes are corrected and then the data-to-MC simulation ratio is applied to account for residual differences between the identification selection efficiencies. For this purpose, the data-to-MC simulation ratios from the three measurement methods described above are combined using a weighted average in each \et{}--$\eta$ bin, assuming the statistical and systematic uncertainties to be uncorrelated among the methods. For the statistical uncertainties, this is an approximation, but one that largely holds: the electron extrapolation uses an independent data sample, whilst there is some overlap between the sample of radiative $Z$-boson decays used and the inclusive sample of photons, which contains the former. However most of the events in the inclusive photon sample stem from photon+jet production.

The result of the combination of the efficiency ratios is also shown in the lower panels of Figures~\ref{fig:photon_unconv} and~\ref{fig:photon_conv}. The total uncertainty of the combined ratio ranges between 5\% at low \et and about 0.4\% at high \et for unconverted photons, and between 7\% (low \et) and about 0.7\% (high \et) for converted photons.
The over-all improvement is of the order of 30\%--40\%, slightly larger than what is expected from the increase in statistics alone, as some aspects of the determination of the systematic uncertainty are also statistically limited.

\begin{figure}[t!]
\centering
\subfloat[]{
\includegraphics[width=0.45\textwidth]{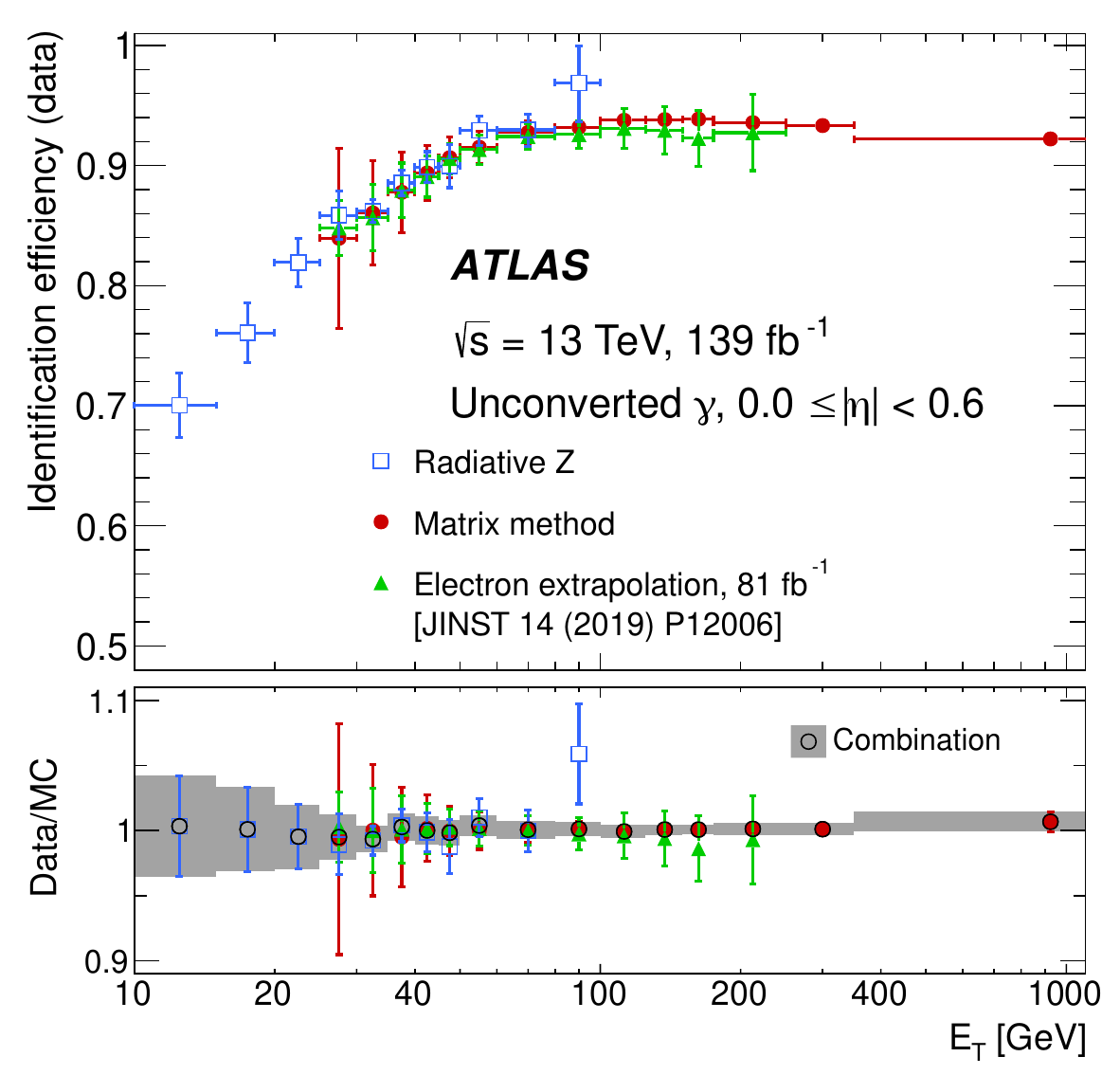}
}
\subfloat[]{
\includegraphics[width=0.45\textwidth]{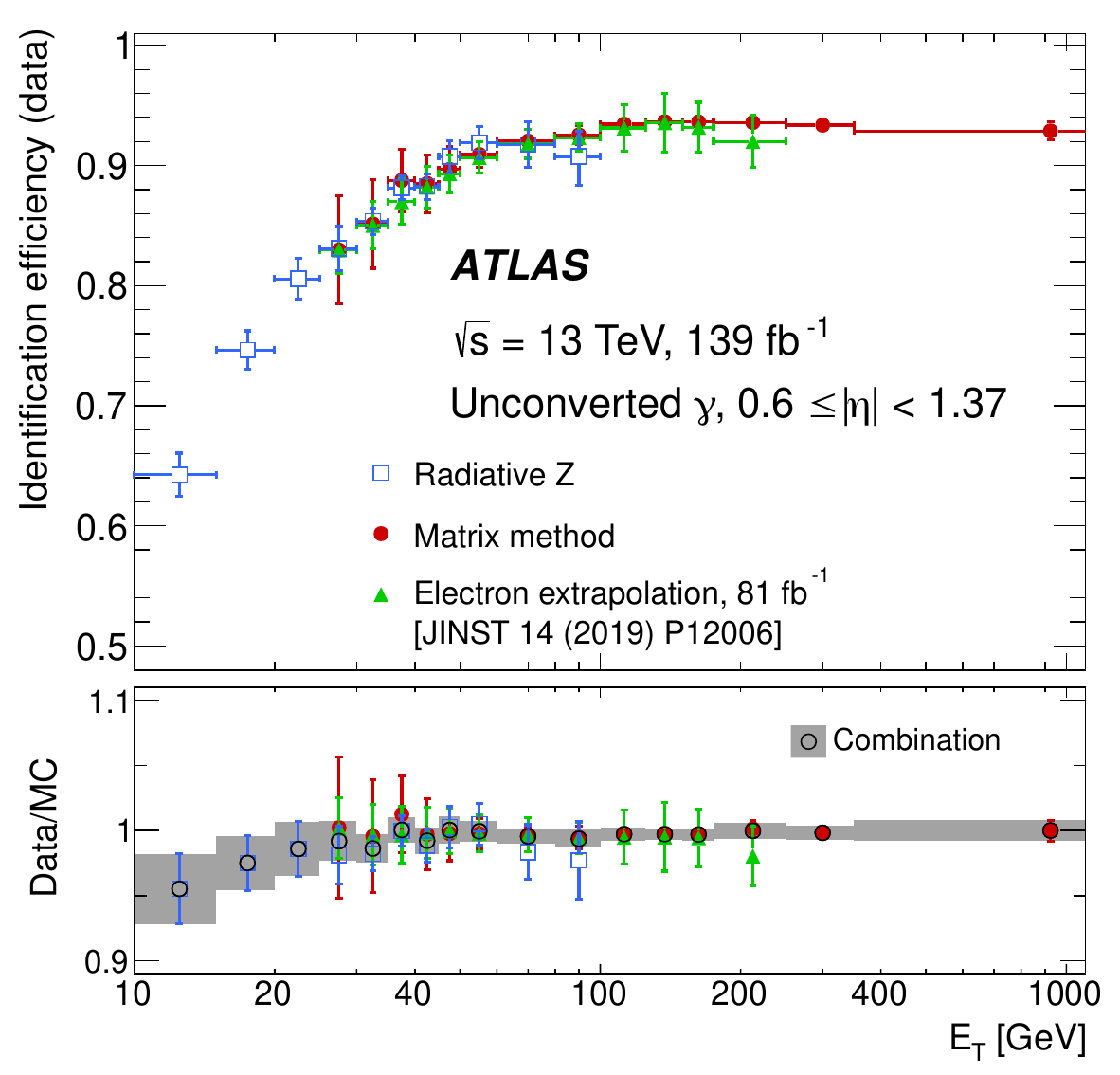}
}
\\
\subfloat[]{
\includegraphics[width=0.45\textwidth]{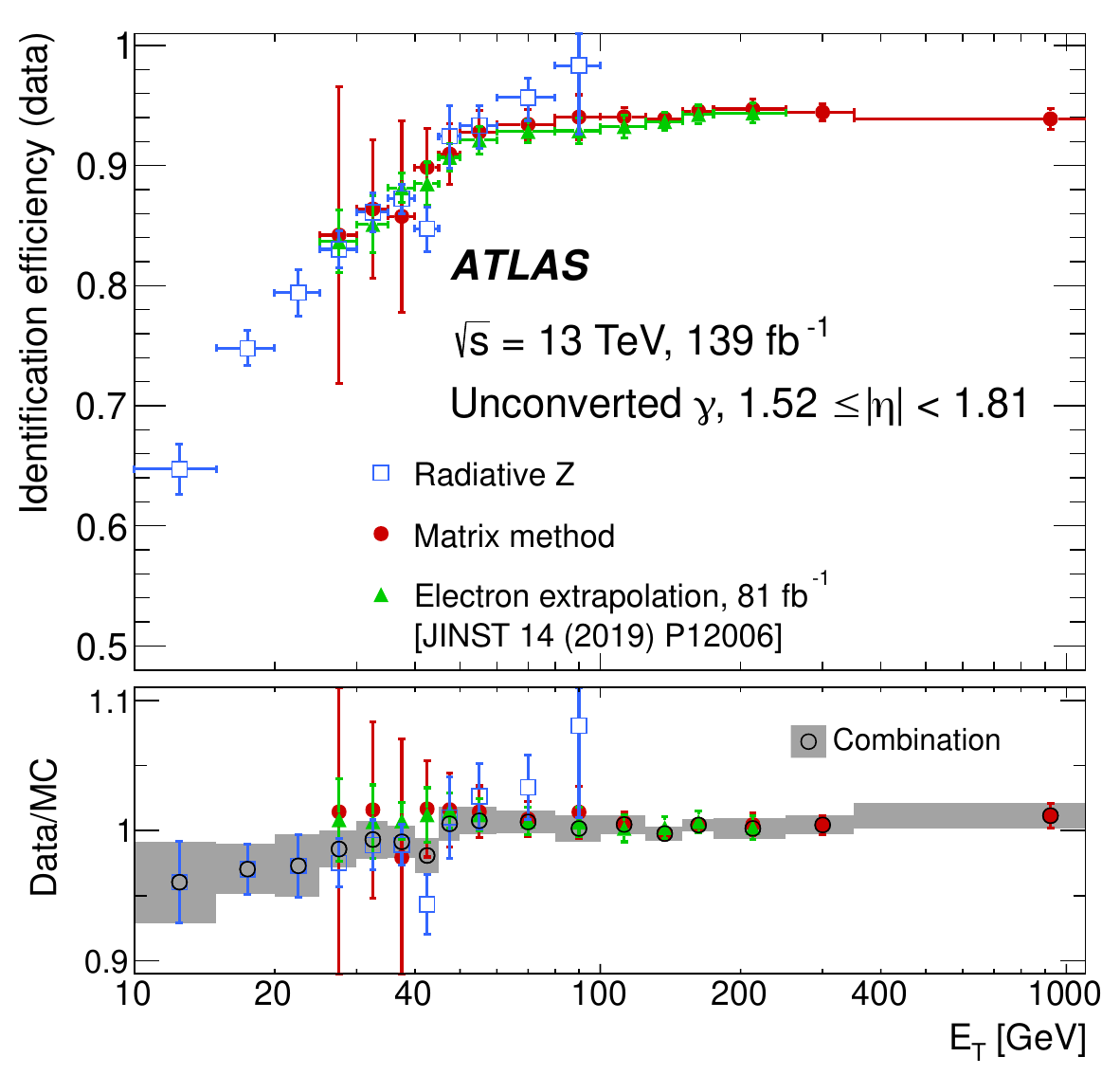}
}
\subfloat[]{
\includegraphics[width=0.45\textwidth]{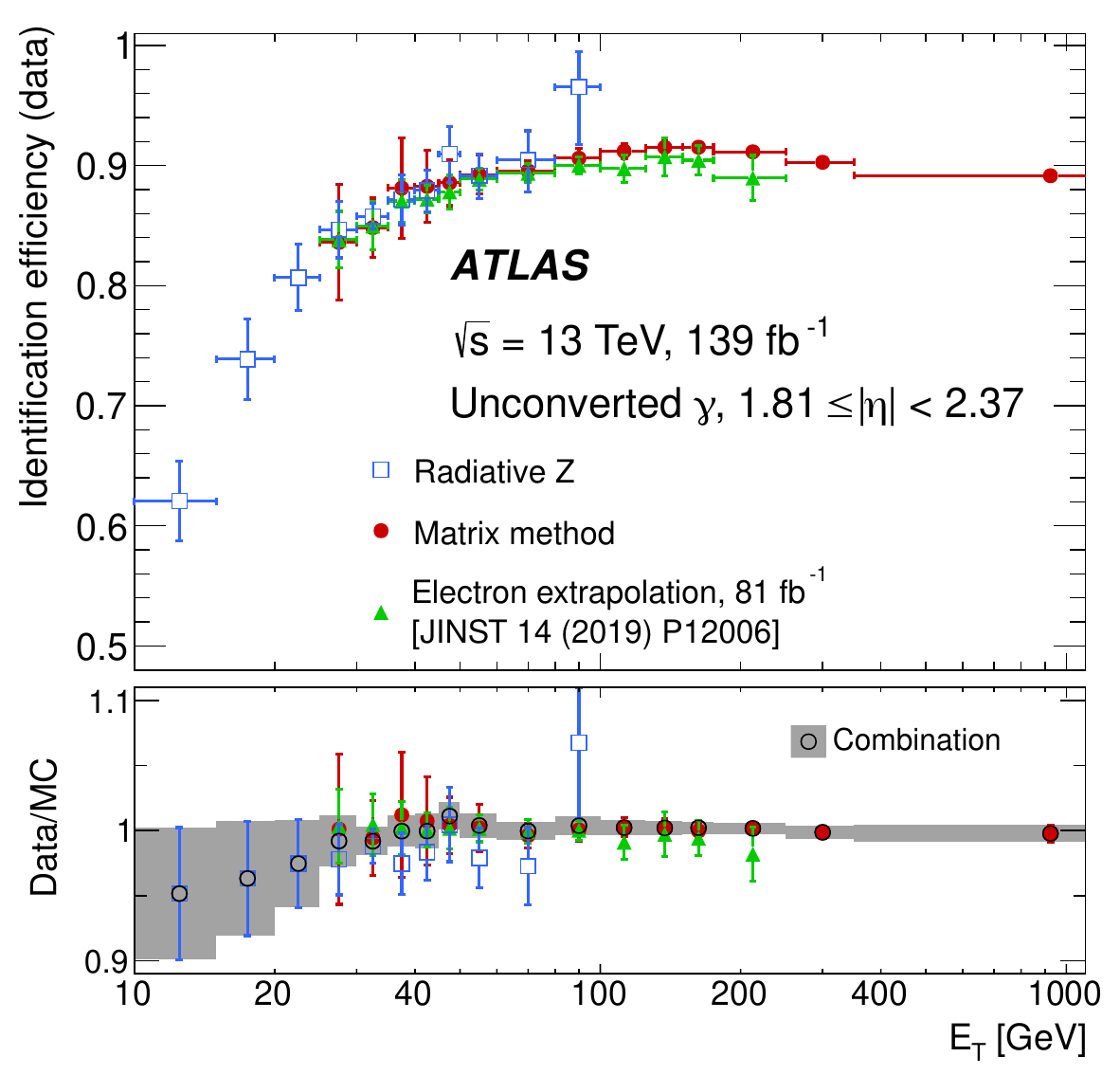}
}
\caption{The photon identification efficiency measured in data, and the ratio of data-to-MC simulation efficiencies, for unconverted photons with a Loose isolation requirement applied as preselection, as a function of \et in four different $|\eta|$ regions. The results for the electron extrapolation were not updated with regards to Ref.~\cite{EGAM-2018-01}. The combined correction factor, obtained using a weighted average of correction factors from the individual measurements, is also presented; the band represents the total uncertainty.
}\label{fig:photon_unconv}
\end{figure}

\begin{figure}[t!]
\centering
\subfloat[]{
\includegraphics[width=0.45\textwidth]{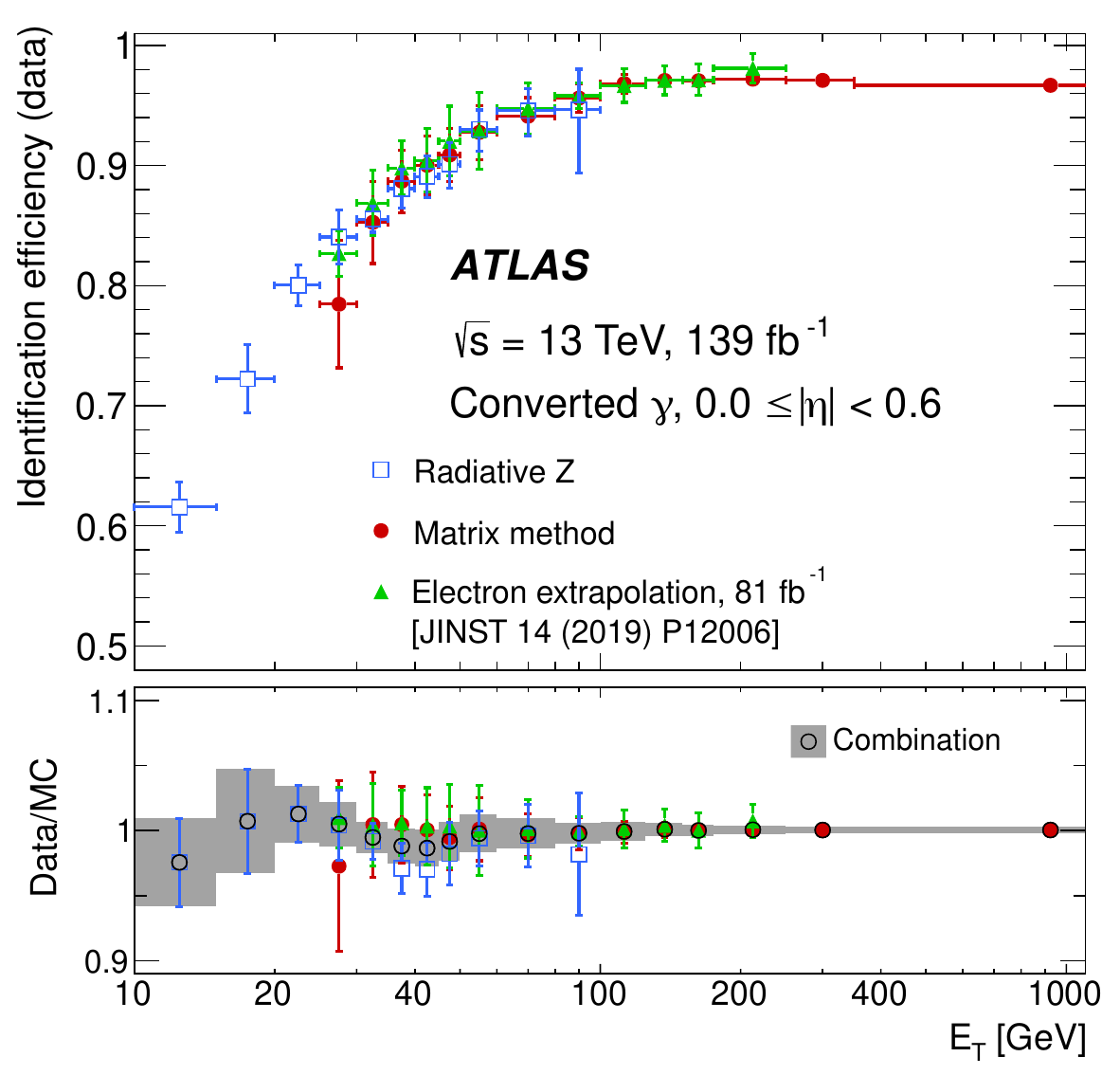}
}
\subfloat[]{
\includegraphics[width=0.45\textwidth]{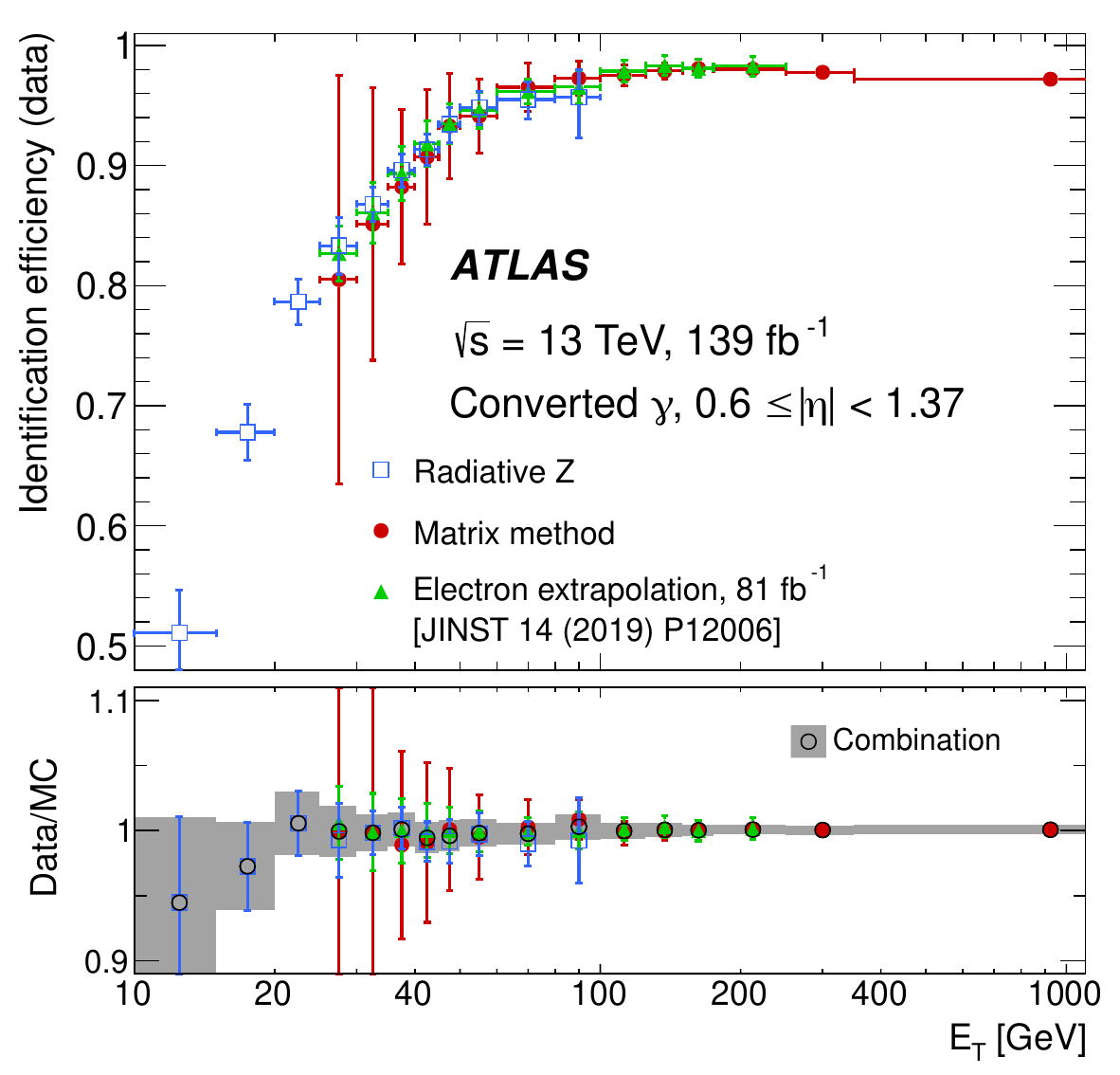}
}
\\
\subfloat[]{
\includegraphics[width=0.45\textwidth]{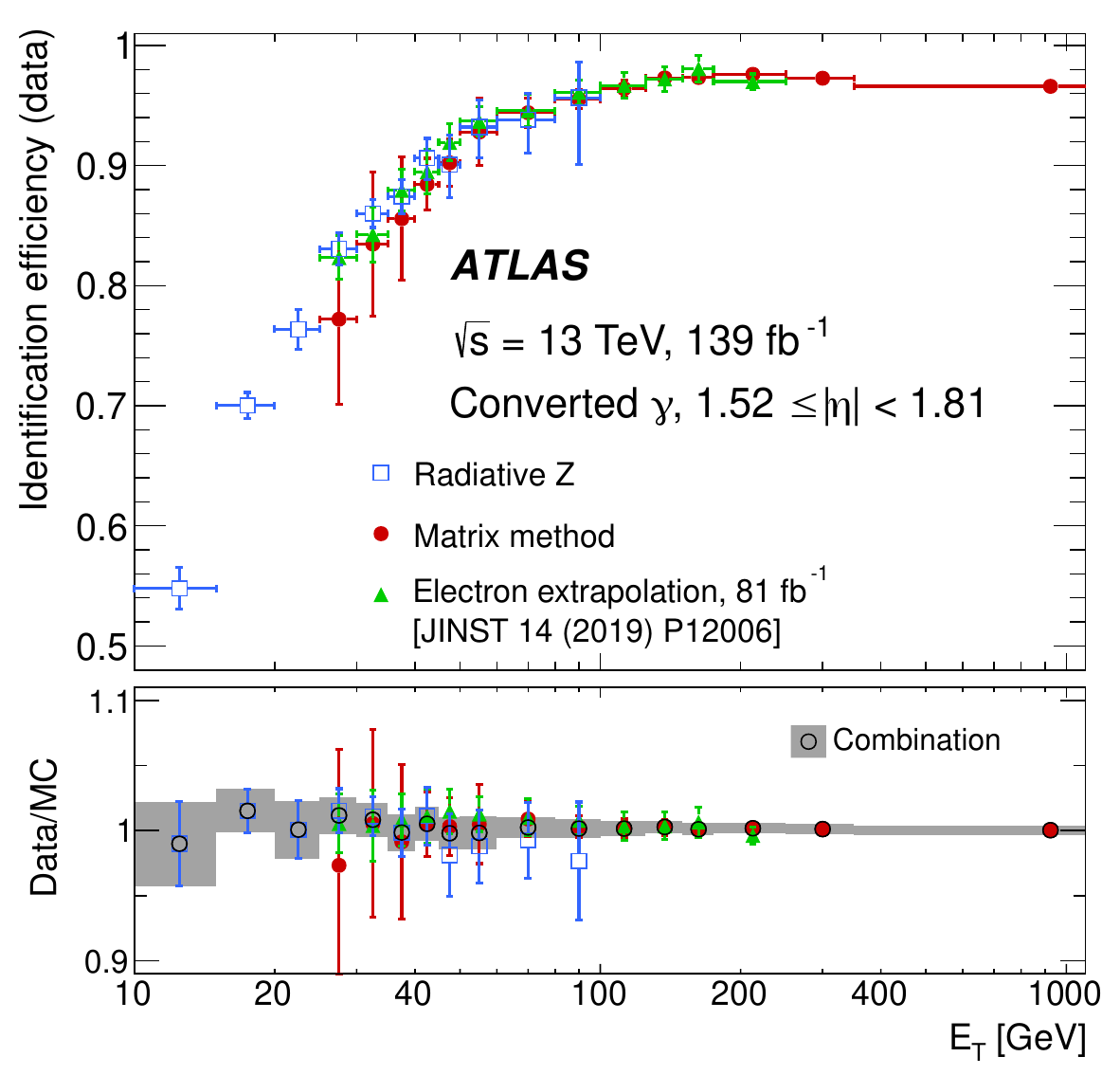}
}
\subfloat[]{
\includegraphics[width=0.45\textwidth]{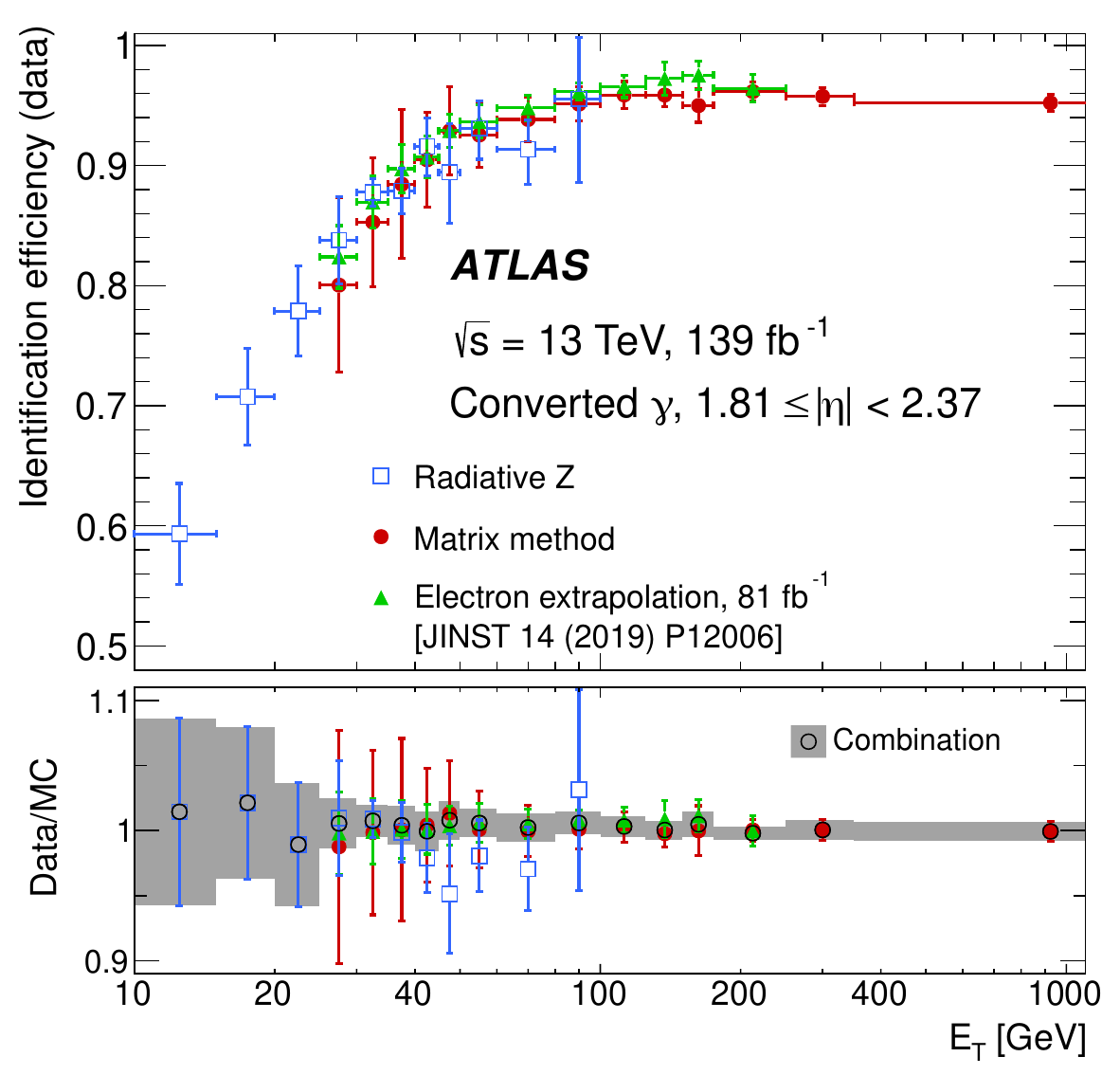}
}
\caption{The photon identification efficiency measured in data, and the ratio of data-to-MC simulation efficiencies, for converted photons with a Loose isolation requirement applied as preselection, as a function of \et in four different $|\eta|$ regions. The result for the electron extrapolation were not updated with regards to Ref.~\cite{EGAM-2018-01}. The combined scale factor, obtained using a weighted average of scale factors from the individual measurements, is also presented; the band represents the total uncertainty. }\label{fig:photon_conv}
\end{figure}


%

%
%
%
\FloatBarrier
\section{Conclusion}
\label{sec:conclusion}

Measurements of electron reconstruction, identification, and isolation efficiencies, and photon efficiencies have been presented. These use the full Run 2 data with a luminosity of 139~$\mathrm{fb}^{-1}$ collected by the ATLAS experiment during the years 2015--2018. The uncertainties on these results are up to 50\% smaller than the previously published measurements. The improvements have been achieved not only by including more data but more importantly by significant updates to the methodology of the measurements used, including optimising the background subtraction and the combination of measurements, and the update of the subtraction of pile-up contamination in the case of selections based on isolation variables. These measurements and the derived correction factors are to-date the most precise measurements using the Run 2 data and are the basis of ATLAS precision analyses in the Higgs boson, electroweak, and top quark sector. Improvements in the definitions of the isolation variables enhanced the stability of the signal efficiencies over the full $\eta$ range and significantly increased the rejection of non-prompt electrons at large transverse momenta. The presented measurements will have a significant impact on future ATLAS precision measurements.


%

%
\section*{Acknowledgements}
%

%
%

%
%

We thank CERN for the very successful operation of the LHC and its injectors, as well as the support staff at
CERN and at our institutions worldwide without whom ATLAS could not be operated efficiently.

The crucial computing support from all WLCG partners is acknowledged gratefully, in particular from CERN, the ATLAS Tier-1 facilities at TRIUMF/SFU (Canada), NDGF (Denmark, Norway, Sweden), CC-IN2P3 (France), KIT/GridKA (Germany), INFN-CNAF (Italy), NL-T1 (Netherlands), PIC (Spain), RAL (UK) and BNL (USA), the Tier-2 facilities worldwide and large non-WLCG resource providers. Major contributors of computing resources are listed in Ref.~\cite{ATL-SOFT-PUB-2023-001}.

We gratefully acknowledge the support of ANPCyT, Argentina; YerPhI, Armenia; ARC, Australia; BMWFW and FWF, Austria; ANAS, Azerbaijan; CNPq and FAPESP, Brazil; NSERC, NRC and CFI, Canada; CERN; ANID, Chile; CAS, MOST and NSFC, China; Minciencias, Colombia; MEYS CR, Czech Republic; DNRF and DNSRC, Denmark; IN2P3-CNRS and CEA-DRF/IRFU, France; SRNSFG, Georgia; BMBF, HGF and MPG, Germany; GSRI, Greece; RGC and Hong Kong SAR, China; ISF and Benoziyo Center, Israel; INFN, Italy; MEXT and JSPS, Japan; CNRST, Morocco; NWO, Netherlands; RCN, Norway; MEiN, Poland; FCT, Portugal; MNE/IFA, Romania; MESTD, Serbia; MSSR, Slovakia; ARRS and MIZ\v{S}, Slovenia; DSI/NRF, South Africa; MICINN, Spain; SRC and Wallenberg Foundation, Sweden; SERI, SNSF and Cantons of Bern and Geneva, Switzerland; MOST, Taipei; TENMAK, T\"urkiye; STFC, United Kingdom; DOE and NSF, United States of America.

Individual groups and members have received support from BCKDF, CANARIE, CRC and DRAC, Canada; CERN-CZ, PRIMUS 21/SCI/017 and UNCE SCI/013, Czech Republic; COST, ERC, ERDF, Horizon 2020, ICSC-NextGenerationEU and Marie Sk{\l}odowska-Curie Actions, European Union; Investissements d'Avenir Labex, Investissements d'Avenir Idex and ANR, France; DFG and AvH Foundation, Germany; Herakleitos, Thales and Aristeia programmes co-financed by EU-ESF and the Greek NSRF, Greece; BSF-NSF and MINERVA, Israel; Norwegian Financial Mechanism 2014-2021, Norway; NCN and NAWA, Poland; La Caixa Banking Foundation, CERCA Programme Generalitat de Catalunya and PROMETEO and GenT Programmes Generalitat Valenciana, Spain; G\"{o}ran Gustafssons Stiftelse, Sweden; The Royal Society and Leverhulme Trust, United Kingdom.

In addition, individual members wish to acknowledge support from Chile: Agencia Nacional de Investigación y Desarrollo (FONDECYT 1190886, FONDECYT 1210400, FONDECYT 1230987); China: National Natural Science Foundation of China (NSFC - 12175119, NSFC 12275265); European Union: European Research Council (ERC - 948254), Horizon 2020 Framework Programme (MUCCA - CHIST-ERA-19-XAI-00), Italian Center for High Performance Computing, Big Data and Quantum Computing (ICSC, NextGenerationEU), Marie Sklodowska-Curie Actions (EU H2020 MSC IF GRANT NO 101033496); France: Agence Nationale de la Recherche (ANR-20-CE31-0013, ANR-21-CE31-0013, ANR-21-CE31-0022), Investissements d'Avenir Idex (ANR-11-LABX-0012), Investissements d'Avenir Labex (ANR-11-LABX-0012); Germany: Baden-Württemberg Stiftung (BW Stiftung-Postdoc Eliteprogramme), Deutsche Forschungsgemeinschaft (DFG - CR 312/5-1); Italy: Istituto Nazionale di Fisica Nucleare (FELLINI G.A. n. 754496, ICSC, NextGenerationEU); Japan: Japan Society for the Promotion of Science (JSPS KAKENHI 22H01227, JSPS KAKENHI JP21H05085, JSPS KAKENHI JP22H04944); Netherlands: Netherlands Organisation for Scientific Research (NWO Veni 2020 - VI.Veni.202.179); Norway: Research Council of Norway (RCN-314472); Poland: Polish National Agency for Academic Exchange (PPN/PPO/2020/1/00002/U/00001), Polish National Science Centre (NCN 2021/42/E/ST2/00350, NCN UMO-2019/34/E/ST2/00393, UMO-2020/37/B/ST2/01043, UMO-2021/40/C/ST2/00187); Slovenia: Slovenian Research Agency (ARIS grant J1-3010); Spain: BBVA Foundation (LEO22-1-603), Generalitat Valenciana (Artemisa, FEDER, IDIFEDER/2018/048), La Caixa Banking Foundation (LCF/BQ/PI20/11760025), Ministry of Science and Innovation (RYC2019-028510-I, RYC2020-030254-I), PROMETEO and GenT Programmes Generalitat Valenciana (CIDEGENT/2019/023, CIDEGENT/2019/027); Sweden: Swedish Research Council (VR 2022-03845, VR 2022-04683), Knut and Alice Wallenberg Foundation (KAW 2017.0100, KAW 2018.0157, KAW 2019.0447); Switzerland: Swiss National Science Foundation (SNSF - PCEFP2\_194658); United Kingdom: Leverhulme Trust (Leverhulme Trust RPG-2020-004); United States of America: Neubauer Family Foundation.

%
%


%
%
%
\printbibliography

\clearpage
 
\begin{flushleft}
\hypersetup{urlcolor=black}
{\Large The ATLAS Collaboration}

\bigskip

\AtlasOrcid[0000-0002-6665-4934]{G.~Aad}$^\textrm{\scriptsize 102}$,
\AtlasOrcid[0000-0002-5888-2734]{B.~Abbott}$^\textrm{\scriptsize 120}$,
\AtlasOrcid[0000-0002-1002-1652]{K.~Abeling}$^\textrm{\scriptsize 55}$,
\AtlasOrcid[0000-0001-5763-2760]{N.J.~Abicht}$^\textrm{\scriptsize 49}$,
\AtlasOrcid[0000-0002-8496-9294]{S.H.~Abidi}$^\textrm{\scriptsize 29}$,
\AtlasOrcid[0000-0002-9987-2292]{A.~Aboulhorma}$^\textrm{\scriptsize 35e}$,
\AtlasOrcid[0000-0001-5329-6640]{H.~Abramowicz}$^\textrm{\scriptsize 151}$,
\AtlasOrcid[0000-0002-1599-2896]{H.~Abreu}$^\textrm{\scriptsize 150}$,
\AtlasOrcid[0000-0003-0403-3697]{Y.~Abulaiti}$^\textrm{\scriptsize 117}$,
\AtlasOrcid[0000-0002-8588-9157]{B.S.~Acharya}$^\textrm{\scriptsize 69a,69b,m}$
% ...
\bigskip
\\

% ...
$^{29}$Physics Department, Brookhaven National Laboratory, Upton NY; United States of America.\\
% ...
$^{35}$$^{(a)}$Facult\'e des Sciences Ain Chock, R\'eseau Universitaire de Physique des Hautes Energies - Universit\'e Hassan II, Casablanca;$^{(b)}$Facult\'{e} des Sciences, Universit\'{e} Ibn-Tofail, K\'{e}nitra;$^{(c)}$Facult\'e des Sciences Semlalia, Universit\'e Cadi Ayyad, LPHEA-Marrakech;$^{(d)}$LPMR, Facult\'e des Sciences, Universit\'e Mohamed Premier, Oujda;$^{(e)}$Facult\'e des sciences, Universit\'e Mohammed V, Rabat;$^{(f)}$Institute of Applied Physics, Mohammed VI Polytechnic University, Ben Guerir; Morocco.\\
% ...
$^{49}$Fakult\"{a}t Physik , Technische Universit{\"a}t Dortmund, Dortmund; Germany.\\
% ...
$^{55}$II. Physikalisches Institut, Georg-August-Universit\"{a}t G\"ottingen, G\"ottingen; Germany.\\
% ...
$^{69}$$^{(a)}$INFN Gruppo Collegato di Udine, Sezione di Trieste, Udine;$^{(b)}$ICTP, Trieste;$^{(c)}$Dipartimento Politecnico di Ingegneria e Architettura, Universit\`a di Udine, Udine; Italy.\\
% ...
$^{102}$CPPM, Aix-Marseille Universit\'e, CNRS/IN2P3, Marseille; France.\\
% ...
$^{117}$Department of Physics, New York University, New York NY; United States of America.\\
% ...
$^{120}$Homer L. Dodge Department of Physics and Astronomy, University of Oklahoma, Norman OK; United States of America.\\
% ...
$^{150}$Department of Physics, Technion, Israel Institute of Technology, Haifa; Israel.\\
$^{151}$Raymond and Beverly Sackler School of Physics and Astronomy, Tel Aviv University, Tel Aviv; Israel.\\
% ...
$^{a}$ Also Affiliated with an institute covered by a cooperation agreement with CERN.\\
$^{b}$ Also at An-Najah National University, Nablus; Palestine.\\
$^{c}$ Also at Borough of Manhattan Community College, City University of New York, New York NY; United States of America.\\
$^{d}$ Also at Center for High Energy Physics, Peking University; China.\\
$^{e}$ Also at Center for Interdisciplinary Research and Innovation (CIRI-AUTH), Thessaloniki; Greece.\\
$^{f}$ Also at Centro Studi e Ricerche Enrico Fermi; Italy.\\
% ...
$^{m}$ Also at Department of Physics, King's College London, London; United Kingdom.\\
% ...

\end{flushleft}


%
%

%

\end{document}